\newif\ifarxiv\arxivtrue
\ifarxiv\pdfoutput=1\fi
{\def\usepackage{ws-procs9x6}}
\documentclass[12pt,a4paper]{article}

\ifarxiv\ifnum\pdfoutput=1\else
\PassOptionsToPackage{hypertex}{hyperref}
\PassOptionsToPackage{draft}{graphicx}
\usepackage{showkeys}
\fi\fi

\usepackage{amsmath,amssymb}

\usepackage[bookmarks=true,hyperfigures=true]{hyperref}
\usepackage{graphicx}
\usepackage[nosort]{cite}
\usepackage[bulletsep]{collref}
\usepackage{color}

\providecommand{\hypersetup}[1]{}
\hypersetup{pdftitle={Long-Range Deformations for Integrable Spin Chains}}
\hypersetup{pdfauthor={Till Bargheer, Niklas Beisert, Florian Loebbert}}
\hypersetup{pdfsubject={}}
\hypersetup{pdfkeywords={Integrable Spin Chains, Long-Range Interactions, Boost Operators}}
\hypersetup{plainpages=false}
\hypersetup{pdfpagemode=UseNone}
\hypersetup{bookmarksnumbered=true}
\hypersetup{pdfstartview=FitH}
\hypersetup{colorlinks=false}
\hypersetup{citebordercolor={.5 1 .5}}
\hypersetup{urlbordercolor={.5 1 1}}
\hypersetup{linkbordercolor={1 .7 .7}}
\hypersetup{linktocpage=true}

\usepackage[a4paper,left=2.5cm,right=2.5cm,top=2.5cm,bottom=2.5cm]{geometry}

\usepackage[font=small,labelfont=bf,width=0.85\textwidth]{caption}

\numberwithin{equation}{section}

\expandafter\def\expandafter\bfseries\expandafter{\bfseries\ifmmode\else\boldmath\fi}
\expandafter\def\expandafter\mdseries\expandafter{\mdseries\ifmmode\else\unboldmath\fi}
\expandafter\def\expandafter\normalfont\expandafter{\normalfont\ifmmode\else\unboldmath\fi}

\allowdisplaybreaks[3]

\usepackage{fixmath}

\definecolor{gray}{gray}{.4}
\newcommand{\gray}[1]{{\color{gray}#1}}

\newcommand{\op}[1]{\mathcal{#1}}
\newcommand{\spc}[1]{\mathbb{#1}}
\newcommand{\ham}{\op{H}}
\newcommand{\rmat}{\op{R}}
\newcommand{\transfer}{\op{T}}
\newcommand{\shift}{\op{U}}
\newcommand{\charge}{\op{Q}}
\newcommand{\yang}{\op{Y}}
\newcommand{\liegen}{\op{J}}
\newcommand{\loc}{\op{L}}
\newcommand{\idop}{\op{I}}
\newcommand{\permop}{\op{P}}
\newcommand{\lengthop}{\op{N}}
\newcommand{\crossings}[1]{\mathopen{\rangle}#1\mathclose{\langle}}
\newcommand{\range}[1]{|#1|}
\newcommand{\bigrange}[1]{\bigl|#1\bigr|}
\newcommand{\boost}[1]{\op{B}[#1]}
\newcommand{\boostop}{\op{P}}
\newcommand{\biloc}[2]{[#1|#2]}
\newcommand{\bigbiloc}[2]{\big[#1\big|#2\big]}
\newcommand{\biop}{\op{Y}}
\newcommand{\rotgen}{\mathfrak{G}}
\newcommand{\basgen}{\mathfrak{E}}
\newcommand{\defop}{\op{X}}
\newcommand{\superN}{\mathcal{N}}
\newcommand{\cder}{\op{D}}

\newcommand{\defform}{\Xi}
\newcommand{\boostform}{\Pi}
\newcommand{\biform}{\Upsilon}
\newcommand{\rotform}{\Gamma}
\newcommand{\locform}{\Lambda}

\newcommand{\perm}[1]{[#1]}
\newcommand{\permB}[1]{\boost{\perm{#1}}}

\newcommand{\Tr}{\mathop{\mathrm{Tr}}}
\newcommand{\ad}{\mathop{\mathrm{ad}}}

\newcommand{\step}{\mathrm{\Theta}}

\newcommand{\order}[1]{\mathcal{O}(#1)}

\newcommand{\Integers}{\mathbb{Z}}
\newcommand{\Complex}{\mathbb{C}}

\newcommand{\cp}{\mathbb{CP}}

\newcommand{\dd}{d}

\newcommand{\specleg}{%
\begin{picture}(4,11)(0,2)%
\put(2,1){\line(0,1){9}}%
\put(2,1){\circle*{2}}%
\put(2,10){\circle*{2}}%
\end{picture}%
}

\newcommand{\MMM}[2]{{\arraycolsep0pt\begin{array}[b]{c}\makebox[0cm]{$\atopfrac{#2}{\downarrow}$}\\#1\end{array}}}

\makeatletter
\newlength{\apb@width}
\newcommand{\autoparbox}[2][c]{\settowidth{\apb@width}{#2}\parbox[#1]{\apb@width}{#2}}
\newcommand{\includegraphicsbox}[2][]{\autoparbox{\includegraphics[#1]{#2}}}
\makeatother

\ifx\genfrac\sdflkaj
\newcommand{\atopfrac}[2]{{{#1}\above0pt{#2}}}
\else
\newcommand{\atopfrac}[2]{\genfrac{}{}{0pt}{}{#1}{#2}}
\fi
\newcommand{\sfrac}[2]{{\textstyle\frac{#1}{#2}}}
\newcommand{\half}{\sfrac{1}{2}}
\newcommand{\ihalf}{\sfrac{i}{2}}
\newcommand{\quarter}{\sfrac{1}{4}}
\newcommand{\iquarter}{\sfrac{i}{4}}

\newcommand{\indup}[1]{_{\mathrm{#1}}}

\newcommand{\supup}[1]{^{\mathrm{#1}}}
\newcommand{\supups}[1]{^{\mathrm{\scriptscriptstyle #1}}}

\newcommand{\brk}[1]{(#1)}
\newcommand{\lrbrk}[1]{\left(#1\right)}
\newcommand{\bigbrk}[1]{\bigl(#1\bigr)}

\newcommand{\biggbrk}[1]{\biggl(#1\biggr)}
\newcommand{\brc}[1]{\{#1\}}
\newcommand{\lrbrc}[1]{\left\{#1\right\}}

\newcommand{\floor}[1]{\lfloor#1\rfloor}

\newcommand{\ceil}[1]{\lceil#1\rceil}

\newcommand{\vev}[1]{\langle#1\rangle}

\newcommand{\comm}[2]{[#1,#2]}

\newcommand{\bigcomm}[2]{\big[#1,#2\big]}
\newcommand{\acomm}[2]{\{#1,#2\}}
\newcommand{\bigacomm}[2]{\big\{#1,#2\big\}}

\newcommand{\PTerm}[1]{[#1]}

\newcommand{\set}[1]{\{#1\}}
\newcommand{\state}[1]{\mathopen{|}#1\mathclose{\rangle}}

\newcommand{\ket}[1]{|#1\rangle}

\newcommand{\alg}[1]{\mathfrak{#1}}
\newcommand{\grp}[1]{\mathrm{#1}}

\newcommand{\nn}{\nonumber}
\newcommand{\nln}{\nonumber\\}
\newcommand{\nl}[1][0pt]{\nonumber\\[#1]&\hspace{-4\arraycolsep}&\mathord{}}

\newcommand{\earel}[1]{\mathrel{}&\hspace{-2\arraycolsep}#1\hspace{-2\arraycolsep}&\mathrel{}}
\newcommand{\eq}{\earel{=}}

\newcommand{\defeq}{\mathrel{\mathop:}=}

\def\[{\begin{equation}}
\def\]{\end{equation}}
\def\<{\begin{eqnarray}}
\def\>{\end{eqnarray}}

\makeatletter
\def\mr@ignsp#1 {\ifx\:#1\@empty\else #1\expandafter\mr@ignsp\fi}%
\newcommand{\multiref}[1]{\begingroup
\xdef\mr@no@sparg{\expandafter\mr@ignsp#1 \: }%
\def\mr@comma{}%
\@for\mr@refs:=\mr@no@sparg\do{\mr@comma\def\mr@comma{,}\ref{\mr@refs}}%
\endgroup}
\makeatother

\newcommand{\sectionname}{Section}

\newcommand{\hypref}[2]{\ifx\href\asklfhas #2\else\href{#1}{#2}\fi}
\newcommand{\Secref}[1]{\sectionname~\multiref{#1}}
\newcommand{\Appref}[1]{\appendixname~\multiref{#1}}
\newcommand{\Tabref}[1]{\tablename~\multiref{#1}}
\newcommand{\Figref}[1]{\figurename~\multiref{#1}}
\renewcommand{\eqref}[1]{(\multiref{#1})}

\providecommand{\href}[2]{#2}
\newcommand{\arxivlink}[1]{\href{http://arxiv.org/abs/#1}{arxiv:#1}}

\begin{document}


\thispagestyle{empty}
\begin{flushright}\footnotesize
\texttt{\arxivlink{0902.0956}}\\
\texttt{AEI-2009-009}%
\end{flushright}
\vspace{1cm}

\begin{center}%
{\Large\textbf{\mathversion{bold}%
Long-Range Deformations\\for Integrable Spin Chains%
}\par}
\vspace{1cm}%

\textsc{Till Bargheer, Niklas Beisert, Florian Loebbert}\vspace{5mm}%

\textit{Max-Planck-Institut f\"ur Gravitationsphysik\\%
Albert-Einstein-Institut\\%
Am M\"uhlenberg 1, 14476 Potsdam, Germany}\vspace{3mm}%

\{\texttt{bargheer,nbeisert,loebbert}\}\texttt{@aei.mpg.de}
\par\vspace{1cm}

\includegraphics[scale=.9]{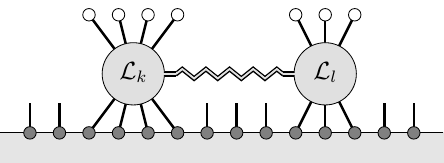}

\vspace{1cm}

\textbf{Abstract}\vspace{7mm}

\begin{minipage}{12.7cm}
We present a recursion relation for the explicit construction of
integrable spin chain Hamiltonians with long-range interactions. 
Based on arbitrary short-range (e.g.\
nearest-neighbor) integrable spin chains, it allows to construct an
infinite set of  conserved long-range charges. 
We explain the moduli space of deformation parameters by different classes of
generating operators. The rapidity map and dressing phase in the long-range Bethe equations 
are a result of these deformations.
The closed chain asymptotic Bethe equations for long-range
spin chains transforming under a generic symmetry algebra are derived.
Notably, our construction applies to generalizations of standard
nearest-neighbor chains such as alternating spin chains. We also discuss
relevant properties for its application to planar $D=4$, $\superN=4$ and
$D=3$, $\superN=6$ supersymmetric gauge theories. 
Finally, we present a map between long-range and inhomogeneous spin chains 
delivering more insight into the structures of these models as well as 
their limitations at wrapping order.
\end{minipage}

\end{center}

\newpage

\setcounter{tocdepth}{1}
\hrule height 0.75pt
\tableofcontents
\vspace{0.8cm}
\hrule height 0.75pt
\vspace{1cm}

\setcounter{tocdepth}{2}


\section{Introduction and Overview}

Much progress has been made in the last few years towards a verification
of the proposed duality \cite{Maldacena:1998re,Witten:1998qj}
between planar $\superN=4$ super
Yang--Mills theory (SYM) and type IIB string theory on
$\grp{AdS}_5\times\grp{S}^5$.
In this venture, integrability turned out
to be an important feature of the spin chain structure underlying both
theories in the planar limit
\cite{Minahan:2002ve,Beisert:2003tq,Bena:2003wd,Beisert:2003yb}. 
A novel class of \emph{long-range} spin chains had to be considered to reflect
the complexity of the proposedly dual theories
\cite{Beisert:2003tq,Beisert:2003yb,Beisert:2004hm}.
A similar class of integrable chains \cite{Minahan:2008hf} 
appears to be the key to rapid progress
within the recently conjectured duality 
\cite{Aharony:2008ug} between $\superN=6$ superconformal Chern--Simons
theory (SCS) and type IIA
string theory on $\grp{AdS}_4\times\cp^3$.

It is remarkable that the only well studied examples of long-range
spin chains come from a completely different branch of physics: 
The condensed matter models described by Haldane--Shastry
\cite{Haldane:1988gg,SriramShastry:1988gh} and Inozemtsev
\cite{Inozemtsev:1989yq,Inozemtsev:2002vb} incorporate interactions of
well-separated as opposed to nearest-neighbor sites of the chain. These
interactions, however, involve only two spins at a time, while the
more general long-range operators arising from gauge/string duality
act on several sites at the same time. Extensions of the Haldane--Shastry chain to multi-site interactions were also studied
\cite{BasuMallick:1996ki,BasuMallick:1999aa}. Their nature, however, is different from the 
gauge/string theory inspired interactions. Nevertheless, intersections
among the condensed matter and high-energy spin chain models are
inherent and give rise to a fruitful overlap of interests.

In this work we investigate 
integrable long-range spin chain models
from a general point of view.
Our class of models includes as special cases the 
above mentioned spin chains.
The structure of the models can be motivated best by its origin in gauge theory:
Trace operators providing a basis for all local
gauge-invariant
operators of $\superN=4$ super Yang-Mills theory, are mapped to spin chain states, e.g.\
\begin{equation}
\mathrm{Tr}\,[\phi_1\phi_2\phi_2\phi_1\phi_1\phi_2]
\to\ket{\uparrow\,\downarrow\,\downarrow\,\uparrow\,\uparrow\,\downarrow\,}\,.
\end{equation}
That is, \emph{fields} transforming under some spacetime symmetry are mapped to
\emph{spins} transforming under the same symmetry algebra. It was shown that
the one-loop Hamiltonian of planar $\superN=4$ SYM is equivalent to a
nearest-neighbor spin chain Hamiltonian \cite{Minahan:2002ve,Beisert:2003jj}. 
The spectral problem of the gauge theory therefore becomes
equivalent to the spectral problem of the spin chain.
Excitingly the Hamiltonian turns out to be integrable \cite{Minahan:2002ve,Beisert:2003yb}
leading to remarkable simplifications in the computation of the spectrum.
Similar observations were made for $\superN=6$ SCS theory 
\cite{Minahan:2008hf,Bak:2008cp,Zwiebel:2009vb,Minahan:2009te}. 
For several subsectors of $\superN=4$ SYM,
the correspondence was generalized to higher loop orders
\cite{Beisert:2003tq,Beisert:2003ys,Zwiebel:2005er}.
Higher powers of the 't Hooft coupling constant $\lambda$ arising from vertices in planar Feynman
diagrams indicate an increasing interaction range of the spin chain Hamiltonian:
\[\ham(\lambda)=
\includegraphicsbox[scale=.8]{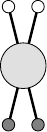}
+\lambda
\includegraphicsbox[scale=.8]{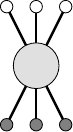}
+\lambda^2
\includegraphicsbox[scale=.8]{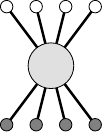}
+\lambda^3\dots.
\label{eq:pichamlambda}
\]
While the nearest-neighbor Hamiltonian $\ham(\lambda=0)$ only acts on
two neighboring spin sites at a time, a contribution at order
$\lambda^k$ is allowed to have interactions among at most $k+2$ spins.
For finite coupling $\lambda$, the Hamiltonian would actually be of
infinite range which led to the notion of \emph{long-range} spin
chains. This class of chains can be considered as a long-range
deformation of the prime example of a spin chain, the Heisenberg model 
\cite{Heisenberg:1928aa}.

Generally, the integrability of a spin chain system manifests itself in the
existence of an infinite tower of local conserved charges $\charge_r$, 
the first of which is usually the Hamiltonian
$\charge_2(\lambda)=\ham(\lambda)$:
\begin{equation}
 	\comm{\charge_r(\lambda)}{\charge_s(\lambda)}=0\,,\quad r,s=2,3,\dots.
\label{eq:chargecomm}
\end{equation}
In the magnon excitation picture around the ferromagnetic vacuum
this implies the factorization of multi-magnon scattering into
two-magnon scattering. Admissible eigenstates of the Hamiltonian on a
finite periodic chain of length $N$ can then be constructed by means of the
Bethe ansatz \cite{Bethe:1931hc}, which was first introduced in 1931 for the Heisenberg
model with $\alg{su}(2)$ symmetry \cite{Heisenberg:1928aa}. 
It represents a periodicity condition on the magnon momenta $p_k$ 
or rapidities $u_k=\half\cot(\half p_k)$, 
in terms of the two-magnon scattering factor
$S\supups{NN}(u_k-u_j)=(u_k-u_j+i)/(u_k-u_j-i)$
which altogether quantizes the spectrum:
\begin{equation}
\frac{(u_{k}+\ihalf)^N}{(u_{k}-\ihalf)^N}
=\prod_{\textstyle\atopfrac{j=1}{j\neq k}}^{M}S\supups{NN}(u_k-u_j)\,,
\qquad
k=1,\dots,M\,.
\end{equation}
Similar to the spin chain charges, this ansatz was perturbatively generalized to the class of
long-range chains by deformations in the coupling constant 
\cite{Serban:2004jf,Beisert:2004hm}.
The two-magnon scattering factor was deformed by the so-called
dressing phase $\theta(u_k,u_j;\lambda)$ and a rapidity map
$x(u;\lambda)$ deforming the term $u_k\pm\ihalf$ was introduced.
The resulting long-range $\alg{su}(2)$ equations read
\begin{equation}
\frac{x(u_{k}+\ihalf)^N}{x(u_{k}-\ihalf)^N}
=\prod_{\textstyle\atopfrac{j=1}{j\neq k}}^{M}S\supups{NN}(u_k-u_j)\exp\bigbrk{2i\theta(u_{k},u_{j})}\,,\qquad k=1,\dots,M\,.
\label{eq:bethedeformedsu2}
\end{equation}
They were later extended to the complete spectrum of $\superN=4$ SYM
\cite{Beisert:2005fw} and $\superN=6$ SCS
\cite{Gromov:2008qe}.
These asymptotic Bethe equations describe the spectrum of the Hamiltonian
for long chains. For finite chains the interaction range of the
Hamiltonian necessarily exceeds the length of the spin chain at some
order of the coupling $\lambda$. Beyond this order it is not known if
or how the Bethe ansatz can be modified to incorporate these so called
wrapping interactions. 
For recent progress on the understanding of the wrapping problem 
in the special context of gauge/string duality see
\cite{Ambjorn:2005wa,Janik:2007wt,Arutyunov:2007tc,Minahan:2008re,Heller:2008at,
Gromov:2008ie,Fiamberti:2008sh,Bajnok:2008bm,Bajnok:2008qj,Gromov:2009tv,Beccaria:2009eq}. 

Recently, a first step was taken towards a more general exploration of
long-range integrable spin chains. In \cite{Beisert:2005wv} a study of the
$\alg{gl}(K)$ chain incorporating the gauge theory $\alg{su}(2)$
sector as a special case was presented: Starting with a generic ansatz for the
first two integrable charges and imposing \eqref{eq:chargecomm} on
this ansatz, the form of the charges was determined up to some
perturbative order in the coupling. 
The moduli space for closed chains was found to
be characterized by four different types of parameters: 
The parameters $\alpha_k$ govern the degrees of freedom corresponding 
to deformations of the dispersion relation by means of the rapidity map $x(u)$. 
The parameters $\beta_{r,s}$ account for deformations by means of the dressing phase
$\theta(u_k,u_j)$, cf.\ \eqref{eq:bethedeformedsu2}. The additional moduli
$\gamma_{m,n}$ and $\varepsilon_l$ correspond to linear combinations
of commuting charges and similarity
transformations of the integrable system, respectively, and have no impact on the spectrum. 
The most general integrable Hamiltonian 
$\ham=\ham(\alpha_k,\beta_{r,s},\gamma_{r,s},\varepsilon_k)$ then takes the schematic form
\[
\ham=
\gamma_{22}\lrbrk{\,\,\,
\includegraphicsbox[scale=.7]{PicTwoLeg}
+\alpha_3
\includegraphicsbox[scale=.7]{PicThreeLeg}
+\alpha_4
\includegraphicsbox[scale=.7]{PicFourLeg}
+\beta_{23}
\includegraphicsbox[scale=.7]{PicFourLeg}
+\varepsilon_1
\includegraphicsbox[scale=.7]{PicFourLeg}
+\alpha_3^2\dots+\ldots}
+\ldots.
\]
Generalizing the long-range gauge theory model to this chain, the single
deformation parameter $\lambda$ was replaced by four different sets of moduli
with different physical interpretations. A certain choice of the new
parameters as functions of $\lambda$ results in the special expansion
\eqref{eq:pichamlambda}. While the assumption of integrability of the
$\alg{gl}(K)$ chain in \cite{Beisert:2005wv} was based on the
existence of two commuting charges, a perturbative proof for its
integrability up to $\order{\lambda^3}$ was later presented in
\cite{Beisert:2007jv} by the construction of a Yangian representation. 
However, a formal setting for the understanding of the all-order integrable
long-range spin chain is still lacking. A rigorous mathematical
construction might provide a scenery for the understanding of
wrapping effects in a general context.

In this work we develop a framework for the construction of
long-range integrable spin chains of 
arbitrary Lie (super)algebra symmetry. We present a
recursion relation for the charges
$\charge_t(\alpha_k,\beta_{r,s},\gamma_{r,s},\varepsilon_k)$ whose
solutions are manifestly integrable to all orders in the deformation
parameters and cover the whole moduli space explored in
\cite{Beisert:2005wv}. This proves the all-order existence of an
integrable long-range model on infinite spin chains. Furthermore the recursion
allows the explicit construction of the integrable charge operators.
The four different types of moduli are related to four different types
of deformation operators. Analyzing their impact on short-range spin
chains explains the emergence of the dressing phase and the rapidity map in the
long-range Bethe equations. This work is an extension of the
considerations sketched in the letter \cite{Bargheer:2008jt}. 
The paper is structured as follows:

\emph{\Secref{sec:intspinchain}}: 
We introduce the framework of integrable long-range spin chains of infinite extent.

\emph{\Secref{sec:generalconst}}: 
A recursion relation is defined which induces a set of manifestly integrable long-range 
charges as deformations of a short-range system. Four different kinds
of deformation generators corresponding to the four different moduli
discussed above are presented.

\emph{\Secref{sec:geom}}:  
The geometry of the moduli space is analyzed. 
We investigate the curvature associated to the generating equation
and derive flatness conditions for the deformations.

\emph{\Secref{sec:intrange}}: 
We present a parametrization that minimizes the interaction range of
the charges at each order of the deformation moduli
and at the same time renders the space of deformed charges flat. 
This parametrization yields integrable charges as they occur 
in the gauge/string correspondence.  

\emph{\Secref{sec:betheansatz}}: 
It is demonstrated how the four different types of generators 
deform the nearest-neighbor Bethe ansatz. Most notably, 
the appearance of the rapidity map as well as the dressing phase is explained 
by classes of so called boost and bilocal operators, respectively. 
The result is given by the well known form of the asymptotic Bethe equations for finite periodic chains.

\emph{\Secref{sec:altchains}}: We demonstrate how to apply the
recursion relation to alternating spin chains. The alternating
$\alg{su}(2)\times\alg{su}(2)$ long-range Bethe equations as well as
the first orders of the $\alg{gl}(K\indup{e})\times\alg{gl}(K\indup{o})$ long-range
Hamiltonian are explicitly given.

\emph{\Secref{sec:inhchain}}: Finally and somewhat outside the main theme of this
paper we present a map relating \emph{homogeneous} 
long-range chains without dressing phase to \emph{inhomogeneous} chains.
The latter can be defined consistently even beyond wrapping order,
however, in almost arbitrary ways.

Most parts of the paper focus on deformations of operators on spin chains with \emph{infinite}
extent. In \Secref{sec:betheansatz}, parts of \Secref{sec:altchains} and \Secref{sec:inhchain}, we consider
implications of these deformations on \emph{finite} chains. 

\section{Integrable Spin Chain Models}
\label{sec:intspinchain}

\subsection{Integrable Spin Chains and Local Operators}
\label{sec:intchainloc}

A spin chain is a physical model based on a Hilbert space $\spc{H}$
which is a tensor product of identical vector spaces $\spc{V}_a=\spc{V}$:
\begin{equation}
\spc{H}=\dots\,\otimes \spc{V}_a\otimes \spc{V}_{a+1}\otimes \spc{V}_{a+2}\otimes \,\dots\,,
\qquad
\spc{V}_a=\spc{V}.
\end{equation}
The vector spaces $\spc{V}_a$ are labeled by consecutive integers $a$ 
describing the position along the chain;
neighboring sites of the chain have adjacent integer positions.
The chain can be finite and have open or periodic boundary conditions.
Alternatively, it can have infinite extent which is the case we shall consider
in the first half of this paper.
A basis of the Hilbert space $\spc{H}$ is given by states for 
which the spin at each site $a$ has a definite orientation $v_a$
being a basis vector of $\spc{V}$:
\begin{equation}
\ket{\dots,v_a,v_{a+1},v_{a+2},\dots}\in \spc{H}.
\end{equation}
Typically, there is also a Lie symmetry algebra $\alg{g}$
which can be represented on the vector space $\spc{V}$,
i.e.\ $\spc{V}$ can be considered as a (generalized) spin of $\alg{g}$.

The physical model is furthermore defined by a set of observables
which are linear operators $\op{A}$
acting on the Hilbert space
\[
\op{A}:\spc{H}\to\spc{H}.
\]
Typically, the operators have some well-defined transformation properties
under the symmetry algebra $\alg{g}$, i.e.\ they may be invariant
or transform in a certain representation.

\begin{figure}\centering
\includegraphics{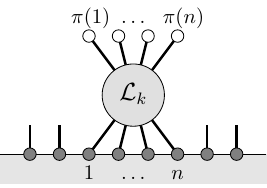}
\caption{A local operator $\loc_k$ acting on a spin chain. Its
position on the chain is summed over, see
\protect\eqref{eq:homogeneousaction}.}
\label{fig:local}
\end{figure}

We are mainly interested in operators that act \emph{locally} and \emph{homogeneously} 
on the spin chain and are \emph{invariant} under the symmetry $\alg{g}$.
We call these simply \emph{local operators} and denote them by
\begin{equation}
\loc_k \defeq \sum_a \loc_k(a),
\qquad
\loc_k(a):\spc{V}_a\otimes\ldots\otimes\spc{V}_{a+n-1}\to
\spc{V}_a\otimes\ldots\otimes\spc{V}_{a+n-1}.
\label{eq:homogeneousaction}
\end{equation}
Here $\loc_k(a)$ is some linear operator which acts on several consecutive
spins starting with site $a$, cf.\ \Figref{fig:local}. 
The number $n$ of interacting sites is called the interaction range 
of the operator $\loc_k$ and will be denoted by $\range{\loc_k}=n$.
These operators furnish the basic framework for the following considerations.

An \emph{integrable} spin chain model is defined
by an infinite tower of commuting charges
acting locally and homogeneously on an infinite chain
\begin{equation}\label{eq:intcond}
	\comm{\charge_r}{\charge_s}=0,\quad r,s=2,3,\dots.
\end{equation}
Integrability can also be defined for finite chains,
cf.\ \Secref{sec:betheansatz,sec:inhchain}, 
but for the time being we shall restrict ourselves to infinite chains.
Integrable spin chains are generally based on
some Lie algebra $\alg{g}$ or a quantum deformation thereof.
The algebra is a symmetry of the charges $\charge_r$. 
In fact, the symmetry typically extends to an
infinite-dimensional algebra
of Yangian or quantum-affine type \cite{Drinfeld:1985rx,Drinfeld:1986in}.

\paragraph{Local Operator Identifications.}

\begin{figure}\centering
\includegraphics{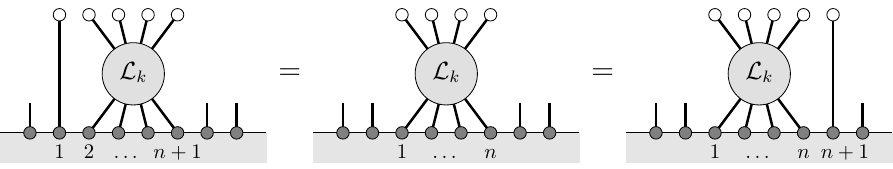}
\caption{On a spin chain without boundaries, local homogeneous
operators that differ only by spectator legs can be identified. The
position on the spin chain at which the operators act is implicitly
summed over.}
\label{fig:ibtl}
\end{figure}

The definition of homogeneous local 
operators in \eqref{eq:homogeneousaction}
allows for identifications of certain operators
on infinite chains.
For a local operator $\loc_k$ of range $n$ define
the local operators $\specleg\loc_k$ and $\loc_k\specleg$ 
of range $n+1$ as follows
\[\label{eq:spectator}
\specleg\loc_k(a):=\idop(a)\otimes \loc_k(a+1),
\qquad
\loc_k\specleg(a):=\loc_k(a)\otimes \idop(a+n),
\]
where $\idop(a)$ is the identity operator acting on site $a$.
The additional identity operators are called \emph{spectator legs}
of the local operator because their action is trivial.
Clearly, all three operators are equivalent
after the position $a$ is summed over in \eqref{eq:homogeneousaction},
see also \Figref{fig:ibtl},
\[
\label{eq:loceq}
\specleg\loc_k\simeq\loc_k\simeq \loc_k\specleg.
\]
Note that on finite chains these operators are equivalent only 
up to boundary terms.

\paragraph{Example.}

For illustration purposes we will make a specific choice of
the symmetry algebra and its representation at certain points of the text.
We will choose the fundamental representation of
$\alg{g}=\alg{gl}(K)$
on the vector space $\spc{V}=\Complex^K$. 
Spin orientations can then simply be denoted by numbers
corresponding to one of the $K$ directions, 
e.g.\ a basis vector of the Hilbert space is given by
\begin{equation}
	\ket{\dots,3,2,4,1,2,4,\dots}.	
\end{equation}

For this specific choice of algebra and representation,
a basis of invariant local operators $\loc_\pi$ consists of permutations $\pi$
of nearby spins: 
A permutation $\pi\in S_n$ of range $n$, mapping the spin sites
$(a+1,a+2,\dots,a+n)$ to the sites $(a+\pi(1),a+\pi(2),\dots,a+\pi(n))$, is denoted by
\begin{equation}
	\loc_\pi(a+1)=[\pi(1),\pi(2),\dots,\pi(n)]_{a+1}.
	\label{eq:local}
\end{equation}
The building blocks for $\alg{gl}(K)$-invariant operators are then
given by permutation symbols acting homogeneously on the whole chain
\begin{equation}
	\loc_\pi=[\pi(1),\pi(2),\dots,\pi(n)]=\sum_a[\pi(1),\pi(2),\dots,\pi(n)]_a.
\end{equation}
As an example, the operator $\PTerm{2,1}=\sum_a \permop_{a,a+1}$ 
sums over all pairs of nearest neighbors, $(a,a+1)$, and permutes them $(\permop_{a,a+1})$
\begin{equation}
	\PTerm{2,1}\ket{\dots,1,2,3,\dots}=\dots\,+\ket{\dots,2,1,3,\dots}+\ket{\dots,1,3,2,\dots}+\dots.
\end{equation}
Adding spectator legs to $\PTerm{2,1}$ according to \eqref{eq:spectator} 
yields the following two operators
\[
\specleg\PTerm{2,1}=\PTerm{1,3,2},\qquad
\PTerm{2,1}\specleg=\PTerm{2,1,3}.
\]

Note that $\alg{g}=\alg{gl}(K)$ only serves as an illustrative example
for the following ideas, most of which are valid for a generic
Lie symmetry algebra $\alg{g}$ and arbitrary spin representations.

\subsection{Nearest-Neighbor Models}
\label{sec:nnmodels}

An ordinary nearest-neighbor spin chain is characterized by
a Hamiltonian $\ham=\charge_2\supups{NN}$, which acts on two
adjacent sites at a time only. The set of commuting charges can then
be ordered by their interaction range starting with the Hamiltonian.
The charges are given by some linear combination of local
operators
\begin{equation}
	\charge_r\supups{NN}=\sum_k c_{r,k} \loc_k.
\end{equation}
The coefficients $c_{r,k}$ can be uniquely fixed
by a suitable normalization condition
but only modulo identification of spectator legs 
\eqref{eq:loceq}

\begin{figure}\centering
\includegraphics{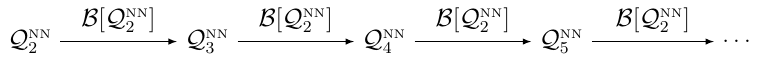}
\caption{The short-range charges are generated iteratively by the
so-called boost-operator $\boost{\charge_2\supups{NN}}$, cf.\
\protect\eqref{eq:nngen} and \protect\cite{Sogo:1983aa}.}
\label{fig:nnboost}
\end{figure}

Note that there exists an iterative definition of
the commuting charges \cite{Sogo:1983aa}:
Based on the nearest-neighbor Hamiltonian $\charge_2\supups{NN}$ 
one can define a so called boost operator
$\boost{\charge_2\supups{NN}}$ such that the integrable system is
generated by a single equation (cf.\ \Figref{fig:nnboost})
\begin{equation}\label{eq:nngen}
\charge_{r+1}\supups{NN} = 
-\frac{i}{r}\bigcomm{\boost{\charge_2\supups{NN}}}{\charge_r\supups{NN}}.
\end{equation}
The precise definition of $\boost{\cdot}$ 
will be discussed in \Secref{sec:gen.boostops}.
The interaction range of the charges following from this iteration relation
reads for a nearest-neighbor Hamiltonian with $\range{\charge\supups{NN}_2}=2$,
\[\label{eq:nngenrange}
\range{\charge\supups{NN}_r}=(r-1)\range{\charge\supups{NN}_2}-(r-2)=r.
\]
%

\paragraph{Example.}

For $\alg{g}=\alg{gl}(K)$ the first few charges take the form,
see e.g.\ \cite{Beisert:2005wv}
\begin{align}
	\charge_2\supups{NN}&=[1]-[2,1],\nln
	\charge_3\supups{NN}&=\ihalf ([3,1,2]-[2,3,1]),\nln
	\charge_4\supups{NN}&=\sfrac{1}{3} (-[1]+2[2,1]-[3,2,1]+[2,3,4,1]\nln
				&\qquad-[2,4,1,3]-[3,1,4,2]+[4,1,2,3]).
\label{eq:nncharges}
\end{align}

\subsection{Long-Range Models}
\label{sec:longrangemod}

Perturbatively long-ranged spin chains are formally defined as
deformations of the above nearest-neighbor chains. 
The nearest-neighbor charges $\charge_r\supups{NN}$ 
are taken to be the leading order $\charge_r^{(0)}$
in a formal power series
\begin{equation}
\charge_r(\lambda)
=\charge_r^{(0)}
+\lambda\charge_r^{(1)}
+\lambda^2\charge_r^{(2)}
+\order{\lambda^3},
\qquad
\charge_r^{(0)}=\charge_r\supups{NN},
\label{eq:expansionlambda}
\end{equation}
such that the interaction range of the charges grows linearly 
with the perturbative order in $\lambda$,
e.g.\ $\range{\charge_{r}^{(k)}}=r+k$ \cite{Beisert:2003tq}. 
Hence, assuming that $\lambda$ can take finite values, 
the charges $\charge_r(\lambda)$ would be of infinite (i.e.\ long) range. 

The long-range charges can still be written as linear combinations 
of local operators
\begin{equation}
	\charge_r(\lambda)=\sum_k c_{r,k}(\lambda) \loc_k,
\end{equation}
but now with coefficients $c_{r,k}(\lambda)$
which are formal power series in $\lambda$
starting at a certain order,
e.g.\ $c_{r,k}(\lambda)=\order{\lambda^{\range{\loc_k}-r}}$.
The charges have to obey the integrability condition \eqref{eq:intcond}
order by order in $\lambda$.

It is the aim of this paper to present an equation 
(similar to the recursion relation \eqref{eq:nngen}) 
that generates the long-range system through
deformations of the nearest-neighbor charges. The resulting long-range
spin chain model will be manifestly integrable.

\paragraph{Example.}

For fundamental $\alg{g}=\alg{gl}(K)$ chains, the charges take the form
\cite{Beisert:2005wv}
\begin{align}
	\charge_2(\lambda)&=[1]-[2,1]\nln
				&\qquad+\alpha_3\lambda\bigbrk{-3[1]+4[2,1]-[3,2,1]}+\order{\lambda^2},\nln
	\charge_3(\lambda)&=\ihalf ([3,1,2]-[2,3,1])\nln
				&\qquad+\ihalf \alpha_3\lambda\bigl(6[2,3,1]-6[3,1,2]+[4,1,3,2]\nln
				&\qquad\quad+[4,2,1,3]-[2,4,3,1]-[3,2,4,1]\bigr)+\order{\lambda^2}.
\label{eq:gllongcharge}
\end{align}
Note that in contrast to the nearest-neighbor chain not all
coefficients $c_{r,k}(\lambda)$ are fixed by integrability but some free
parameters $\xi_{k}=\set{\alpha_3,\ldots}$ remain. 
The latter are thus moduli of the long-range integrable system.

\section{Algebra-Preserving Deformations}
\label{sec:generalconst}

In the following a general mechanism for the construction of
long-range spin chains is presented. 
The key idea is a differential equation that generates long-range
charges as deformations of some short-range (e.g.\ nearest-neighbor) charges.
We present transformations of the integrable charges of an
\emph{infinite} short-range spin chain that do not leave the space of
local, homogeneous operators and preserve integrability. Later
(\Secref{sec:betheansatz}), we study finite chains, i.e.\ the impact
of the presented deformations on the boundary conditions.

The construction is applicable to short-range spin chains, 
which usually have a Lie symmetry algebra $\alg{g}$
under which the integrable charges are invariant. 
In the present work we make the assumption that the representation
of $\alg{g}$ on the spin chain remains undeformed.
We shall illustrate the various deformations
by means of examples for the
specific case of $\alg{g}=\alg{gl}(K)$
with spins transforming in the fundamental representation.

\subsection{Generating Equation}

Consider an algebra of charges $\charge_r(0)$
(not necessarily abelian)
and a deformation $\charge_r(\lambda)$ 
which obeys the differential equation
\begin{equation}
\frac{\dd}{\dd\lambda}\,\charge_r(\lambda) 
=
i\bigcomm{\defop(\lambda)}{\charge_r(\lambda)}.
\label{eq:deformationl}
\end{equation}
Here, $\defop(\lambda)$ is some yet-to-be specified operator
that has well-defined commutation relations 
with the charges $\charge_r(\lambda)$ for all $\lambda$,
in particular in the sense of a formal power 
series in $\lambda$.

The deformation \eqref{eq:deformationl} leaves the algebra of
charges $\charge_r(\lambda)$ unchanged: 
It implies by the Jacobi identity
\begin{equation}
\frac{\dd}{\dd\lambda}\,\bigcomm{\charge_r(\lambda)}{\charge_s(\lambda)} 
= i\bigcomm{\defop(\lambda)}{\comm{\charge_r(\lambda)}{\charge_s(\lambda)}}.
\label{eq:algebraconservation}
\end{equation}
For a generic algebra of independent operators $\charge_r$, this shows 
that the structure constants $f_{rst}$ do not change under the
deformation
\begin{equation}
\label{eq:fixstruct}
\bigcomm{\charge_r(\lambda)}{\charge_s(\lambda)} = f_{rst}\,\charge_t(\lambda),
\qquad
\frac{\dd}{\dd\lambda}\, f_{rst} = 0 \,.
\end{equation}

In particular, 
if the algebra of the initial charges $\charge_r(0)$ is abelian, $f_{rst}=0$, 
also the algebra of the deformed charges 
defined by \eqref{eq:deformationl} 
is abelian for all $\lambda$ and any $\defop(\lambda)$:
\begin{equation}
\bigcomm{\charge_r(0)}{\charge_s(0)}=0 \qquad\Longrightarrow\qquad 
\bigcomm{\charge_r(\lambda)}{\charge_s(\lambda)}=0 \,.
\end{equation}
Thus the differential equation \eqref{eq:deformationl} 
preserves the existence of commuting charges
which represents the most important property 
\eqref{eq:intcond} of any integrable system.
However, for integrable spin chains 
including our class of models
one requires stronger properties: 
The charges $\charge_r(\lambda)$ must act \emph{locally} and \emph{homogeneously}
\eqref{eq:homogeneousaction} on the spin chain. 
Hence, if the deformation \eqref{eq:deformationl} 
is to describe such models, the resulting charges 
$\charge_r(\lambda)$ must not violate these properties. 
That is, the operators $\defop(\lambda)$ must be chosen such that
\begin{equation}
\comm{\defop(\lambda)}{\charge_r(\lambda)} \quad\text{is local and homogeneous}
\label{eq:lochom}
\end{equation}
for all $\lambda$. 
The possible choices for $\defop(\lambda)$ that satisfy these requirements are
discussed in the following subsections.

Before we continue, we will construct a perturbative solution to
the generating equation \eqref{eq:deformationl}.
We first integrate it,
\begin{equation}
\charge_r(\lambda)
=\charge_r(0) 
+\int_0^\lambda \dd\lambda'\, i\bigcomm{\defop(\lambda')}{\charge_r(\lambda')}.
\end{equation}
Expansion into a power series in $\lambda$ then straightforwardly yields
\begin{eqnarray}
\charge_r(\lambda)\eq\charge_r^{(0)}+\lambda\charge_r^{(1)}+\lambda^2\charge_r^{(2)}+\lambda^3\charge_r^{(3)}+\ldots\,,
\nln
\defop(\lambda)\eq\defop^{(0)}+\lambda\defop^{(1)}+\lambda^2\defop^{(2)}+\lambda^3\defop^{(3)}+\ldots\,,
\nln
\charge_r^{(1)}\eq i\bigcomm{\defop^{(0)}}{\charge_r^{(0)}},
\nln
\charge_r^{(2)}\eq
\ihalf\bigcomm{\defop^{(1)}}{\charge^{(0)}_r}+\ihalf\bigcomm{\defop^{(0)}}{\charge^{(1)}_r},
\nln
\charge_r^{(3)}\eq\ldots\,.
\label{eq:recsol}
\end{eqnarray}
This shows that the higher orders of the charges
are completely determined through the lower orders by iteration.
The generating equation is thus a recursion relation
for the construction of an integrable system.

\subsection{Local Operators}
\label{sec:gen.localops}

The commutator of two local operators is again local and
homogeneous. Hence any local operator $\loc_l$
results in an admissible deformation $\defop=\loc_l$ 
\eqref{eq:deformationl} of the charges
\begin{equation}
\frac{\dd}{\dd\lambda}\,\charge_r(\lambda) = i\bigcomm{\loc_l}{\charge_r(\lambda)}\,.
\label{eq:deformationepsilon}
\end{equation}
Note that this deformation can be integrated exactly:
\begin{align}
\charge_r(\lambda)
&=\exp(+i\lambda\loc_l)\, \charge_r(0)\, \exp(-i\lambda\loc_l)\nln
&= (1+i\lambda\loc_l)\charge^{(0)}_r(1-i\lambda\loc_l)+\order{\lambda^2} \nln
&= \charge^{(0)}_r+i\lambda\bigcomm{\loc_l}{\charge_r^{(0)}}+\order{\lambda^2} \,.
\label{eq:simtrafo}
\end{align}
It thus merely constitutes a similarity transformation of the charges
$\charge_r$ by the operator $\exp(i\lambda\loc_l)$.
Importantly, as we shall see in \Secref{sec:betheansatz}, 
this deformation does not change any of the quantities
we are ultimately interested in. 
In the following we shall therefore disregard transformations
by local operators. In any case, one can easily 
reintroduce them at the very end through the similarity transformation
\eqref{eq:simtrafo}.

\paragraph{Degrees of Freedom.} 

When counting all degrees of freedom of the system, 
i.e.\ the total number of possible deformation operators $\defop$
(of a given maximum range), 
it has to be taken into account that the
conserved charges $\defop=\charge_r$ generate trivial deformations
due to the integrability condition \eqref{eq:intcond}.
Moreover, on a spin chain without boundaries, local operators that
differ by spectator legs only \eqref{eq:spectator}, 
must be identified, cf.\ \eqref{eq:loceq} and \Figref{fig:ibtl}.

\paragraph{Example.}

As an example of a deformation of the type
\eqref{eq:deformationepsilon}, consider the $\alg{gl}(K)$ spin chain.
As mentioned above \eqref{eq:local}, the only invariant
operators on this spin chain are permutations
\begin{equation}
\loc_k=\perm{\pi(1),\pi(2),\dots,\pi(n)}\,,
\end{equation}
which act homogeneously on the chain. Deforming the nearest-neighbor
Hamiltonian $\charge_2^{(0)}=\perm{1}-\perm{2,1}$ of this
chain by the local homogeneous operator $\loc=\perm{3,2,1}$ yields the
deformed Hamiltonian
\begin{align}
\charge_2
&=\charge_2^{(0)}+\lambda\charge_2^{(1)}+\order{\lambda^2},\nln
\charge_2^{(1)}
&=i\comm{\loc}{\charge_2^{(0)}}
 =i\bigcomm{\perm{3,2,1}}{\perm{2,1}}\nln
&=i\bigbrk{\perm{4,2,1,3}-\perm{3,2,4,1}+\perm{2,4,3,1}-\perm{4,1,3,2}}
\,.
\end{align}

\subsection{Boost Operators}
\label{sec:gen.boostops}

As mentioned above \eqref{eq:nngen}, 
it is well known \cite{Sogo:1983aa} that a set of
mutually commuting charges $\charge_r^{(0)}$ 
can be generated iteratively, 
starting from a nearest-neighbor Hamiltonian $\charge_2^{(0)}$. The higher
charges are constructed through commutation with the ``boosted''
Hamiltonian $\boost{\charge_2^{(0)}}$:
$\charge_{r+1}^{(0)}\sim\comm{\boost{\charge_2^{(0)}}}{\charge_r^{(0)}}$.
We define the boost $\boost{\loc_k}$ of a local operator 
$\loc_k$ as
\begin{equation}
\loc_k = \sum_a \loc_k(a) \qquad\Longrightarrow\qquad 
\boost{\loc_k} \defeq \sum_a a\,\loc_k(a) \,,
\label{eq:boostdef}
\end{equation}
where the local operator $\loc_k(a)$ acts on a set of adjacent
spins, starting at site $a$. Boost operators defined in this way act
locally, but \emph{inhomogeneously} on the spin chain.
For the purpose of a concise notation, we shall reserve the term
\emph{local operator} for local homogeneous operators; 
local inhomogeneous operators will be called \emph{boost operators}.

In general, the commutator of a boost operator with a local operator 
yields a combination of boost and local operators (cf.\ \Figref{fig:boostcomm}):
\begin{figure}\centering
\includegraphics{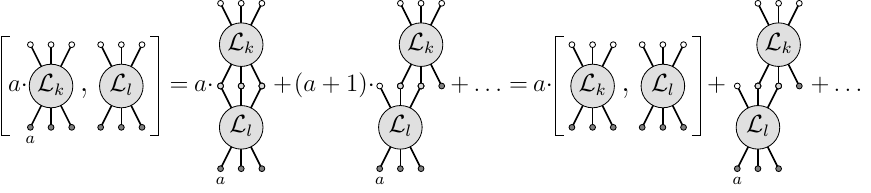}
\caption{Graphical representation of the commutator between a boost
operator $\boost{\loc_k}$ and a local operator $\loc_l$. The
contributions combine into the boost of the commutator
$\comm{\loc_k}{\loc_l}$ between the two local operators plus local
terms, cf.\ \protect\eqref{eq:boostcomm,eq:boostcommloc}. When the local operators commute,
only the local terms remain.}
\label{fig:boostcomm}
\end{figure}%
\begin{equation}
\bigcomm{\boost{\loc_k}}{\loc_l} = \boost{\comm{\loc_k}{\loc_l}} + \loc_r \,,
\label{eq:boostcomm}
\end{equation}
with $\loc_r$ some local operator.
However, if the underlying local operators commute, the
commutator becomes local:
\begin{equation}
\comm{\loc_k}{\loc_l} = 0
\qquad\Longrightarrow\qquad
\bigcomm{\boost{\loc_k}}{\loc_l} = \loc_r \,.
\label{eq:boostcommloc}
\end{equation}
Therefore, the boosts of the commuting charges themselves are
operators that yield admissible deformations \eqref{eq:deformationl}
\begin{equation}
\frac{\dd}{\dd\lambda}\,\charge_r(\lambda)
=i\bigcomm{\boost{\charge_k(\lambda)}}{\charge_r(\lambda)},
\quad k=3,\dots,\infty \,,
\label{eq:deformationboost}
\end{equation}
which result in charges $\charge_r(\lambda)$ that are homogeneous, as
desired. 

Note that the deformation operator
$\defop(\lambda)=\boost{\charge_k(\lambda)}$ directly depends on the charges $\charge_k(\lambda)$
which are themselves the solution of the generating equation \eqref{eq:deformationl}.
This implies, in particular, 
that boost deformations are not simply
exponential similarity transformations \eqref{eq:simtrafo}
of the undeformed charges; 
the dependence on the deformation parameter $\lambda$ is 
more involved.
Nevertheless, the equation as well as its 
recursive solution \eqref{eq:recsol}
with $\defop^{(n)}=\boost{\charge^{(n)}_k}$
remains perfectly well-defined.

\paragraph{Boost Operator Identifications.}

\begin{figure}\centering
\includegraphics{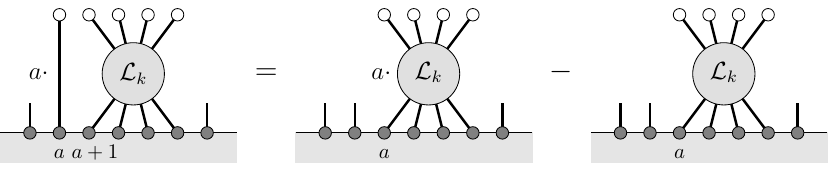}\par\vspace{0.5cm}
\includegraphics{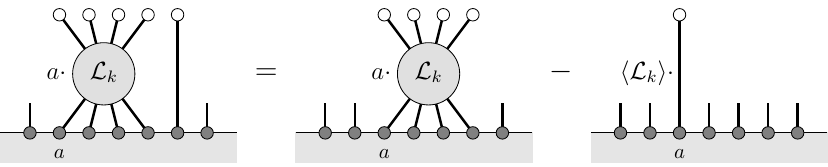}
\caption{Illustration of the boost operator identifications
\protect\eqref{eq:specboundboost} on an infinite chain.
As in the previous pictures, the position of
the local operators on the spin chain is summed over. The upper
identification is evident by \protect\eqref{eq:boostorigin}:
Putting a spectator leg to the left of the operator amounts to
shifting the origin (site $0$) of the chain by $-1$. The subtraction of the length
operator in the lower identity is required for a correct
normalization and stems from the different action of the two boost
operators at the right boundary of a finite chain. Using the symmetric
regularization of the boost operator \protect\eqref{eq:boostreg}, the
identifications become \protect\eqref{eq:ibtb}.}
\label{fig:ibtb}
\end{figure}

As illustrated in \Figref{fig:ibtl}, local operators on an infinitely
long chain that differ only by spectator legs can be identified
\eqref{eq:loceq}. But according to \eqref{eq:boostdef}, the
corresponding boost operators would differ by a local operator. This
is related to the fact that their definition depends on the location
of the spin chain origin (site 0). Namely, shifting the origin by $n$
sites leads to a shift of $\boost{\loc_k}$ by the local operator
$n\loc_k$
\begin{equation}
\boost{\loc_k}\to\boost{\loc_k}+n\loc_k.
\label{eq:boostorigin}
\end{equation}
When deforming with a boosted charge $\boost{\charge_k}$
\eqref{eq:deformationboost}, the location of the spin chain origin is
irrelevant because $\charge_k$ commutes with $\charge_r$.
On an open chain of length $N$ whose leftmost site is labeled by
$a=1$, we can thus regularize the boost operator \eqref{eq:boostdef} as
\begin{equation}
\boost{\loc_k}=\sum_a a\loc_k(a)-\half\bigbrk{N-\range{\loc_k}}\loc_k\,,
\label{eq:boostreg}
\end{equation}
where $\range{\loc_k}$ denotes the range of the operator $\loc_k$. 
This choice puts the origin in the middle of the finite chain 
such that $\boost{\loc_k}$ has exactly opposite 
parity of $\loc_k$.
On such an open chain, spectator legs on local operators yield boundary
terms:
\begin{equation}
\specleg\loc_k=\loc_k-\loc_k(1)\,,\qquad
\loc_k\specleg=\loc_k-\loc_k(N-\range{\loc_k}+1)\,.
\label{eq:specboundloc}
\end{equation}
Accordingly, boost operators are modified by spectator legs as (cf.\
also \Figref{fig:ibtb})
\begin{align}
\sum_aa(\specleg\loc_k)(a)&=\sum_aa\loc_k(a)-\loc_k\,,\nln
\sum_aa(\loc_k\specleg)(a)&=\sum_aa\loc_k(a)-(N-\range{\loc_k}+1)\loc_k(N-\range{\loc_k}+1)\,.
\label{eq:specboundboost}
\end{align}
In the limit $N\to\infty$, we can effectively replace the boundary operators $\loc_k(1)$,
$\loc_k(N-\range{\loc_k}+1)$ by their expectation values $\vev{\loc_k}$ 
on a ferromagnetic vacuum. Combining \eqref{eq:specboundloc} and
\eqref{eq:specboundboost}, we find that the relation between
equivalent boost operators on an infinite chain is given by%
\footnote{Without the regularization \eqref{eq:boostreg}, we would
have arrived at similar, but less symmetric identifications. 
For an elementary permutation
$\loc_\pi$ without a prefactor on a $\alg{gl}(K)$ chain, $\vev{\loc_\pi}=1$.}
\begin{equation}
\begin{aligned}
\boost{\specleg\loc_k} &= \boost{\loc_k} -\half\bigbrk{\loc_k -\vev{\loc_k} \lengthop}\,,\\
\boost{\loc_k\specleg} &= \boost{\loc_k} +\half\bigbrk{\loc_k -\vev{\loc_k}\lengthop}\,,
\end{aligned}
\qquad\qquad
\lengthop \defeq \sum_a \idop(a) \,.
\label{eq:ibtb}
\end{equation}
Here, the symbol $\idop(a)$ denotes the
identity operator acting at spin site $a$
and $\lengthop$ is thus the operator which measures
the length of the chain. The combination 
$\brk{\loc_k -\vev{\loc_k} \lengthop}$
is an operator annihilating the ferromagnetic vacuum.

\paragraph{Degrees of Freedom.}

A deformation with the boost of $\charge_2$ only leads to a
shift of the charge $\charge_r$ by the charge $\charge_{r+1}
\sim\comm{\boost{\charge_2}}{\charge_r}$
\eqref{eq:nngen}. The corresponding degrees of freedom are accounted
for by a different set of deformations (cf.\ \Secref{sec:rotations}) and
are hence not included in \eqref{eq:deformationboost}. 

Moreover, also the charges $\charge_k$ are defined only modulo
identification of spectator legs \eqref{eq:loceq}. Consequently, due
to the identifications \eqref{eq:ibtb}, also $\boost{\charge_k}$ is
defined modulo some local operators. This ambiguity is not a problem
though because all degrees of freedom resulting from deformations
through local operators have already been taken into account in
\eqref{eq:deformationepsilon}.

\paragraph{Example.}

As an example for the deformation \eqref{eq:deformationboost},
consider again the $\alg{gl}(K)$ spin chain with nearest-neighbor
Hamiltonian $\charge_2^{(0)}=\perm{1}-\perm{2,1}$. By virtue
of \eqref{eq:nngen}, the next higher commuting charge
$\charge_3^{(0)}$ is given by
\begin{align}
\charge_3^{(0)}
&= -\ihalf\bigcomm{\boost{\charge_2^{(0)}}}{\charge_2^{(0)}} \nln
&= -\ihalf\bigcomm{\boost{\perm{1}}-\boost{\perm{2,1}}}{\perm{1}-\perm{2,1}} \nln
&= -\ihalf\bigbrk{\perm{2,3,1} -\perm{3,1,2}} \,,
\end{align}
where for example $\boost{\perm{2,1}}$ is the boost of the permutation
operator $\perm{2,1}$.
Now, deforming $\charge_2^{(0)}$ with
$\boost{\charge_3^{(0)}}$, yields
\begin{align}
	\charge_2^{(1)}
	&=i\bigcomm{\boost{\charge_3^{(0)}}}{\charge_2^{(0)}}
\nln
	&=\half\bigcomm{\boost{\perm{2,3,1}}-\boost{\perm{3,1,2}}}{\perm{1}-\perm{2,1}}
\nln
	&=\half\bigbrk{-2\permB{1,3,2} +2\permB{2,1,3} -\perm{2,3,4,1} +\perm{2,4,1,3} +\perm{3,1,4,2} -\perm{4,1,2,3}}
\nln
	&=\half\bigbrk{-2\perm{1}+2\perm{2,1} -\perm{2,3,4,1}+\perm{2,4,1,3}+\perm{3,1,4,2}-\perm{4,1,2,3}}
\,,
\label{eq:boostexample}
\end{align}
where in the last line the two boost operators $\permB{1,3,2}$ and
$\permB{2,1,3}$ were identified according to the prescription
\eqref{eq:ibtb}. Deforming $\charge_3^{(0)}$ in the same
fashion results in
\begin{align}
\charge_3^{(1)}
&= i\bigcomm{\boost{\charge_3^{(0)}}}{\charge_3^{(0)}}
\nln
&= \iquarter\bigl(\perm{2,4,3,1}+\perm{3,2,4,1}-\perm{4,1,3,2}-\perm{4,2,1,3} 
\nln
		&\qquad -2\perm{2,3,4,5,1}+2\perm{2,3,5,1,4}+2\perm{2,4,1,5,3}-2\perm{2,5,1,3,4} 
\nln
		&\qquad +2\perm{3,1,4,5,2}-2\perm{3,1,5,2,4}-2\perm{4,1,2,5,3}+2\perm{5,1,2,3,4}\bigr)
 \,,
\end{align}
and the deformed $\charge_2(\lambda)$ and $\charge_3(\lambda)$ indeed commute up to
terms of order $\order{\lambda^2}$. Reinserting these
expressions into the differential equation \eqref{eq:deformationboost}
and further expanding in the deformation parameter $\lambda$
recursively yields the higher order terms of the deformed charges
$\charge_2(\lambda)$ and $\charge_3(\lambda)$.

\subsection{Bilocal Operators}

\begin{figure}\centering
\includegraphics{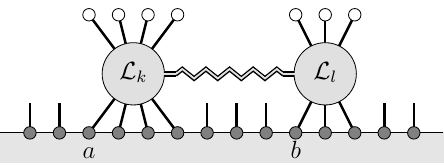}
\caption{A bilocal operator $\biloc{\loc_k}{\loc_l}$ can be
constructed from two local operators by summing over all positions
with $\loc_k$ acting on the left of $\loc_l$, $a\leq b$.}
\label{fig:biloc}
\end{figure}
\begin{figure}\centering
\includegraphics{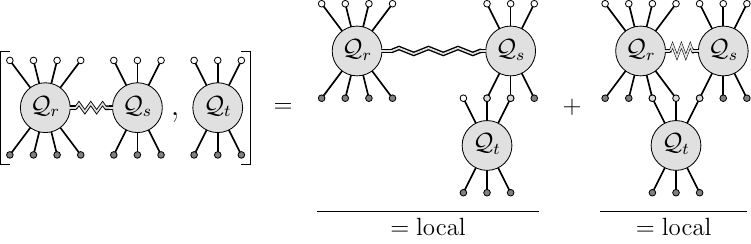}
\caption{The commutator of a bilocal operator composed of two local
charges $\charge_r$ and $\charge_s$ with a local charge $\charge_t$
gives a local result: When the two parts of the bilocal operator
are well separated, the commutator either vanishes
(if the two local charges commute with each other), or yields a bilocal
operator in which one part consists of boundary terms only
(if the two local charges commute up to boundary terms), which
on an infinite chain amounts to a local operator. Also when both parts
of the bilocal operator interact with the local charge at the same
time, one gets a nonvanishing local contribution.}
\label{fig:commbiloc}
\end{figure}

Further candidates for deformation generators are \emph{bilocal operators}, 
which can be constructed from any two local operators, see
\Figref{fig:biloc}:%
\footnote{We define the step function $\step(x)=0,1/2,1$ for $x<0$, $x=0$, $x>0$, respectively.}
\begin{equation}
\biloc{\loc_j}{\loc_k}\defeq\sum_{a,b}
	\step\bigbrk{(b+\half\range{\loc_k})-(a+\half\range{\loc_j})}
\,\half\bigacomm{\loc_j(a)}{\loc_k(b)}\,.
\label{eq:defbiloc}
\end{equation}
Here, $\acomm{\loc_j(a)}{\loc_k(b)}$ denotes the anticommutator of the
local operators $\loc_j(a)$ and $\loc_k(b)$, 
which act starting at spin sites $a$ and $b$. 
The bilocal operator $\biloc{\loc_j}{\loc_k}$
is constructed such that the sum of terms where $\loc_j$ 
acts on either side of $\loc_k$ equals the anticommutator
\begin{equation}
\biloc{\loc_j}{\loc_k}+\biloc{\loc_k}{\loc_j}=\half\acomm{\loc_j}{\loc_k}.
\label{eq:bilocsym}
\end{equation}
Similar to the case of boost operators, commuting a bilocal with a local
operator in general yields a combination of bilocal and local
operators:
\begin{equation}
\bigcomm{\biloc{\loc_j}{\loc_k}}{\loc_l} = 
\bigbiloc{\comm{\loc_j}{\loc_l}}{\loc_k} 
+\bigbiloc{\loc_j}{\comm{\loc_k}{\loc_l}} 
+\loc_m \,.
\end{equation}
However, if the underlying local operators commute, the commutator
becomes local, cf.\ \Figref{fig:commbiloc}. 
Hence, bilocal operators that are constructed from the
commuting charges $\charge_r$ yield admissible deformations,%
\footnote{Note that the commutator of higher multilocal operators with local
operators always yields multilocal operators again. Thus higher multilocal operators cannot
be used to generate local structures.}
\begin{equation}
\frac{\dd}{\dd\lambda}\,\charge_t(\lambda) = 
i\bigcomm{\biloc{\charge_r(\lambda)}{\charge_s(\lambda)}}{\charge_t(\lambda)}, \quad s>r=2,\dots,\infty \,,
\label{eq:deformationbiloc}
\end{equation}
which result in deformed charges $\charge_t$ that are local. 
As for the boost operators, 
the deformation $\defop(\lambda)=\biloc{\charge_r(\lambda)}{\charge_s(\lambda)}$
directly depends on the charges $\charge_{r,s}$. Here the dependence
is even quadratic and hence
\[
\defop^{(0)}=\biloc{\charge_r^{(0)}}{\charge_s^{(0)}},
\qquad
\defop^{(1)}=\biloc{\charge_r^{(1)}}{\charge_s^{(0)}}+\biloc{\charge_r^{(0)}}{\charge_s^{(1)}},
\qquad
\defop^{(2)}=\ldots\,.
\]
%

\paragraph{Bilocal Operator Identifications.}

As is the case for boost operators, bilocal operators are defined
modulo local contributions. 
Again, this is due to the equivalence of local homogeneous
operators that differ only by spectator legs 
\eqref{eq:spectator,eq:loceq}. More precisely, the
following identifications can be made, where $\specleg$ again denotes
a spectator leg:%
\footnote{The floor and ceiling functions are defined as
$\floor{x}\defeq\max\{z\in\Integers:z\leq x\}$,
$\ceil{x}\defeq\min\{z\in\Integers:z\geq x\}$.}
\begin{align}
\biloc{\loc_j\specleg}{\loc_k}
&=\biloc{\loc_j}{\loc_k}
 -\half\sum\nolimits_a\half\bigacomm
 		{\loc_j(a)}
 		{\loc_k\bigbrk{a-\floor{\half\range{\loc_k}-\half\range{\loc_j}}}}\,,\nln
\biloc{\loc_j}{\loc_k\specleg}
&=\biloc{\loc_j}{\loc_k}
 +\half\sum\nolimits_b\half\bigacomm
 		{\loc_j\bigbrk{b-\floor{\half\range{\loc_j}-\half\range{\loc_k}}}}
 		{\loc_k(b)}
 -\vev{\loc_k}\loc_j\,,\nln
\biloc{\loc_j}{\specleg\loc_k}
&=\biloc{\loc_j}{\loc_k}
 -\half\sum\nolimits_b\half\bigacomm
 		{\loc_j\bigbrk{b-\ceil{\half\range{\loc_j}-\half\range{\loc_k}}}}
 		{\loc_k(b)}\,,\nln
\biloc{\specleg\loc_j}{\loc_k}
&=\biloc{\loc_j}{\loc_k}
 +\half\sum\nolimits_a\half\bigacomm
 		{\loc_j(a)}
 		{\loc_k\bigbrk{a-\ceil{\half\range{\loc_k}-\half\range{\loc_j}}}}
 -\vev{\loc_j}\loc_k\,.
\label{eq:ibtbi}
\end{align}
While the first correction terms on the right hand side of the
identifications follow straightforwardly from the definition
\eqref{eq:defbiloc}, the addition/subtraction of $\loc_{j,k}$
is necessary for a correct
regularization: Similar to the case of boost operators do for example
$\biloc{\loc_j}{\loc_k\specleg}$ and $\biloc{\loc_j}{\loc_k}$ act
differently at the right boundary of a finite chain. When the length
of the chain is taken to infinity, the difference becomes the local
operator $\vev{\loc_k}\cdot\loc_j$. Here, the $\vev{\loc_k}$ are the same constants as in
\eqref{eq:ibtb}.

\paragraph{Degrees of Freedom.}

The identifications \eqref{eq:ibtbi} show that two pairs of local
operators that differ only by spectator legs yield the same bilocal
operator, up to local terms,
\begin{equation}
\loc_i\simeq\loc_k ,\; \loc_j\simeq\loc_l \text{ (up to spectator legs)}
\quad\Longrightarrow\quad
\biloc{\loc_i}{\loc_j} = \biloc{\loc_k}{\loc_l} +\loc_r \,. 
\end{equation}
Due to this ambiguity, there remains some arbitrariness in the definition of bilocal operators
\eqref{eq:defbiloc}: The ``overlap''
between the two local operators for instance can be adjusted through the addition
of local operators. As in the case of boost operators,
this ambiguity is not troublesome, since all deformations by local homogeneous operators
$\loc_r$ can be absorbed into the similarity transformations
\eqref{eq:deformationepsilon}. The definition \eqref{eq:defbiloc} is
chosen in favor of the identity \eqref{eq:bilocsym}, which immediately implies that
\begin{equation}
\bigcomm{\biloc{\charge_r}{\charge_s}+\biloc{\charge_s}{\charge_r}}{\charge_t} = 0 \,.
\label{eq:bilocswitch}
\end{equation}
Hence, $\biloc{\charge_r}{\charge_s}$ and
$\biloc{\charge_s}{\charge_r}$ generate only one degree of freedom.

\paragraph{Note on Boost Operators.}

Observe that the boost operators \eqref{eq:boostreg} can formally 
be written as particular bilocal operators,
\begin{equation}
\boost{\loc_k}=\half\biloc{\lengthop}{\loc_k}
-\half\biloc{\loc_k}{\lengthop}\,,
\label{eq:boostasbiloc}
\end{equation}
where $\lengthop$ denotes the length operator as introduced in \eqref{eq:ibtb}. 
The equivalence is due to the fact that in $\biloc{\lengthop}{\loc_k}$
the operator $\lengthop$ counts the spin sites to the left of $\loc_k$.%
\footnote{For infinite chains this number is infinite. 
The infinity cancels in the antisymmetric definition in \eqref{eq:boostasbiloc},
but an unspecified finite shift remains as a regularization parameter. 
It is related to the freedom of shifting the origin (site $0$)
of the chain for the definition of boost operators,
cf.\ \eqref{eq:boostdef,eq:boostorigin}.}
Correspondingly, the operator identifications 
in \eqref{eq:ibtb} and \eqref{eq:ibtbi} are compatible. 
The identification would allow us to work with 
bilocal operators only and thus simplify the framework slightly.

\paragraph{Example.}

As an example for a deformation through bilocal operators,
consider once more the $\alg{gl}(K)$ spin chain. As presented in the previous
example, the first two commuting nearest-neighbor charges are
\begin{equation}
\charge_2^{(0)} = \perm{1}-\perm{2,1},\qquad
\charge_3^{(0)} = -\ihalf\bigbrk{\perm{2,3,1}-\perm{3,1,2}} \,.
\end{equation}
Following \eqref{eq:deformationbiloc}, the first-order deformation of
the charge $\charge_2$ is given by
\begin{align}
\charge_2^{(1)}
&=i\bigcomm{\biloc{\charge_2^{(0)}}{\charge_3^{(0)}}}{\charge_2^{(0)}}\nln
&=\half\bigcomm{\biloc{\perm{1}-\perm{2,1}}{\perm{2,3,1}-\perm{3,1,2}}}{\perm{1}-\perm{2,1}}\nln
&=\quarter(-4\biloc{\perm{1}}{\perm{1,3,2}}+4\biloc{\perm{1}}{\perm{2,1,3}}+4\biloc{\perm{2,1}}{\perm{1,3,2}}\nln
		&\qquad -4\biloc{\perm{2,1}}{\perm{2,1,3}}+4\perm{1}-2\perm{2,3,1}-2\perm{3,1,2}-2\perm{2,1,4,3}-2\perm{2,3,4,1}\nln
		&\qquad +2\perm{2,4,1,3}+\perm{2,4,3,1}+2\perm{3,1,4,2}+\perm{3,2,4,1}-2\perm{3,4,1,2}-2\perm{4,1,2,3}\nln
		&\qquad +\perm{4,1,3,2}+\perm{4,2,1,3})\nln
&=\quarter(-4\perm{1}+8\perm{2,1}-2\perm{2,3,1}-2\perm{3,1,2}-2\perm{2,1,4,3}\nln
		&\qquad -2\perm{2,3,4,1}+2\perm{2,4,1,3}+\perm{2,4,3,1}+2\perm{3,1,4,2}+\perm{3,2,4,1}-2\perm{3,4,1,2} \nln
		&\qquad -2\perm{4,1,2,3}+\perm{4,1,3,2}+\perm{4,2,1,3})\,,
\label{eq:bilocexample}
\end{align}
where in the last line the bilocal operators were identified according
to the rules \eqref{eq:ibtbi}. Deforming $\charge_3^{(0)}$ in
the same fashion, the resulting charges $\charge_2$ and $\charge_3$
commute up to terms of order $\order{\lambda^2}$. The
higher order terms of $\charge_2$ and $\charge_3$ can be obtained by
successive reinsertion of the deformed charges into
\eqref{eq:deformationbiloc} and further expansion in
$\lambda$.

\subsection{Basis of Charges}
\label{sec:rotations}

The operators presented above generate almost all
admissible deformations of the form \eqref{eq:deformationl}.
However, taking linear combinations of the
charges $\charge_r$ certainly does not change their algebra and
therefore yields
another type of allowed deformations
\begin{equation}
\frac{\dd}{\dd\lambda}\,\charge_r(\lambda) = \charge_s(\lambda) , \quad r,s=2,\dots,\infty \,.
\label{eq:deformationrotpre}
\end{equation}
While the transformations \eqref{eq:deformationepsilon} describe a
change of basis within the space of local homogeneous operators, this
type of deformation represents a change of basis within the algebra of
charges $\charge_r$.

In order to analyze all admissible deformations in a common framework,
we consider $\charge_r$ as the $r$'th component of a vector $\charge=\basgen_r\charge_r$. 
A rotation generator $\rotgen_{r,s}$ acts on the basis vector $\basgen_k$ as
\[
\comm{\rotgen_{r,s}}{\basgen_{t}}=-i\delta_{r,t}\basgen_s.
\label{eq:rotgen}
\]
In other words, the $\rotgen_{r,s}$ generate
general linear transformations on the space of commuting charges.
The deformation \eqref{eq:deformationrotpre} can then be written as
\begin{equation}
\frac{\dd}{\dd\lambda}\,\charge(\lambda) = 
i\bigcomm{\rotgen_{r,s}}{\charge(\lambda)} , \quad r,s=2,\dots,\infty \,.
\label{eq:deformationrot}
\end{equation}

\subsection{Symmetry Generators}
\label{sec:symgen}

The charges $\charge(\lambda)$ of a symmetric spin chain 
should transform in a particular representation of the symmetry algebra.
Assume that the generator $\liegen_a$ is represented on spin chain states by 
the operators $\liegen_a(\lambda)$.
It is straight-forward to deform the representation by means of
the same differential equation as for the charges
\begin{equation}
\frac{\dd}{\dd\lambda}\,\liegen_a(\lambda)=i\bigcomm{\defop(\lambda)}{\liegen_a(\lambda)}\,.
\label{eq:liedeformation}
\end{equation}
Equation \eqref{eq:fixstruct} guarantees that the structure constants 
of the commutation relations $\comm{\liegen_a}{\liegen_b}$ 
and $\comm{\liegen_a}{\charge}$ 
are preserved under the deformation. 
Since \eqref{eq:liedeformation} generates similarity transformations,
$\liegen_a(\lambda)$
is a one-parameter family of equivalent representations
of a single undeformed algebra.

Let us first consider the symmetry to be a Lie algebra $\alg{g}$
with commutation relations
\begin{equation}
\comm{\liegen_a}{\liegen_b}=F_{a,b}^c\liegen_c\,,
\label{eq:lietrans}
\end{equation}
and invariant charges
\[
\comm{\liegen_a}{\charge}=0\,.
\label{eq:lieinvar}
\]
Obviously these
commutation relations are preserved by the deformation
with undeformed structure constants $F^c_{a,b}$.
This work focuses on deformations for which \eqref{eq:liedeformation} is trivial:
The deformations discussed above are invariant under the symmetry,
$\comm{\defop_k(\lambda)}{\liegen_a(\lambda)}=0$,
which leads to 
$\liegen_a(\lambda)=\liegen_a(0)$.

For integrable spin chains the Lie algebra 
furthermore extends to \emph{Yangian symmetry}.
Yangian symmetry implies an additional set of generators $\yang_a$
which transform in the adjoint representation of the Lie algebra $\alg{g}$,
\begin{equation}
\comm{\liegen_a}{\yang_b}=F_{a,b}^c\yang_c\,,
\label{eq:yangtrans}
\end{equation}
and which leave the conserved charges invariant:
\begin{equation}
\comm{\yang_a}{\charge}=0\,.
\label{eq:yanginvariance}
\end{equation}
Additionally, the Yangian
generators and the Lie algebra generators $\liegen_a$ have to satisfy
the Serre relations \cite{Drinfeld:1985rx}. These are given by
\begin{equation}
\bigcomm{\yang_a}{\comm{\liegen_b}{\yang_c}}+\bigcomm{\yang_b}{\comm{\liegen_c}{\yang_a}}+\bigcomm{\yang_c}{\comm{\liegen_a}{\yang_b}}
=\sfrac{1}{6}A_{abc}^{def}\lrbrc{\liegen_d,\liegen_e,\liegen_f},
\label{eq:serre}
\end{equation}
where $\lrbrc{\ldots}$ represents the sum over all six permutations of
the enclosed generators and the
coefficients $A_{abc}^{def}$ are given in terms of the structure constants
and the Cartan matrix $C^{a,b}$
\begin{equation}
A_{abc}^{def}=\sfrac{1}{4} F_{a,g}^d F_{b,h}^e F_{c,j}^f F_{d,e}^{j} C^{g,d} C^{h,e} \,.
\label{eq:serrecoeffs}
\end{equation}
Since the deformation \eqref{eq:deformationl} in general preserves the
algebra between the deformed quantities, the invariance property
\eqref{eq:yanginvariance} is preserved if the Yangian generators are
deformed in the same way as the charges \eqref{eq:deformationall}:
\begin{equation}
\frac{\dd}{\dd\lambda}\,\yang_a(\lambda)=i\bigcomm{\defop(\lambda)}{\yang_a(\lambda)}\,.
\label{eq:yangdeformation}
\end{equation}
Because the change of basis deformation \eqref{eq:deformationrot} only mixes the
charges $\charge_r$ among themselves, it does not affect
\eqref{eq:yanginvariance}. Since the Lie algebra generators
$\liegen_a$ are deformed \eqref{eq:liedeformation} in the same way as
the Yangian generators \eqref{eq:yangdeformation}, also the Serre
relations are preserved by the deformation.

The fact that the generating equation \eqref{eq:deformationl}
preserves any algebra among the deformed operators
$\charge$ might be particularly interesting for extending our
deformation method to models in which also the Lie algebra representation
$\liegen_a(\lambda)$ is deformed non-trivially. 
One such case was recently studied by Zwiebel \cite{Zwiebel:2008gr}, 
who found a differential equation reminiscent of ours.

\section{Geometry of the Moduli Space}
\label{sec:geom}

In the previous section we have found several admissible
deformations $\defop_k$ which generate long-range integrable spin chains of infinite extent
by means of a simple differential equation.
These are deformations by local operators $\loc_l$, 
boosted charges $\boost{\charge_k}$,
bilocal charges $\biloc{\charge_r}{\charge_s}$
as well as changes $\rotgen_{m,n}$
of the basis of local operators.
Taking all these operators into account results in multi-parameter deformations.
In this section we shall discuss the dependence
of the charges $\charge_r$ on the moduli.
Qualitatively it will depend crucially on whether 
the differential equation obeys a flatness condition.

\subsection{Multi-Parameter Deformations}

There is nothing that prevents us from combining the various deformations
into a system with multiple moduli
$\set{\xi_j}=\set{\alpha_k,\beta_{r,s},\gamma_{m,n},\varepsilon_l}$.
This is done by choosing the deformation $\defop$ to be 
some linear combination%
\footnote{The linear combination may have $\lambda$-dependent coefficients.}
of the 
$\set{\defop_j}=\set{\boost{\charge_k},\biloc{\charge_r}{\charge_s},\rotgen_{m,n},\loc_l}$
(we sum over repeated indices),
\[
\label{eq:multidefX}
\defop(\lambda)= \frac{\dd\xi_j(\lambda)}{\dd\lambda}\,\defop_j(\lambda).
\]
We shall investigate how the charges $\charge_r$ depend on the moduli.
Substituting the linear combination of deformation operators
\eqref{eq:multidefX} into the generating equation \eqref{eq:deformationl} 
yields the differential equation
\begin{equation}
\dd\charge(\lambda) 
=
i\bigcomm{\defop_j(\lambda)}{\charge(\lambda)}\,\dd \xi_j(\lambda).
\label{eq:deformationX}
\end{equation}
For later purposes it will be very convenient to use
the language of differential forms $\defform=\defop_j \dd\xi_j$ on moduli space.
Including all operators discussed in \Secref{sec:generalconst}, 
we find the most general generating equation
\begin{equation}
\label{eq:deformationall}
\dd\charge
=
i\bigcomm{\boost{\charge_k}}{\charge}\boostform_k 
+i\bigcomm{\biloc{\charge_r}{\charge_s}}{\charge}\biform_{r,s} 
+i\bigcomm{\rotgen_{m,n}}{\charge}\rotform_{m,n} 
+i\bigcomm{\loc_l}{\charge}\locform_l .
\end{equation}
The coefficients $\boostform_k,\biform_{r,s},\rotform_{m,n},\locform_l$
are one-forms on moduli space which parametrize
the desired deformation.
They can depend arbitrarily on any of the moduli $\xi_j$. 

Note that, as before, the generating equation defines merely a one-parameter family 
of charges $\charge_r(\lambda)$. 
However, now we have the additional freedom to specify 
the functions $\xi_j(\lambda')$ which define a curve on moduli space.
This means that the charges $\charge_r$ depend not
only on a point $\xi_j(\lambda)$ in moduli space, 
but also on the shape of the curve $\xi_j(\lambda')$ connecting
the undeformed model at $\lambda'=0$ to the deformed model at $\lambda'=\lambda$.
In general one cannot expect the dependence on the shape of the curve 
to be trivial.

A further complication is that the deformation operators $\defop_j(\lambda')$ 
are neither constants nor proper functions of the moduli $\xi_j$: 
For the boost and bilocal deformations,
$\defop=\boost{\charge_k}, \biloc{\charge_r}{\charge_s}$,
they actually depend on the solution $\charge_r(\lambda')$
of the differential equation itself.
This unusual feature complicates the treatment,
but we can at least make use of a weaker fact:
The differential of the deformations $\dd\defop_j(\lambda')$ 
(which is what is needed in practice)
can be expressed through the differential of the charges $\dd\charge_k(\lambda')$
which in turn is determined through the differential equation \eqref{eq:deformationX}.

One can find a perturbative solution 
for the generating equation \eqref{eq:deformationX}
with multiple parameters.
The deformation curve $\xi_j(\lambda')$ is assumed to start at the origin $\xi_j(0)=0$ 
of the moduli space with the undeformed charges $\charge^{(0)}_r$.
It is furthermore assumed to be confined 
to a small neighborhood of the origin. 
Then one can expand the solution $\charge(\lambda)$ in 
terms of small $\xi_j(\lambda)$ as follows
\begin{eqnarray}
\charge(\lambda)\eq
\charge^{(0)} 
+\xi_k(\lambda)i\bigcomm{\defop^{(0)}_k}{\charge^{(0)}}
+\int_0^\lambda \dd\xi_k(\lambda')\, 
i\bigcomm{\defop^{(1)}_{k}}{\charge^{(0)}}
\nl
+\int_0^\lambda \dd\xi_k(\lambda')\, \xi_l(\lambda')\, 
i\bigcomm{\defop_k^{(0)}}{i\comm{\defop^{(0)}_l}{\charge^{(0)}}}
+\order{\xi^3}.
\end{eqnarray}
Here we have expanded the deformation 
$\defop_k(\lambda)=\defop^{(0)}_{k}+\defop^{(1)}_{k}+\ldots$
in powers of $\xi$.
As explained above, we actually know the 
partial derivatives of $\defop_k$ in all directions 
of the moduli space. We can therefore write
$\defop^{(1)}_k=\xi_l(\lambda)\defop^{[l]}_k$.
Splitting the integrals into their symmetric and antisymmetric part
allows to rewrite the expansion in a more illuminating fashion
\begin{eqnarray}
\charge(\lambda)\eq
\charge^{(0)} 
+\xi_k(\lambda)i\bigcomm{\defop^{(0)}_k}{\charge^{(0)}}
+\half\xi_k(\lambda) \xi_l(\lambda)\, 
\bigbrk{
i\bigcomm{\defop^{[l]}_{k}}{\charge^{(0)}}
+i\bigcomm{\defop_k^{(0)}}{i\comm{\defop^{(0)}_l}{\charge^{(0)}}}
}
\nl
+\half\int_0^\lambda \dd\xi_k(\lambda')\, \xi_l(\lambda')\, 
i\bigcomm{\defop^{[l]}_{k}-\defop^{[k]}_{l}+i\comm{\defop^{(0)}_k}{\defop_l^{(0)}}}{\charge^{(0)}}
+\order{\xi^3}.
\label{eq:chargeexpansion}
\end{eqnarray}
%

\subsection{Connection and Curvature}

The generating equation \eqref{eq:deformationX}
can be interpreted as a parallel transport equation
for the vector of commuting charges $\charge$
\[
\cder_{\ad}\charge=0.
\label{eq:parallel}
\]
Here $\cder_{\ad}$ is the covariant derivative $\cder$
in the adjoint representation
\begin{equation}
\label{eq:covder}
\cder \defeq \dd-i\defform,
\qquad
\cder_{\ad}\defeq \dd-i\ad(\defform),
\qquad
\ad(\defform)\charge\defeq \comm{\defform}{\charge}.
\end{equation}
The operator-valued connection $\defform$
which includes all the admissible deformations reads
\begin{equation}
\label{eq:connectall}
\defform\defeq
\defop_j\,\dd\xi_j=\boost{\charge_k}\,\boostform_k 
         +\biloc{\charge_r}{\charge_s}\,\biform_{r,s}
	 +\rotgen_{m,n}\,\rotform_{m,n} 
         +\loc_l\locform_l \,.
\end{equation}
This connection may or may not be flat:
Flatness would imply that the deformed charges $\charge(\lambda)$
are independent of the shape of the path $\xi(\lambda')$
along which they are parallel transported.
They would only depend on the endpoint $\xi(\lambda)$ of the path and
thus they could be defined as proper functions $\charge(\xi)$ on moduli space.
In the expansion \eqref{eq:chargeexpansion} one can observe the
influence of flatness: The terms on the first line exclusively
depend on the endpoint $\xi(\lambda)$ while the term on 
the second line requires some integrals over $\xi(\lambda')$.
Importantly, the latter term is proportional 
to the curvature of the connection $\defform$
and consequently it vanishes for a flat connection.

\begin{figure}\centering
\includegraphics{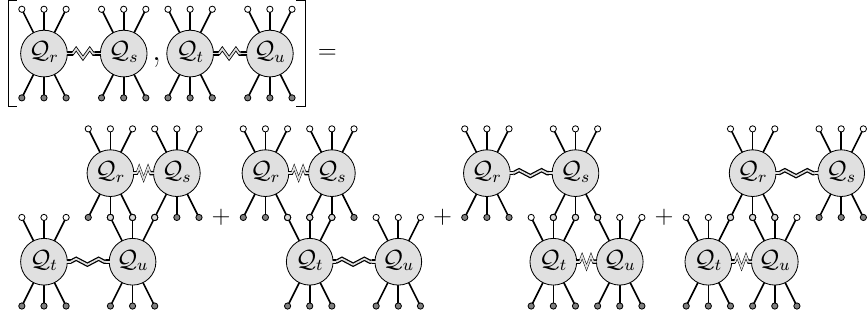}
\caption{Graphical representation of the commutator
\protect\eqref{eq:biloccomm} of two bilocal operators. Since the local
charges commute with each other, the only contributing terms (up to
local operators) are
those where both local parts of one bilocal operator overlap with one
of the local charges of the other bilocal operator.}
\label{fig:biloccomm}
\end{figure}

Let us now calculate the curvature of the connection $\defform$.
According to \eqref{eq:covder} it reads
\[
i\cder^2=\dd \defform-i \defform\wedge\defform.
\]
In the following we shall neglect the deformations $\locform_l$ 
by local operators. 
This is favorable for several reasons:
Firstly, these deformations turn out
to form an ideal, they do not influence the
curvature components associated to boosts, 
bilocal and basis change transformations.
Secondly, it allows us to discard local contributions 
originating from the other deformations
at most steps of the calculation. 
These would be hard to treat quantitatively and in full generality.
And last but not least the deformations by local operators
are irrelevant in the sense that they have no impact 
on the spectrum of finite chains.
We are thus left with a connection
\begin{equation}
\defform = \boostform+\biform+\rotform\,.
\label{eq:conn}
\end{equation}
The operator-valued one-forms $\boostform$, $\biform$ and $\rotform$ are
given by
\[
\boostform \defeq \boostop_k\boostform_k,
\qquad
\biform \defeq \biop_{r,s}\biform_{r,s},
\qquad
\rotform \defeq \rotgen_{m,n}\rotform_{m,n}\,.
\]
with the abbreviations $\boostop_k,\biop_{r,s}$ for the boost and bilocal charges
\[
\boostop_k \defeq \boost{\charge_k},
\qquad
\biop_{r,s} \defeq \biloc{\charge_r}{\charge_s}.
\]
The curvature is now given by
\begin{equation}
i\cder^2 =\dd\boostform +\dd\biform +\dd\rotform -i(\boostform+\biform+\rotform)\wedge(\boostform+\biform+\rotform) \,.
\label{eq:curvform}
\end{equation}
First, consider the bilocal connection $\biform$. Given that
$\cder_{\ad}\charge=0$, a short calculation shows that its exterior
derivative is given by
\begin{align}
\dd\biform
&=\bigbrk{\biloc{\dd\charge_r}{\charge_s}+\biloc{\charge_r}{\dd\charge_s}}\wedge\biform_{r,s}+\biop_{r,s}\,\dd\biform_{r,s}\nln
&=i\bigbrk{\biloc{\comm{\biop_{t,u}}{\charge_r}}{\charge_s}+\biloc{\charge_r}{\comm{\biop_{t,u}}{\charge_s}}}\biform_{t,u}\wedge\biform_{r,s}\nln
&\phantom{\,=\,}+i\bigbrk{\biloc{\comm{\boostop_k}{\charge_r}}{\charge_s}+\biloc{\charge_r}{\comm{\boostop_k}{\charge_s}}}\boostform_k\wedge\biform_{r,s}\nln
&\phantom{\,=\,}+(\delta_{m,r}\biop_{n,s}+\delta_{m,s}\biop_{r,n})\,\rotform_{m,n}\wedge\biform_{r,s}\nln
&\phantom{\,=\,}+\biop_{r,s}\,\dd\biform_{r,s}\,,
\label{eq:dbiform}
\end{align}
where it was used that $\comm{\rotgen_{m,n}}{\charge}_r=-i\delta_{m,r}\charge_n$.
Similarly, the exterior derivative of the boost connection $\boostform$ reads
\begin{align}
\dd\boostform
&=\boost{\dd\charge_k}\wedge\boostform_k 
  +\boostop_k\,\dd\boostform_k\nln
&=i\boost{\comm{\boostop_l}{\charge_k}}\boostform_l\wedge\boostform_k
	+i\boost{\comm{\biop_{r,s}}{\charge_k}}\biform_{r,s}\wedge\boostform_k\nln
&\phantom{\,=\,}+\delta_{m,k}\boostop_n\,\rotform_{m,n}\wedge\boostform_k +\boostop_k\,\dd\boostform_k\,.
\label{eq:dboostform}
\end{align}
Because the operators $\rotgen_{m,n}$ that generate basis changes do not depend on the coordinates
$\xi_j$, the exterior derivative of the connection $\rotform$
simply reads
\begin{equation}
\dd\rotform =\rotgen_{m,n}\,\dd\rotform_{m,n}\,.
\label{eq:drot}
\end{equation}
Using the 
Jacobi identity and \eqref{eq:bilocswitch}, one finds
\begin{align}
i\biform\wedge \biform
&=\ihalf\bigcomm{\biloc{\charge_r}{\charge_s}}{\biloc{\charge_t}{\charge_u}}\biform_{r,s}\wedge\biform_{t,u}\nln
&=-i\bigbrk{\biloc{\comm{\biop_{t,u}}{\charge_r}}{\charge_s}+\biloc{\charge_r}{\comm{\biop_{t,u}}{\charge_s}}}\biform_{r,s}\wedge\biform_{t,u}\,,
\label{eq:biwbi}
\end{align}
where it was further used that the commutator of two bilocal operators
evaluates to (cf.\ \Figref{fig:biloccomm}) 
\begin{align}
\bigcomm{\biloc{\charge_r}{\charge_s}}{\biloc{\charge_t}{\charge_u}}
&=\bigbiloc{\comm{\charge_r}{\biloc{\charge_t}{\charge_u}}}{\charge_s}
 +\bigbiloc{\charge_r}{\comm{\charge_s}{\biloc{\charge_t}{\charge_u}}}\nln&\phantom{\,=\,}
 +\bigbiloc{\comm{\biloc{\charge_r}{\charge_s}}{\charge_t}}{\charge_u}
 +\bigbiloc{\charge_t}{\comm{\biloc{\charge_r}{\charge_s}}{\charge_u}}
 +\mathrm{local}\,.
\label{eq:biloccomm}
\end{align}
Similarly, one finds
\begin{align}
i\boostform\wedge\boostform
&=\ihalf\bigcomm{\boost{\charge_k}}{\boost{\charge_l}}\boostform_k\wedge\boostform_l
=-i\boost{\comm{\boostop_l}{\charge_k}}\boostform_k\wedge\boostform_l \label{eq:boostwboost} \,,\\
i\boostform\wedge\biform+i\biform\wedge\boostform
&=-\ihalf\comm{\boostop_k}{\biop_{r,s}}\,\boostform_k\wedge\biform_{r,s}
	-\ihalf\comm{\biop_{r,s}}{\boostop_k}\biform_{r,s}\wedge\boostform_k\nln
&=\bigbrk{-i\boost{\comm{\biop_{r,s}}{\charge_k}}
	+i\biloc{\charge_r}{\comm{\boostop_k}{\charge_s}}
     +i\biloc{\comm{\boostop_k}{\charge_r}}{\charge_s}}\boostform_k\wedge\biform_{r,s} \,.
\label{eq:boostwbi} 
\end{align}
Since the generators $\rotgen_{m,n}$ commute with the boost
and bilocal
operators $\boostop_k$ and $\biop_{r,s}$, the remaining terms of the
curvature \eqref{eq:curvform} are
\begin{align}
i\rotform\wedge \rotform
&=\ihalf\comm{\rotgen_{m,n}}{\rotgen_{p,q}}\,\rotform_{m,n}\wedge\rotform_{p,q}
 =\delta_{n,p}\rotgen_{m,q}\,\rotform_{m,n}\wedge\rotform_{p,q},\nln
i\boostform\wedge\rotform+\rotform\wedge\boostform
&=-i\comm{\boostop_k}{\rotgen_{m,n}}\boostform_k\wedge\rotform_{m,n} =0,\nln
i\biform\wedge\rotform+\rotform\wedge\biform
&=-i\comm{\biop_{r,s}}{\rotgen_{m,n}}\biform_{r,s}\wedge\rotform_{m,n} =0\,.
\end{align}
Hence, the curvature reduces to
\begin{align}
i\cder^2
&=\boostop_k(\dd\boostform_k+\rotform_{p,k}\wedge\boostform_p)\nln
&\phantom{\,=\,}+\biop_{r,s}(\dd\biform_{r,s}+\rotform_{p,r}\wedge\biform_{p,s}+\rotform_{p,s}\wedge\biform_{r,p})\nln
&\phantom{\,=\,}+\rotgen_{m,n}(\dd\rotform_{m,n}-\rotform_{m,p}\wedge\rotform_{p,n}) \,.
\label{eq:curvformfinal}
\end{align}
It is curious to see that the non-linear components of the curvature
are all due to basis change deformations $\rotform_{m,n}$.
The group structure underlying the connection consists of 
a general linear group ($\rotform_{m,n}$) 
and two abelian ideals ($\boostform_k$, $\biform_{r,s}$).

\subsection{Flatness}

According to \eqref{eq:curvformfinal} the 
covariant derivative defined in \eqref{eq:covder} 
is flat if
\begin{align}
0&=\dd\boostform_k+\rotform_{p,k}\wedge\boostform_p\,,
\label{eq:flatconnboost}
\\
0&=\dd\biform_{r,s}+\rotform_{p,r}\wedge\biform_{p,s}+\rotform_{p,s}\wedge\biform_{r,p}\,,
\label{eq:flatconnbi}
\\
0&=\dd\rotform_{m,n}-\rotform_{m,p}\wedge\rotform_{p,n} \,.
\label{eq:flatconnrot}%
\end{align}
\begin{figure}\centering
\includegraphics{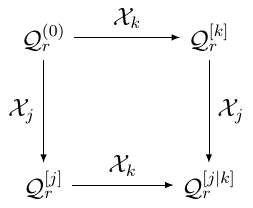}
\caption{The deformed charges are independent of the shape 
of the path within the moduli space only if the connection is flat.
In this case, first deforming in direction $\xi_l$ and then in direction $\xi_k$ 
yields the same result as 
first deforming in direction $\xi_k$ and then in direction $\xi_l$. 
Hence, the charges can be directly expanded 
\protect\eqref{eq:flatexpansion} in the deformation parameters $\xi_j$.}
\label{fig:FlatConn}%
\end{figure}%
Flatness implies that the charges $\charge_r(\lambda)$ 
depend only on the final position $\xi(\lambda)$ in moduli space;
they are independent of the shape of the path $\xi(\lambda')$. 
In other words, the charges are single-valued functions
$\charge_r(\xi)$ on moduli space. 
An expansion of the charges in the deformation parameters therefore exists,
see \Figref{fig:FlatConn},
\begin{equation}
\charge(\xi) = \charge^{(0)} +\xi_j\charge^{[j]} +\half\xi_j\xi_k\charge^{[j,k]}+\order{\xi^3}\,.
\label{eq:flatexpansion}
\end{equation}
According to \eqref{eq:chargeexpansion}
the first two expansion coefficients read
\begin{eqnarray}
\charge^{[k]}\eq
i\bigcomm{\defop^{(0)}_k}{\charge^{(0)}},
\nln
\charge^{[k,l]}\eq
i\bigcomm{\defop^{[k]}_{l}}{\charge^{(0)}}
+i\bigcomm{\defop_l^{(0)}}{\charge^{[k]}}.
\nln\eq
i\bigcomm{\defop^{[l]}_{k}}{\charge^{(0)}}
+i\bigcomm{\defop_k^{(0)}}{\charge^{[l]}}
\label{eq:flatexpansionterms}
\end{eqnarray}

\begin{figure}\centering
\includegraphics{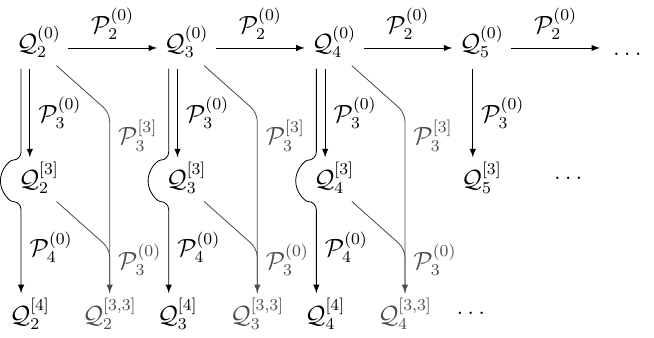}
\caption{The boost deformations commute among themselves. Thus if for
instance
$\rotform\equiv\biform\equiv0$ and $\boostform_k=\dd\alpha_k$, the boost-deformed
charges can be directly expanded 
in the deformation parameters $\alpha_k$, cf.\ \protect\eqref{eq:nngen}. The zeroth
order higher charges are generated by the boost
$\boostop_2^{(0)}$ while the coefficient
$\charge_r^{[k]}$ of $\alpha_k$ is given by the commutator of
$\boostop_k^{(0)}$ with the corresponding charge
$\charge_r^{(0)}$. Note in particular that coefficients of higher powers
of the $\alpha_k$ are generated by the higher order terms of the
charge and boost operators drawn in gray, e.g.\ $\charge_r^{[3,3]}
\sim\bigcomm{\boostop_3^{[3]}}{\charge_r^{(0)}}
+\bigcomm{\boostop_3^{(0)}}{\charge_r^{[3]}}$.}
\label{fig:BoostTree}
\end{figure}

Note that if the deformation \eqref{eq:deformationall} does not
incorporate a change of basis, i.e.\ $\rotform\equiv0$, 
exact boost and bilocal connections
$\boostform_k=\dd\alpha_k$, $\biform_{r,s}=\dd\beta_{r,s}$ 
lead to a flat connection and expansion 
in terms of $\alpha_k$ and $\beta_{r,s}$, cf.\ \Figref{fig:BoostTree}.
On the other hand, when $\rotform$ is chosen to be nonzero but such that it
satisfies \eqref{eq:flatconnrot}, the boost and
bilocal connections $\boostform$, $\biform$ must be modified in
order to keep the connection flat, i.e.\ in order to satisfy
\eqref{eq:flatconnboost,eq:flatconnbi}. In the next section, explicit
forms $\boostform$, $\biform$ in terms of moduli $\alpha_k$,
$\beta_{r,s}$ will be constructed for a specific, nonzero choice of
$\rotform$. It will turn out that these forms yield a flat connection
$\boostform+\biform+\rotform$.

\section{Interaction Range}
\label{sec:intrange}

In this section we will investigate the change of the interaction range 
of the integrable charges due to the various deformations.
We present a parametrization that leads to a
definite, minimal increase of the range by each deformation.
A minimal range is necessary to make a comparison 
to earlier studies \cite{Beisert:2005wv} where this feature is manifest.
As previously the interaction range of a local operator $\loc_k$ will
be denoted by $\range{\loc_k}$.

\subsection{First Comparison to \texorpdfstring{$\alg{gl}(K)$}{gl(K)} Chains}

The set of deformation moduli discussed in \Secref{sec:generalconst}
qualitatively agrees with the set of moduli
$\{\alpha_k,\beta_{r,s},\gamma_{m,n},\varepsilon_l\}$ 
of long-range integrable $\alg{gl}(K)$
spin-chains proposed in \cite{Beisert:2005wv}.%
\footnote{Also the deformation of the Yangian generators 
agrees qualitatively with the results of \cite{Beisert:2007jv}:
The bi-local terms remain undeformed
while there are local deformations of $\yang_a$.
The Serre relations defining the Yangian algebra are unmodified.}
This is a useful indication that both constructions
describe the same system and that we have not missed any
admissible deformations. 
Nevertheless, the precise form of the deformations,
cf.\ \eqref{eq:boostexample,eq:bilocexample}, 
does not match with the ones obtained in \cite{Beisert:2005wv},
there are two main differences:

The qualitative comparison suggests that the 
deformation of the lowest boost $\boostop_3$
should correspond to the lowest rapidity parameter $\alpha_0$ in \cite{Beisert:2005wv}.
Nevertheless the leading-order deformation \eqref{eq:boostexample}
does not appear among the deformations in \cite{Beisert:2005wv}.
In particular, each deformation by $\boostop_3$ 
increases the range by two units such that 
$\range{\charge_2^{(1)}}=4$ in \eqref{eq:boostexample}
as opposed to \cite{Beisert:2005wv} where each power of $\alpha_0$ 
increases the range by merely one unit.
Nevertheless the explorative study in \cite{Beisert:2005wv}
is complete up to range $6$, 
and therefore our deformation by $\boostop_3$ must be among 
the deformations in \cite{Beisert:2005wv}.
The resolution to the puzzle is that the $\alpha_0$ deformation in 
\cite{Beisert:2005wv} is a combination of the boost deformations
$\boostop_3$ and the change of basis deformation $\rotgen_{2,4}$.
The combination must be chosen such that the range decreases by one step,
which will be the topic of \Secref{sec:range.boost}.
Furthermore higher order deformations should have a consistent pattern of ranges,
see \Secref{sec:range.boosthigher,sec:range.biloc}.
In \Secref{sec:Range.Flatness} it will be shown that this 
actually leads to a flat connection as discussed in the previous section.

The leading-order bilocal deformation \eqref{eq:bilocexample}
indeed agrees literally
with the structure multiplying the coefficient $\gamma_{2,2}\beta_{2,3}$ 
in \cite{Beisert:2005wv,Beisert:2007jv}.
The coefficient $\gamma_{2,2}$ appears because in \cite{Beisert:2005wv}
the final charges $\charge_r$ are given as linear combinations $\gamma_{r,s}\bar\charge_{s}$
of normalized charges $\bar\charge_{s}$. 
The latter contain only the deformation moduli $\alpha_k,\beta_{r,s},\varepsilon_l$.
Conversely, in our differential equation the change of basis deformations
act at all points of the deformation path and thus mix with the other moduli.
We should therefore aim to reproduce merely the normalized charges 
$\bar\charge_{s}$ of \cite{Beisert:2005wv} using the moduli $\alpha_k,\beta_{r,s},\varepsilon_l$
and at the end apply a finite change of basis $\charge_r\to\gamma_{r,s}\charge_s$

Finally, as in the previous section, 
we will disregard deformations by local operators
$\loc_l$ and the corresponding moduli $\varepsilon_l$.
As mentioned earlier and as to be explained in \Secref{sec:betheansatz},
they do not affect the quantities we are
interested in and can be reintroduced easily through the similarity
transformation \eqref{eq:simtrafo}.

\subsection{Boost Connection at Leading Order}
\label{sec:range.boost}

First, consider the boost connection alone,
\begin{equation}
\boostform=\boostop_k\boostform_k=\boost{\charge_k}\boostform_k\,.
\end{equation}
As one would expect, the first-order deformation terms
$i\comm{\boostop_k^{(0)}}{\charge_r^{(0)}}$ generically have range
$ \range{\charge_k^{(0)}}
+\range{\charge_r^{(0)}} -1$. In our examples, however, 
we observe that the longest contributions of some boosted charge
precisely match with the longest contributions of another, undeformed charge,%
\begin{equation}
\label{eq:longestmatch}
i(s-1)\bigcomm{\boostop_s}{\charge_r}
\simeq
-(s+r-2)\charge_{s+r-1}\,.
\end{equation}
Note that according to \eqref{eq:nngenrange} 
both sides have a coincident range.
In \Appref{app:cancellation} we prove \eqref{eq:longestmatch} by using
the generating equation \eqref{eq:nngen}.
Consequently we can reduce the range by fixing the connection
$\rotform_{m,n}$ to
\begin{equation}
 	\rotform_{m,n}=\frac{n-1}{n-m}\,\boostform_{n-m+1} \,.
\label{eq:gammafix}
\end{equation}
Up to local similarity transformations, the connection \eqref{eq:conn}
and the corresponding general deformation \eqref{eq:deformationall} then
become
\begin{equation}
\Xi=\tilde\boostform+\biform\,,
\qquad
\dd\charge=i\comm{\tilde\boostop_k}{\charge}\,\tilde\boostform_k+i\comm{\biop_{r,s}}{\charge}\,\biform_{r,s}\,,\qquad
\label{eq:deformallgammafix}
\end{equation}
where
\begin{equation}
\tilde\boostform\defeq\boostform+\rotform=\tilde\boostop_k\tilde\boostform_k\,,
\qquad
\tilde\boostform_k=\frac{1}{k-1}\,\boostform_k\,,
\qquad
\tilde\boostop_k\defeq(k-1)\boostop_k+\rotgen_k\,.
\label{eq:boostconnwithgamma}
\end{equation}
The matrix $\rotgen_k$ is defined by (cf.\ \eqref{eq:deformationrot})
\begin{align}
	\rotgen_k&=\sum_n(n+k-2)\,\rotgen_{n,n+k-1}\,,\quad\mbox{that is}\nln
	\comm{\rotgen_k}{\charge}_n&=-i(n+k-2)\,\charge_{n+k-1}\,.
\label{eq:rotgenfix}
\end{align}
The factor $(k-1)$ in the definition of $\tilde\boostop_k$ and
$\tilde\boostform_k$ is
introduced for later convenience. 
Computationally, we find that the first-order deformation terms
for the $\alg{gl}(K)$ chain agree with the ones obtained in
\cite{Beisert:2005wv} and have the range
\begin{equation}
\bigrange{i\comm{\tilde\boostop_k^{(0)}}{\charge_r^{(0)}}}
=r+k-2\,.
\label{eq:rangefirstorder}
\end{equation}
In other words, deforming with $\tilde\boostop_k(0)$ increases the range
by $k-2$. It
appears that this is the minimal range one can achieve by correcting
the boost deformation with the connection $\rotform$.

\paragraph{Example.}

For the $\alg{gl}(K)$ chain, the operators of range four in the
first-order term $\charge_2^{(1)}$ \eqref{eq:boostexample} match the
longest terms in the undeformed charge $\charge_4^{(0)}$
\eqref{eq:nncharges}. With the choice \eqref{eq:gammafix}, 
the new term $\charge_2^{(1)}$ becomes
\begin{align}
\label{eq:leadboostdef}
\charge_2^{(1)}
&= 2\cdot\half(-2\perm{1} +2\perm{2,1} -\perm{2,3,4,1} +\perm{2,4,1,3} +\perm{3,1,4,2} -\perm{4,1,2,3})\nln
	&\phantom{\,=\,}+3\cdot\sfrac{1}{3}(-\perm{1} +2\perm{2,1} -\perm{3,2,1}\nln
		&\qquad\qquad\qquad+\perm{2,3,4,1} -\perm{2,4,1,3} -\perm{3,1,4,2} +\perm{4,1,2,3})\nln
&= -3\perm{1} +4\perm{2,1} -\perm{3,2,1} \,;
\end{align}
its range is reduced by one unit.
It also agrees literally with the corresponding deformation in \cite{Beisert:2005wv,Beisert:2007jv}.

\subsection{Boost Connection at Higher Orders}
\label{sec:range.boosthigher}

While the first-order terms of the boost deformation now have minimal range, 
the range of higher-order terms
in the expansion \eqref{eq:chargeexpansion} can be further reduced.
In particular, with the simplest choice
$\tilde{\boostform}_k=\dd\alpha_k$, we find that for a deformation along any path
$\alpha_k(\lambda)$, the range of higher-order terms is not additive
in the powers of the $\alpha_k(\lambda)$. For example the term
of $\charge_{r=2}(\lambda)$ of the $\alg{gl}(K)$ chain that is of order
$\alpha_3(\lambda)\alpha_4(\lambda)$ has range six;%
\footnote{It is possible that a naive application 
of the recursion relation does not yield
the desired range.
Note, however, that we do not display explicitly 
deformations by local operators, which are required
to reduce the length in most cases, see \Appref{app:example}.}
additivity
would require it to have range five ($r+3$), since
the leading-order terms proportional to
$\alpha_3(\lambda)$, $\alpha_4(\lambda)$ have range three ($r+1$) and
four ($r+2$), cf.\ \eqref{eq:rangefirstorder}.

Experimentally, we find that the following expansion 
for the connection $\tilde\boostform_k$ minimizes the range
\begin{equation}
\tilde\boostform_k
=\dd\alpha_k
+\sum_{M=1}^\infty\sum_{\ell_1,\ldots,\ell_M=3}^\infty
	\lrbrk{\prod_{j=1}^M(\ell_j-2)\alpha_{\ell_j}}\dd\alpha_{k+M-\sum_j\ell_j}\,,
\label{eq:sf}
\end{equation}
where we set $\dd\alpha_{k<3}=0$.
This choice also renders the range of higher-order terms additive in
the above sense.
While this result appears ad hoc, we also found
an equivalent implicit definition which makes a connection
to earlier results: Define first the function
$u(x)$ as in \cite{Beisert:2005wv}
\begin{equation}
 	u(x) = x+\sum_{n=3}^\infty\frac{\alpha_n}{x^{n-2}} \,,
\label{eq:rapmap}
\end{equation}
and its inverse $x(u)=u+\order{\alpha}$.
The connection $\tilde\boostform$ is then implicitly defined by the relation
\begin{equation}
\dd x(u)=-\sum_{n=3}^\infty\frac{\tilde\boostform_n}{x^{n-2}}\,.
\label{eq:alphamap}
\end{equation}
The relation should be interpreted in the following way: 
The function $x(u)$ depends implicitly on the moduli $\alpha_k$
and the differential acts on these only. 
Thus $\dd x(u)$ is a function of $\alpha_k,\dd\alpha_k$ and $u$. 
Replacing the latter by the inverse function $u(x)$ and casting
the result in the form on the right hand side of \eqref{eq:alphamap}
defines the $\tilde\boostform_n$.

While the choice \eqref{eq:alphamap} seems not well-motivated at first
sight, we
observe that the parametrization \eqref{eq:sf} gives the deformed
charges a definite, canonical range pattern: For the $\alg{gl}(K)$
chain, we find that with a given path $\alpha_k(\lambda)$ the range of each term in the
expansion \eqref{eq:chargeexpansion} depends additively on the
moduli $\alpha_k(\lambda)$ in its coefficient -- for each power of each
$\alpha_k(\lambda)$, the range increases by $k-2$,
cf.\ \eqref{eq:rangefirstorder}.
Furthermore, this range apparently is maximally reduced by
\eqref{eq:sf} and does not depend on the specific path $\alpha(\lambda)$ in
moduli space. As we will see below,
the choice \eqref{eq:sf} moreover guarantees that the connection
$\tilde\boostform$ is flat, which turns the deformed charges into
proper functions on moduli space.

\subsection{Bilocal Connection at Higher Orders}
\label{sec:range.biloc}

It remains to consider the bilocal deformations: As long as we
restrict to \emph{only bilocal deformations}, i.e.\
$\boostform=\rotform=0$, the simple choice
\begin{equation}
\biform_{r,s}=2\dd\beta_{r,s}\,
\label{eq:bilocsf}
\end{equation}
apparently results in a definite, minimal increase of the charges'
ranges. Here, the factor of two is introduced for consistency with
the existing literature. Our explicit computations for the
$\alg{gl}(K)$ chain show that with \eqref{eq:bilocsf}, as for the
boost connection \eqref{eq:sf}, the range of
each term in the expansion \eqref{eq:chargeexpansion} depends
additively on the moduli $\beta_{r,s}(\lambda)$ in its
coefficient. Here, each power of each $\beta_{r,s}(\lambda)$ increases
the range by $s-1$; at leading order, the range equals
\begin{equation}
\bigrange{i\comm{\biloc{\charge_r^{(0)}}{\charge_s^{(0)}}}{\charge_t^{(0)}}}
=t+s-1\,.
\label{eq:rangefirstorderbeta}
\end{equation}
Moreover, the range of all deformation terms is independent of the
path $\beta(\lambda)$ in moduli space and apparently cannot be further
reduced by a redefinition of $\biop_{r,s}$. As an example, the term of
$\charge_2$ proportional to $\beta_{2,3}(\lambda)$ is given in
\eqref{eq:bilocexample}, up to an overall factor.

If we, however, switch on the boost connection $\tilde\boostform$ with
the parametrization \eqref{eq:sf} and including the change of basis
\eqref{eq:gammafix}, we find that the range of terms that come with
mixed powers of $\alpha$'s and $\beta$'s is not minimal. Again, the
example of the $\alg{gl}(K)$ chain suggests
that the range of each term is minimized by the choice (with
$\beta_{r,s}=-\beta_{s,r}$)
\begin{equation}
\biform_{r,s}
= 2\,\dd\beta_{r,s} +2\beta_{n,s}\,\rotform_{n,r}
 + 2\beta_{r,n}\,\rotform_{n,s}\,.
\label{eq:betamap}
\end{equation}
With this choice, also the range of mixed terms becomes additive in
the powers of each modulus $\alpha_k(\lambda)$,
$\beta_{r,s}(\lambda)$. As for the pure $\alpha$-terms and the pure
$\beta$-terms, the range increases by $k-2$ for each power of
$\alpha_k(\lambda)$ and by $s-1$ for each power of
$\beta_{r,s}(\lambda)$.

Note that since $\rotform_{r,s}$ is parametrized by the moduli
$\alpha_k$, the prescription \eqref{eq:betamap} changes
\eqref{eq:boostconnwithgamma} once more by a non-trivial term. 
Namely, $\tilde\boostop_k$ becomes
\begin{equation}
\tilde\boostop_k=
(k-1)\boostop_k
+\rotgen_k
+2i\beta_{r,s}\bigbiloc{\comm{\rotgen_k}{\charge}_r}{\charge_s}
+\bigbiloc{\charge_r}{\comm{\rotgen_k}{\charge}_s}.
\label{eq:boostopfix}
\end{equation}
Examples for higher order terms of the expansion
\eqref{eq:chargeexpansion} are given in \Appref{app:example}.

\subsection{Flatness}
\label{sec:Range.Flatness}

The redefinition \eqref{eq:gammafix} made in favor of a canonical
interaction range spoils the flatness of the connection
\eqref{eq:connectall}.
As discussed in \Secref{sec:geom}, the boost and bilocal connection
$\boostform+\biform$ alone is flat,%
\footnote{For vanishing $\rotform$, the simple choice
$\boostform_k=\dd\alpha_k$, $\biform_{r,s}=2\dd\beta_{r,s}$ results in
a flat connection.}
while a change of basis $\rotform$
introduces curvature \eqref{eq:curvformfinal}. As we will see in the
following, flatness of the connection is restored by the
parametrization constructed above.

\paragraph{Boost Connection.}

With the definition \eqref{eq:gammafix}, the flatness condition
\eqref{eq:flatconnrot} for $\rotform$ becomes
\begin{align}
0&=\dd\rotform_{m,n} +\rotform_{m,k}\wedge\rotform_{k,n}\nln
 &=(n-1)\dd\tilde\boostform_{n-m+1}
  +\sum_k(k-1)(n-1)\,\tilde\boostform_{k-m+1}\wedge\tilde\boostform_{n-k+1}\,,\
  \quad m,n \geq 2\,,
\label{eq:nonflat}
\end{align}
which is not satisfied for generic $\tilde\boostform$. Note that for
$m=1$, equation \eqref{eq:nonflat} becomes the boost flatness
condition \eqref{eq:flatconnboost}. Thus we can summarize both
formulas if we extend the range of $m$ in \eqref{eq:nonflat} to $m\geq
1$. Setting
\begin{equation}
 	n\to n+m-1\,,\qquad k\to k+m-1\,,
\end{equation}
the flatness conditions \eqref{eq:flatconnboost,eq:flatconnrot} thus
become
\begin{equation}
0=\dd\tilde\boostform_n+\sum_k(k+m-2)\tilde\boostform_{k}\wedge\tilde\boostform_{n-k+1}\,,
\quad n\geq 2, m\geq 1\,.
\label{eq:inftoone}
\end{equation}
If we define the connection $\tilde\boostform_n$ to be zero for $n<3$,
we do not have to change the limits of the sum when shifting the
indices. The antisymmetry of the wedge product in \eqref{eq:inftoone}
implies that under the sum over $k$ the part with a constant coefficient
vanishes. Hence, we have shown that the equations
\eqref{eq:flatconnboost,eq:flatconnrot} for the choice
\eqref{eq:gammafix} are actually independent of the index $m$ and
reduce to
\begin{equation}
0=\dd\tilde\boostform_n+\sum_kk\tilde\boostform_{k}\wedge\tilde\boostform_{n-k+1}\,,\quad n\geq 2\,.
\label{eq:flatnessoneparam}
\end{equation}
We now show that this flatness condition is satisfied by the
definition \eqref{eq:alphamap}. This can be seen as follows:
\begin{align}
 0=-\dd \dd x(u)&=\sum_{n=3}^\infty \left(\frac{\dd\tilde\boostform_n}{x^{n-2}}- (n-2)\frac{\tilde\boostform_n\wedge \dd x(u)}{x^{n-1}}\right)\nln
&=\sum_{n=3}^\infty\left( \frac{\dd\tilde\boostform_n}{x^{n-2}}+\sum_{k=3}^\infty \frac{(n-2)\tilde\boostform_n\wedge \tilde\boostform_{k}}{x^{n+k-3}}\right)\nln
&=\sum_{n=3}^\infty\frac{1}{x^{n-2}}\left( \dd\tilde\boostform_n+\sum_{k=3}^\infty k\tilde\boostform_k\wedge \tilde\boostform_{n-k+1}\right),
\label{eq:boostflatness}
\end{align}
where going to the last line we have renamed the indices
$k\leftrightarrow n$ in the second sum followed by an index shift
$n\to n-k+1$. We furthermore dropped constant factors in front of the
wedge product due to the antisymmetry and did not change the limits of
the sum since $\tilde\boostform_{k<3}=0$ by definition. Now,
\eqref{eq:boostflatness} implies \eqref{eq:flatnessoneparam}.

\paragraph{Bilocal Connection.}

We now show that the remaining flatness condition
\eqref{eq:flatconnbi} is satisfied by the parametrization
\eqref{eq:betamap}, provided that $\rotform$ fulfills
\eqref{eq:flatconnrot}. As we have seen above, this is the case for
the choice \eqref{eq:gammafix}. Plugging \eqref{eq:betamap} into
\eqref{eq:flatconnbi} yields
\begin{align}
\sfrac{1}{2}(\dd\biform_{r,s}+\rotform_{p,r}&\wedge\biform_{p,s}+\rotform_{p,s}\wedge\biform_{r,p})\nln
&=\dd\beta_{n,s}\wedge\rotform_{n,r}+\beta_{n,s}\dd\rotform_{n,r}+\dd\beta_{r,n}\wedge\rotform_{n,s}+\beta_{r,n}\dd\rotform_{n,s}\nln
&\quad+\rotform_{p,r}\wedge \dd\beta_{p,s}+\beta_{n,s}\rotform_{p,r}\wedge\rotform_{n,p}+\beta_{p,n}\rotform_{p,r}\wedge\rotform_{n,s}\nln
&\quad+\rotform_{p,s}\wedge \dd\beta_{r,p}+\beta_{n,p}\rotform_{p,s}\wedge \rotform_{n,r}+\beta_{r,n}\rotform_{p,s}\wedge\rotform_{n,p}\nln
&=0\,,
\end{align}
where we have used equation \eqref{eq:flatconnrot}. Hence, the
definition \eqref{eq:betamap} leads to a flat bilocal connection in
the sense that \eqref{eq:flatconnbi} is satisfied. 

\subsection{Summary}

To summarize, the parametrization 
\eqref{eq:gammafix,eq:alphamap,eq:betamap},
\begin{align}
\rotform_{r,s}&=\frac{s-1}{s-r}\,\boostform_{s-r+1}\,,\nln
\sum_{n=3}^\infty\frac{\boostform_n(\alpha_m)}{(n-1)x^{n-2}}&=-\dd x(u)\,,\nln
\biform_{r,s}&= 2\,\dd \beta_{r,s} +2\,\beta_{n,s}\,\rotform_{n,r}+2\,\beta_{r,n}\,\rotform_{n,s}\,,
\label{eq:fixedforms}
\end{align}
results in deformed charges $\charge_r$ with a definite pattern of ranges. 
Moreover, this parametri\-zation renders the connection
\eqref{eq:connectall} flat.
Hence the charges are proper functions of the moduli space
parameters $\alpha$, $\beta$ up to local similarity transformations
\begin{equation}
\charge_n
=\charge_n^{(0)}
+\alpha_k\charge_n^{[k]}
+\beta_{r,s}\charge_n^{[r|s]}
+\alpha_k\alpha_l\charge_n^{[k,l]}
+\alpha_k\beta_{r,s}\charge_n^{[k,r|s]}
+\beta_{r,s}\beta_{t,u}\charge_n^{[r|s,t|u]}
+\ldots\,.
\label{eq:expansionalphabeta}
\end{equation}
The individual terms $\charge_n^{[\ldots]}$ in the $\alg{gl}(K)$
case agree with the ones obtained in \cite{Beisert:2005wv,Beisert:2007jv}.%
\footnote{Some charge terms $\charge_n^{[\ldots]}$ are calculated
explicitly in \Appref{app:example}.}
As discussed in the previous subsections, they have an interaction range 
\[
\bigrange{\charge_n^{[k_1,\ldots,k_t,r_1|s_1,\ldots,r_u|s_u]}}
=n+\sum_{\ell=1}^t(k_\ell-2)+\sum_{\ell=1}^u(s_\ell-1)\,,
\label{eq:lengthofterm}
\]
i.e.\ each boost deformation $\tilde\boostop_k$ increases the range
by $k-2$ and each bilocal deformation $\biop_{r,s}$ increases the
range by $s-1$.

Note that the definition \eqref{eq:gammafix} fixes a specific change
of basis that accompanies the boost deformation. This specific choice
minimizes the ranges of the deformed charges $\charge$ and renders
the connection flat. The charges $\charge$ thus form a canonical
basis of the space of charges at each point in moduli space. They
reproduce the normalized charges $\bar\charge_r$ of
\cite{Beisert:2005wv} and depend only on the reduced set of moduli
$\brc{\alpha_k,\beta_{r,s},\varepsilon_l}$. However,
nothing prevents us from choosing a different basis
\begin{equation}
\tilde\charge_m=\gamma_{m,n}\charge_n
\label{eq:postrot}
\end{equation}
for the space of charges \emph{after} the charges have been deformed.
For comparison with \cite{Beisert:2005wv}, note that the symbols
$\{\charge,\tilde\charge,\alpha_k,\beta_{r,s},\gamma_{r,s}\}$ here correspond to the quantities
$\{\bar\charge,\charge,\alpha_{k-3},\linebreak\beta_{r,s},\gamma_{r,s}\}$ there. 

\subsection{Properties of Deformations}

We have seen that the interaction ranges of the deformed charges
obey a certain pattern \eqref{eq:lengthofterm}. In this
subsection we discuss additional properties of the deformations as
well as their relation to gauge theory.

\paragraph{Parity.}

We introduce a parity operator $P$ acting on local, boost and bilocal
operators. For a local charge of manifest parity we then have
\begin{equation}
 	P \charge_r P^{-1}=(-1)^{p_r} \charge_r,
\end{equation}
where $p_r$ is even or odd for $\charge_r$ being even or odd under
parity. Since we only consider bilocal charges within commutators with
local charges, we can make use of the fact that due to
\eqref{eq:bilocsym} we have 
\begin{equation}
\bigcomm{\biloc{\charge_r}{\charge_s}}{\charge_t}
=-\bigcomm{\biloc{\charge_s}{\charge_r}}{\charge_t}.
\end{equation}
Further using that
$P\biloc{\charge_r}{\charge_s}P^{-1}=\biloc{P\charge_sP^{-1}}{P\charge_rP^{-1}}$
due to \eqref{eq:defbiloc}, we find 
\begin{align}
P\bigcomm{\biloc{\charge_r}{\charge_s}}{\charge_t}P^{-1}
&=\bigcomm{P\biloc{\charge_r }{ \charge_s}P^{-1}}{P\charge_t P^{-1}}\nln
&=(-1)^{p_s+p_r+p_t}\bigcomm{\biloc{\charge_s}{\charge_r }}{\charge_t }\nln
&=(-1)^{p_s+p_r+p_t-1}\bigcomm{\biloc{\charge_r}{\charge_s}}{\charge_t}\,.
\label{eq:paritybiloc}
\end{align}
The interpretation \eqref{eq:boostasbiloc} of boost operators in terms
of bilocal operators implies that
\begin{equation}
P\bigcomm{\boost{\charge_k}}{\charge_r}P^{-1}=(-1)^{p_k+p_r-1}\bigcomm{\boost{\charge_k}}{\charge_r}\,.
\label{eq:parityboost}
\end{equation}
Assuming that the undeformed Hamiltonian $\charge_2^{(0)}$ has even
parity, equation \eqref{eq:nngen} then implies that the undeformed
even and odd  charges $\charge_{2r}^{(0)}$ and $\charge_{2r+1}^{(0)}$
have even and odd parity, respectively. According to
\eqref{eq:paritybiloc,eq:parityboost} the long-range
deformations can be classified according to
\begin{equation}
P\charge_t^{[k_1,\ldots,k_t,r_1|s_1,\ldots,r_u|s_u]}P^{-1}
=(-1)^{n+\sum_{\ell=1}^t (k_\ell-1)+\sum_{\ell=1}^u(r_\ell+s_\ell -1)}
	\charge_t^{[k_1,\ldots,k_t,r_1|s_1,\ldots,r_u|s_u]}.
\end{equation}

\paragraph{Number of Crossings for Fundamental $\alg{g}=\alg{gl}(K)$.}

Considering the fundamental $\alg{gl}(K)$ spin chain, the building blocks of
invariant operators are given by simple permutations \eqref{eq:local}. 
One characteristic quantity of the interaction terms 
is the \emph{number of elementary permutations} 
(crossings) $\crossings{\loc_k}$ contained in an operator
$\loc_k$, e.g.\
\begin{align}
&\crossings{[2,1]+[2,1,4,3]}=2.\nln
&\hspace{.35cm}\includegraphics{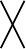}\hspace{.9cm}\includegraphics{PicPermut21}
	\hspace{.4cm}\includegraphics{PicPermut21}
\label{eq:crossings}
\end{align}
As shall be explained below, these numbers are a relevant quantity
for the gauge theory, because of their relation 
to the minimum perturbative order at which they may appear
\cite{Beisert:2003tq,Beisert:2007hz}.
Note that as for the range $\range{\loc_k}$ of 
a linear combination of local operators 
we define $\crossings{\loc_k}$ to be the \emph{maximum}
number of crossings.

In commutators of boost and local operators we add the numbers of
crossings of the boost operator $\boost{\loc_k}$ 
and the local operator $\loc_l$. 
Equation \eqref{eq:nngen} implies that the leading order charges
$\charge_r^{(0)}$ have $r-1$ crossings:
\begin{equation}
 	\crossings{\charge_r^{(0)}}=r-1.
\end{equation}
Consequently, the number of crossings of a \emph{boost deformation}
$\comm{\boostop_k^{(0)}}{\charge_r^{(0)}}$ has $r+k-2$ crossings:
\begin{equation}
 	\crossings{\comm{\boostop_k^{(0)}}{\charge_r^{(0)}}}=r+k-2.
\end{equation}
Adding a term proportional to $\charge_{r+k-1}^{(0)}$ with $r+k-2$
crossings for the minimal interaction range
\eqref{eq:deformallgammafix} apparently does not reduce the crossings.
Hence, the structure $\charge_r^{[k]}$ multiplied by $\alpha_k$
generically contains $r+k-2$ crossings. More generally, the number of
crossings increases by $k-1$ for each 
power of $\alpha_k$.

Also for bilocal deformations, the number of crossings of the
bilocal operator adds up with the number of crossings of the deformed
structure to give the number of crossings of the result. The
commutator of a bilocal operator
$\biloc{\charge_r^{(0)}}{\charge_s^{(0)}}$ which has $r+s-2$ crossings
and a charge $\charge_t^{(0)}$ with $t-1$ crossings thus has $t+r+s-3$
crossings:
\begin{equation}
 \crossings{\comm{\biloc{\charge_r^{(0)}}{\charge_s^{(0)}}}{\charge_t^{(0)}}}=t+r+s-3.
\end{equation}
The number of crossings of a general deformation term is therefore given by
\begin{equation}
\crossings{\charge_t^{[k_1,\ldots,k_t,r_1|s_1,\ldots,r_u|s_u]}}
=t-1+\sum_{\ell=1}^u(r_\ell+s_\ell-2)+\sum_{\ell=1}^t(k_\ell-1)\,,
\label{eq:numbercrossall}
\end{equation}
in full agreement with the prediction of \cite{Beisert:2005wv}.

\paragraph{Number of Crossings for $\alg{g}=\alg{gl}(2)$.}

A particularly interesting case is given for $\alg{gl}(2)$ symmetry
which represents the $\alg{su}(2)$ sector spin chain of $\superN=4$
SYM theory. We observe that for $\alg{gl}(2)$ the 
anti-symmetrizer of \ifarxiv\href{http://www.willkuere.de/}{$2+1$}\else$2+1$\fi{} spins vanishes 
\begin{equation}
 	0=\sum_{i,j,k\in\set{1,2,3}}\varepsilon^{ijk}\PTerm{i,j,k}.
\label{eq:glkeps}
\end{equation}
This allows to reduce 
the number of crossings of boost deformations.
Consider for
instance the Hamiltonian structure proportional to $\alpha_3$:
\begin{equation}
  \charge_2^{[3]}=-3 \PTerm{1} + 4 \PTerm{2, 1} - \PTerm{3, 2, 1}.
\end{equation}
Here we can replace the term with
three vertices by permutations with at most two vertices:
\begin{align}
\PTerm{3,2,1}&=\PTerm{2,3,1}+\PTerm{3,1,2}-2\PTerm{2,1}+\PTerm{1}.\nln
\includegraphicsbox{PicPermut321}
\hspace{.2cm}&=\hspace{0.17cm}
\includegraphicsbox{PicPermut231}
\hspace{.22cm}+\hspace{.20cm}
\includegraphicsbox{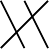}
\hspace{.19cm}-2\hspace{.20cm}
\includegraphicsbox{PicPermut21}
\hspace{.20cm}+\hspace{.21cm}
\includegraphicsbox{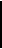}
\label{eq:epsid3}
\end{align}
Note that we have performed boundary identifications. 
Hence, the operator structure $\charge_2^{[3]}$ in fact contains only $2$
elementary permutations  \cite{Beisert:2003tq}. It appears that one can reduce the number of
crossings of other boost deformations in a similar fashion. Also the
terms multiplied by higher powers of $\alpha_k$ appear to be
regularizable in this way \cite{Beisert:2005wv}. 
This observation for
$\alg{g}=\alg{gl}(2)$ suggests a modification of \eqref{eq:numbercrossall} to
\begin{equation}
\crossings{\charge_t^{[k_1,\ldots,k_t,r_1|s_1,\ldots,r_u|s_u]}}_{\alg{gl}(2)}
=t-1+\sum_{\ell=1}^u(r_\ell+s_\ell-2)+\sum_{\ell=1}^t(k_\ell-2).
\label{eq:crossingsallgl2}
\end{equation}
The number of crossings for bilocal deformations does not decrease.

\paragraph{Parameter Restrictions from the
\texorpdfstring{$\alg{su}(2)$}{su(2)} Sector of
\texorpdfstring{$\superN=4$}{N=4} Super Yang-Mills Theory.}

Here we consider gauge theory restrictions on the $\alg{gl}(2)$ spin
chain charges based on their number of crossings \cite{Beisert:2003tq,Beisert:2007hz}. 
The integrable Hamiltonian appearing in the $\alg{su}(2)$ sector of $\superN=4$ SYM
theory obeys a fundamental principle: Its number of crossings does not
exceed the order of the coupling due to the correspondence to Feynman
diagrams. As we will see, this requirement can be satisfied if one
restricts the starting order of the moduli
$\{\alpha_k(\lambda),\beta_{r,s}(\lambda)\}$ as a Taylor series in
$\lambda$.

Within a $\alg{gl}(K)$ invariant operator, an elementary permutation
between two spin chain sites corresponds to a quartic scalar vertex of a Feynman
graph in planar gauge theory. The interaction
$\sim\lambda\comm{\Phi_i}{\Phi_j}^2$ in the $\superN=4$ Lagrangian
(cf.\ \cite{Minahan:2002ve}) translates to the fact that each
elementary permutation comes with a factor of the coupling:
\begin{equation}
\includegraphicsbox{PicPermut21}\sim\lambda\,.
\end{equation}
The number of Feynman vertices in the gauge theory Hamiltonian
$\ham(\lambda)$ must therefore equal the
corresponding power of the coupling constant in perturbation theory.
Translating this to the spin chain picture
one has to take into account that the Hamiltonian
$\ham(\lambda)$ is shifted by one power of $\lambda$
\begin{equation}
\ham(\lambda)=\ham_0+\lambda \charge_2(\lambda).
\label{eq:chargeshift}
\end{equation}
Thus, an operator structure with $C$ crossings is allowed to
contribute to $\charge_2(\lambda)$ at
$\order{\lambda^{C-1}}$. 
Comparison to the formula for the number of
crossings of our generated charge terms \eqref{eq:crossingsallgl2}
shows that one has to restrict the parameter functions as in
\cite{Beisert:2005wv}:
\begin{align}
 	\alpha_k(\lambda)&=\order{\lambda^{k-2}},\nln
	\beta_{r,s}(\lambda)&=\order{\lambda^{r+s-2}}.
\label{eq:paramorder2}
\end{align}
With this prescription the ranges of the deformed charges grow by at
most one site with each power of the coupling. The Hamiltonian at order
$\lambda^{N}$ acts on no more than $N+1$ sites.

The long-range Hamiltonian $\charge_2(\lambda)$ for the $\alg{gl}(K)$
chain in terms of permutation symbols and the coefficients
$\alpha_k(\lambda)$, $\beta_{r,s}(\lambda)$ including terms
of order $\order{\lambda^3}$ is printed at the end of 
this paper in \Tabref{tab:N4Q2}.

\section{Long-Range Bethe Ansatz}\label{sec:betheansatz}

We will now consider the asymptotic spectrum of a conserved charge on
finite periodic chains.
So far, the charges of the integrable model have been defined for
infinite chains only.
For a finite chain we demand that a particular charge matches with
the integrable charge $\charge_r$ for all terms whose range does not exceed
the length of the chain. This provides a proper definition of
commuting long-range charges on finite chains up to a certain order
in the power series.
The asymptotic spectrum is the spectrum in the form of a power series
modulo terms of higher orders where the operator is not uniquely
specified by the above definition. In particular, the longer the
chain, the higher the order at which the asymptotic spectrum
truncates.

We will use the asymptotic Bethe ansatz \cite{Sutherland:1978aa,Staudacher:2004tk} to determine
three basic observables of asymptotic states on the infinite chain:
Vacuum charge density, magnon dispersion relation and scattering
matrix.
These data are sufficient to set up the resulting asymptotic Bethe
equations
providing the asymptotic spectrum of conserved charges on finite
periodic chains.
We shall use the full set of deformations found in \Secref{sec:generalconst} and the
corresponding moduli defined in \Secref{sec:intrange},
and our result will turn out to agree with the earlier proposal in
\cite{Beisert:2005wv}. 

In the first parts of this section we shall assume a
$\alg{sl}(2)$ or $\alg{su}(2)$ spin chain with spin $t/2$ representations on all sites,
i.e.\ a long-range Heisenberg XXX$_{t/2}$ model.
Later we will generalize the results to higher-rank symmetry algebras $\alg{g}$.

\subsection{Ferromagnetic Vacuum}

The ferromagnetic vacuum $\state{0}$ is a pure state in which all spins
are aligned to have an identical orientation $\state{\mathrm{HW}}$
being of highest weight w.r.t.\ the symmetry algebra
\[\label{eq:ferrovac}
\state{0}=\ldots\otimes\state{\mathrm{HW}}\otimes\state{\mathrm{HW}}\otimes\state{\mathrm{HW}}\otimes\ldots\,.
\]
Furthermore this state is assumed to be an eigenstate of $\charge\supups{NN}_r$ 
with vanishing eigenvalue (density). The latter property can always be achieved
by subtracting from $\charge\supups{NN}_r$ the length operator $\lengthop$ 
multiplied by the eigenvalue density.

The long-range spin chain is a deformation of the nearest-neighbor model
induced by the equation \eqref{eq:deformationl}. 
This equation is of parallel transport type 
implying that merely the eigenvectors are deformed but not the spectrum. 
Here also the state remains undeformed because the deformations
respect the symmetry and there is only a single highest-weight state.
Altogether, the vacuum density of all deformed $\charge_r$ is zero.

\subsection{One-Magnon States and Dispersion Relations (\texorpdfstring{$\alpha$}{alpha})}
\label{sec:rapmap}

One-magnon states are excitations of the ferromagnetic vacuum \eqref{eq:ferrovac}
where one spin is replaced by a next-to-highest-weight state $\state{\mathrm{NHW}}$.
Let the spin at position $k$ be flipped
\[\label{eq:spinflip}
\state{k}=\ldots\otimes\state{\mathrm{HW}}\otimes\state{\MMM{\mathrm{NHW}}{k}}\otimes\state{\mathrm{HW}}\otimes\ldots\,.
\]
The magnon state $\state{p}$ is a state with definite momentum $p$ along the chain
\[\label{eq:magnon}
\state{p}=\sum\nolimits_k e^{ipk} \state{k}.
\]

Magnons are eigenstates of the charges $\charge_r$ because the latter
are homogeneous local operators. The eigenvalue of $\charge_r$ on $\state{p}$ 
is called the dispersion relation $q_r(p)$.
We will now study how the dispersion relation $q_r(p)$ changes under
the deformation \eqref{eq:deformationl}. 
Although the latter does not change the spectrum, 
it can deform the eigenstate to one with a different momentum $p$.
Let us therefore act with the various deformations on a magnon state.
Local operators \eqref{eq:local} acting on this state are equivalent to
linear combinations of shift operators
\begin{equation}
 	\shift^j:\ket{k}\mapsto\ket{k-j},
\end{equation}
whose action in momentum space is given by
\begin{equation}
\shift^j \ket{p} = e^{ipj} \ket{p} \,.
\end{equation}
Thus local deformations conserve the momentum.
Likewise boost operators can be represented by 
boosted shift operators $\boost{\shift^j}$ 
which act on one-magnon states as
\begin{equation}
\boost{\shift^j}\ket{p} = \sum_k k e^{ipk} \ket{k-j} = \sum_k (k+j) e^{ip(k+j)} \ket{k}
 = -i\,\frac{\partial}{\partial p} \left( e^{ipj} \ket{p} \right).
\label{eq:boostonone}
\end{equation}
This shows that boost operators change the momentum of a magnon state.
Conversely, for bilocal charges $\biloc{\charge_r}{\charge_s}$
each elementary charge will annihilate the state due to a vanishing
vacuum charge density (see above) unless it acts on the flipped spin. 
Therefore the only non-trivial contributions come from
where both charges overlap with the flipped spin. This is equivalent
to the action of a local operator and thus it cannot change the momentum
of the magnon state.

Let us now act with the deformation equation \eqref{eq:parallel,eq:covder,eq:connectall}
on the one-magnon state $\state{p}$. The only relevant contributions come
from boost and change of basis deformations
which have been combined into $\tilde\boostop_k$ in \Secref{sec:intrange}
\begin{equation}
\dd \charge
=i\comm{\tilde\boostop_k}{\charge}\tilde\boostform_k
=\sum_{k=3}^\infty i\bigcomm{(k-1)\boostop_k+\rotgen_k}{\charge}\tilde\boostform_k.
\end{equation}
For $\comm{\boostop_j}{\charge_\ell}=\comm{\boost{\charge_j}}{\charge_\ell}$ we need to compute 
the commutator of a boosted shift with a shift operator
\begin{equation}
\bigcomm{\boost{\shift^j}}{\shift^\ell} \ket{p}
= 
i\,\frac{\partial}{\partial p} \left( e^{ipj}e^{ip\ell}\ket{p}\right)
-i e^{ip\ell} \frac{\partial}{\partial p} \left( e^{ipj}\ket{p} \right) 
=i e^{ipj}\lrbrk{\frac{\partial}{\partial p}\, e^{ip\ell}}\ket{p} \,,
\end{equation}
such that the commutator of a boosted charge $\boost{\charge_r}$ with
a charge $\charge_s$ acts on a one-magnon state as
\begin{equation}
\bigcomm{\boost{\charge_r}}{\charge_s} \ket{p} = i q_r(p)\, \frac{\partial q_s(p)}{\partial p}\, \ket{p} \,.
\end{equation}
This implies the following differential equation for the
one-magnon eigenvalues $q_r$:
\begin{equation}
\dd q_r(p)
=\sum_{k=3}^\infty\left(-(k-1)q_k\frac{\partial q_r}{\partial p}+(k+r-2)\,q_{r+k-1} \right)\tilde\boostform_k\,.
\label{eq:diffqn}
\end{equation}
We now prove that the solution to the above differential equation
\eqref{eq:diffqn} is given by the well-known form of the 
one-magnon eigenvalues \cite{Beisert:2004hm}
\begin{equation}
q_r(t,u) = \frac{i}{n-1}\left(\frac{1}{x(u+\ihalf t)^{r-1}} -\frac{1}{x(u-\ihalf t)^{r-1}}\right)\,.
\label{eq:qr}
\end{equation}
The rapidity map $x(u)=u+\order{\alpha_k}$ was introduced in \eqref{eq:rapmap}.
The map between the
momentum $p$ and the rapidity $u(p)$ is implicitly defined through
\begin{equation}
\exp\bigbrk{ip(t,u)} = \frac{x(u+\ihalf t)}{x(u-\ihalf t)} \,.
\label{eq:momrap}
\end{equation}
Note that $t$ is a constant of integration that can be
freely chosen in the above equations.%
\footnote{Any other linear combination 
of terms with different values of $t$ is permissible in $p,q_r$ as well.}
We now rewrite the above differential equation using the 
parameter $u$ instead of the momentum $p$. 
For the differential of $q_r$ this implies
\begin{equation}
\dd q_r(u)=\dd q_r(p)+\frac{\partial q_r}{\partial p}\, \dd p(u).
\label{eq:dqn}
\end{equation}
Here, the differential operator $\dd$ acts only on the $\xi_j$ on which
the functions $q_r(u)$, $q_r(p)$ explicitly depend. That is
$u$ is held fixed in $\dd q_r(u)$ while $p$ is held fixed in
$\dd q_r(p)$, as is the case in \eqref{eq:diffqn}.
Using the defining equation of
$\tilde\boostform_k$ \eqref{eq:alphamap} and \eqref{eq:qr,eq:momrap}, we can compute
and simplify $\dd p(u),\dd q_r(u)$
\<
\dd p(u)
\eq-i\left(\frac{\dd x(u+\ihalf t)}{x(u+\ihalf t)}-\frac{\dd x(u-\ihalf t)}{x(u-\ihalf t)}\right)
=\sum_{k=3}^\infty (k-1)\tilde\boostform_k \,q_k\,,
\label{eq:dpexpl}
\\
 \dd q_r(u)\eq -i\left(\frac{\dd x(u+\ihalf t)}{x(u+\ihalf t)^r}-\frac{\dd x(u-\ihalf t)}{x(u-\ihalf t)^r}\right)
=\sum_{k=3}^\infty  (r+k-2)\tilde\boostform_k\, q_{r+k-1}\,.
\label{eq:dqnexpl}
\>
The result for $\dd p(u)$ can also be obtained from the result for $\dd q_r(u)$
by interpreting the momentum operator as the first charge $p=q_1$.
Plugging \eqref{eq:dqnexpl,eq:dpexpl} into \eqref{eq:dqn} then yields
the differential equation \eqref{eq:diffqn}, 
which shows that $p(u),q_r(u)$ provides the correct long-range charge eigenvalues. 

At this point we can observe and disentangle the effect of the boost 
and basis rotation in \eqref{eq:diffqn}:
The boosts are responsible for a deformation of the momentum function $p(u)$
while the deformations of the function $q_r(u)$ are caused solely
by a change of basis.

To understand the role of the integration constant $t$, let us turn off 
the deformation moduli $\alpha_k=0$ such that $x(u)=u$
and consider the resulting dispersion relations \eqref{eq:qr,eq:momrap}
\begin{equation}
\label{eq:rapdefgen}
\exp\bigbrk{ip\supups{NN}(t,u)}=\frac{u+\ihalf t}{u-\ihalf t}\,,
\qquad
q\supups{NN}_r(t,u) = \frac{i}{r-1}\left( \frac{1}{(u+\ihalf t)^{r-1}} -\frac{1}{(u-\ihalf t)^{r-1}} \right).
\end{equation}
It is well-known that the functions $p\supups{NN}(t,u),q\supups{NN}_r(t,u)$ 
define the dispersion relation for a Bethe root $u$ 
where $t/2$ is the spin label of the spin representation.

\subsection{Two-Magnon States and Scattering (\texorpdfstring{$\beta$}{beta})}\label{sec:dressingphase}

Two-magnon states are states where the spin has been flipped at two positions
\[
\state{k,\ell}=\ldots\otimes\state{\MMM{\mathrm{NHW}}{k}}\otimes\ldots
\otimes\state{\MMM{\mathrm{NHW}}{\ell}}\otimes\ldots
\,.
\]
When the two spins are far enough apart,
i.e.\ when the range of the Hamiltonian is smaller
than their separation, it is safe to assume 
partial momentum eigenstates
\[
\state{p<p'}=\sum_{k\ll \ell} e^{ipk+ip'\ell} \state{k,l}.
\]
This state describes two magnons of momenta $p,p'$ 
in the asymptotic region where the magnon with momentum $p$ 
is to the left of the magnon with momentum $p'$.
A two-magnon scattering state $\state{p,p'}$ can be written 
as a linear combination of the two asymptotic regions
$\state{p<p'}$ and $\state{p'<p}$
\[
\state{p,p'}\simeq A(p,p')\state{p<p'}+A(p',p)\state{p'<p}.
\]
This expression is valid in the IR, 
we have not payed attention to UV terms 
$\sum_{k\approx k'}\state{k,k'}$ where the 
two magnons are nearby.
These terms are needed in the computation of the
scattering factor which relates the phase in the two asymptotic regions
\[
S(p,p')=\frac{A(p',p)}{A(p,p')}\,.
\]
Consequently, the scattering factor summarizes the effect of 
local interactions between the spins on IR physics,
and it is the quantity that we need to determine.

Before we continue, let us make a change to the 
labels of magnon states. It is useful to replace
the momentum $p$ by the rapidity $u(p)$ and define
a magnon state with definite rapidity
\[
\state{u}=F(u)\state{p(u)},
\]
where $F(u)$ is a convenient normalization factor.
The main difference between $\state{p}$ and $\state{u}$ is that
the former depends only on $p$ while the latter depends on $u$
and implicitly on the moduli. 
We postulate a differential equation for the normalization factor
\[
\frac{\dd F(u)}{F(u)}\,\state{u}=\boostform_k \, \frac{\partial q_k}{\partial p}\,
\state{u}+i\locform_k\loc_k\state{u}.
\]
Using \eqref{eq:dpexpl,eq:boostonone} one finds a simple differential equation 
for the state
\[
\dd\state{u}
=\dd F(u)\state{p}
+F(u)\,\dd p(u) \frac{\partial}{\partial p}\,\state{p}
=
i(\boostform_k \boostop_k+\locform_k\loc_k )\state{u}
=
i(\boostform+\locform)\state{u}.
\label{eq:du}
\]
It is useful in so far 
as to cancel the effect of local, boost and basis change deformations 
in the differential equation \eqref{eq:parallel,eq:covder,eq:connectall}
acting upon the eigenvalue equation for $\charge$
\<
0\eq
\dd\bigbrk{\charge\state{u}-q(u)\state{u}}
=
\dd\charge \state{u}+\charge\dd\state{u}
-\dd q(u)\state{u}-q(u)\dd\state{u}
\nln
\eq
i\comm{\defform}{\charge}\state{u}
-\dd q(u)\state{u}
+i\charge (\boostform+\locform)\state{u}
-i q(u) (\boostform+\locform)\state{u}
\nln
\eq
i\comm{\defform-\boostform-\locform}{\charge}\state{u}
-\dd q(u)\state{u}
\nln
\eq
i\comm{\defform-\boostform-\rotform-\locform}{\charge}\state{u}
=
i\comm{\biform}{\charge}\state{u}\,,
\>
where in the last line we made use of \eqref{eq:dqnexpl}.
We are thus left with only bilocal deformations $\biform$.

Consider now the two-magnon scattering state with the 
corresponding eigenvalue equation
(we discard contributions where the two magnons are close)
\<
\state{u,u'}\earel{\simeq} A(u,u')\state{u<u'}+A(u',u)\state{u'<u},
\nln
\charge\state{u,u'}\eq \bigbrk{q(u)+q(u')}\state{u,u'}.
\>
Differentiating the eigenvalue equation we are led to the following
equation:
\<
0\eq
\dd \bigl[\bigbrk{\charge-q(u)-q(u')}\state{u,u'}\bigr]
\nln\eq
\dd\charge\state{u,u'}-\dd q(u)\state{u,u'}-\dd q(u')\state{u,u'}
\nl
+\bigbrk{\charge-q(u)-q(u')}\bigbrk{A(u,u')\dd\state{u<u'}+A(u',u)\dd\state{u'<u}}
\nl
+\bigbrk{\charge-q(u)-q(u')}\bigbrk{\dd A(u,u')\state{u<u'}+\dd A(u',u)\state{u'<u}}
\\\nn\eq
i\comm{\biform}\charge\state{u,u'}
+\bigbrk{\charge-q(u)-q(u')}\bigbrk{\dd A(u,u')\state{u<u'}+\dd A(u',u)\state{u'<u}}.
\>
Here we used the equations
\begin{align*}
(\dd q(u) +\dd q(u'))\state{u,u'}&=i\comm{\rotform}{\charge}\state{u,u'}\,,\\
\dd\state{u<u'}&=i(\boostform+\locform)\state{u<u'}\,,
\end{align*}
which are generalizations of equations \eqref{eq:dqnexpl,eq:du} and
are valid up to UV terms in which the two magnons are close.
Again the effect of the boost, local and basis change deformations 
cancels out and we are left with 
a differential equation for the prefactor $A(u,u')$ which only depends on 
the bilocal deformations
\[
\frac{\dd A(u,u')}{A(u,u')}\,\state{u<u'}
=i\biform\state{u<u'}
=i\biform_{r,s}\biop_{r,s}\state{u<u'}.
\]

Partial two-magnon states
are obviously eigenstates of bilocal charges
\[
\biloc{\charge_r}{\charge_s} \state{u<u'}=
\bigbrk{q_r(u)\,q_s(u')+f_{r,s}(u)+f_{r,s}(u')}\state{u<u'},
\label{eq:bionordered}
\]
where $f_{r,s}$ is a local contribution from 
both local charges acting on one of the two magnons.
The differential equation for the prefactor $A$ has the simple solution
\[
A(u,u')=\exp \bigl[2i\beta_{r,s}\bigbrk{q_r(u)\,q_s(u')+f_{r,s}(u)+f_{r,s}(u')}\bigr]A_0(u,u').
\]
For the scattering factor $S(u,u')=A(u',u)/A(u,u')$ it implies the deformation
\[
S(u,u')=\exp \bigbrk{-2i\theta(u,u')}S\supups{NN}(u-u'),
\qquad
S\supups{NN}(u-u')=\frac{u-u'-i}{u-u'+i}\,.
\label{eq:scatteringfacstand}
\]
where $S\supups{NN}(u,u')$ is the undeformed scattering factor
and $\theta(u,u')$ is an overall scattering phase, 
the so-called dressing phase 
\begin{equation}
\theta(u,u')=\sum_{s>r=2}^{\infty}\beta_{r,s}\bigbrk{q_r(u)q_s(u')-q_s(u)q_r(u')}.
\label{eq:dressingphase}
\end{equation}

The form of the dressing phase 
as an antisymmetric combination of two magnon charges \eqref{eq:dressingphase}
was proposed in \cite{Arutyunov:2004vx} based on physical intuition. 
It however remained somewhat unclear why this form applies 
to long-range spin chains \cite{Beisert:2004jw}. 
In fact one can argue that $\theta(u,u')$
provides a basis for generic antisymmetric functions
which vanish at $u=\infty$.
This is true because $q_r(u)$ is a basis for alike functions of a single variable. 
In that sense, the form of the dressing phase is natural, because 
the phase is antisymmetric by construction.
Nevertheless, it does not take
into account that the coefficients $\beta_{r,s}$
satisfy certain perturbative bounds \cite{Beisert:2005wv}
which would not hold in an arbitrary basis for antisymmetric functions
of $u,u'$.
Namely, at each order in the deformation parameter $\lambda$, only
finitely many of the $\beta_{r,s}$ may be non-vanishing.

\subsection{Basis of Charges (\texorpdfstring{$\gamma$}{gamma})}
\label{sec:basisrap}

As indicated above in \eqref{eq:postrot}, 
we are still free to perform a change of basis of the charge vector $\charge$ 
after the long-range deformations have been applied. 
These simply correspond to taking linear combinations of the long-range charges 
which do not affect the scattering matrix or the function $p(u)$
but modify the charge eigenvalues in an obvious way:
\begin{equation}
 \charge_r\mapsto \gamma_{r,0} \lengthop + \gamma_{r,s} \charge_s  
\quad\Longrightarrow\quad 
Q_r\mapsto \gamma_{r,0}N+\gamma_{r,s} Q_s.
\end{equation}

\subsection{Higher Rank and Multiple Magnons}
\label{sec:higherrank}
Now we wish to generalize the above results 
to a Lie (super) algebra $\alg{g}$ of higher rank $R$. 
The Lie algebra has $R$ simple roots, and we shall say that
they are distinguished by their \emph{flavor} $a=1,\ldots,R$.
The algebra is specified by a symmetric Cartan matrix $C_{a,b}$.
The spins transform in a representation 
of the Yangian $\grp{Y}(\alg{g})$ specified through Dynkin labels $t_a$ 
of a highest-weight representation of $\alg{g}$.%
\footnote{To have one Dynkin label for each flavor is merely the simplest case. 
In more complicated cases, e.g.\ tensor product representations,
there can be more than one Dynkin label associated to
each flavor. Also the two shifts $+\ihalf t$ and $-\ihalf t$ appearing in 
\eqref{eq:qr,eq:momrap} can in principle be chosen independently.}
The framework for undeformed nearest-neighbor chains
was developed in \cite{Reshetikhin:1983vw,Reshetikhin:1985vd,Ogievetsky:1986hu}.

The generalization to higher rank consists in adding a flavor to each magnon. 
In a multi-magnon state the number of magnons of flavor $a$ will be denoted by $M_a$. 
Their rapidities will be denoted by $u_{a,k}$, $k=1,\ldots,M_a$.
A multi-magnon state is consequently denoted by
\[\ket{\set{u_{a,k}}}.\]

As usual, due to integrability, the multi-magnon eigenvalues $Q_r$ 
of the charges $\charge_r$ are given by sums over single-magnon eigenvalues,
and similarly for the total momentum $P$
\begin{equation}
P=\sum_{a=1}^R\sum_{k=1}^{M_a} p(t_a,u_{a,k}),\qquad
Q_r=\sum_{a=1}^R\sum_{k=1}^{M_a} q_r(t_a,u_{a,k}).
\end{equation}
Here the spin label $t$ for the dispersion relation of a magnon of flavor $a$ 
is the Dynkin label $t=t_a$ corresponding to the same flavor $a$.
This follows from the differential equation for the
dispersion relations \eqref{eq:diffqn} 
which still holds in the case of higher rank.
The value of the integration constant $t$ follows 
from the known result for undeformed models.

The nested Bethe ansatz ensures that scattering of magnons 
of two flavors is diagonal. Thus there is only
one scattering factor $S_{a,b}(u,u')$ for each pair
of magnon flavors $a,b$. 
The scattering factor obeys the same differential 
equation as before and thus the deformation takes the form
\[\label{eq:sdeformgen}
S_{a,b}(u,u')=\exp\bigbrk{-2i\theta_{a,b}(u,u')}\,S\supups{NN}_{a,b}(u-u').
\]
The undeformed scattering factor for magnons of flavor $a,b$ reads
\cite{Reshetikhin:1983vw,Reshetikhin:1985vd,Ogievetsky:1986hu}
\[
S\supups{NN}_{a,b}(u-u')
=
\frac{u-u'-\ihalf C_{a,b}}{u-u'+\ihalf C_{a,b}}\,.
\label{eq:SNN}
\]
The dressing phase is generated by the bilocal deformation 
and hence we have to use the relevant charge eigenvalues for 
each magnon flavor
\begin{equation}
\theta_{a,b}(u,u')=\sum_{s>r=2}^{\infty}\beta_{r,s}\bigbrk{q_r(t_a,u)q_s(t_b,u')-q_s(t_a,u)q_r(t_b,u')}.
\label{eq:dressinggen}
\end{equation}

\subsection{Closed Chain Bethe Equations}
\label{sec:beteheq}

In order to go from an infinite spin chain 
to a closed or open finite chain, one has to impose boundary conditions on the system. 
For a long-range system on a closed chain of length $N$ 
the periodicity condition for 
the multi-magnon wave function $\state{\set{u_{a,k}}}$ reads
\begin{equation}
\exp\bigbrk{ip(t_a,u_{a,k})N}
= \mathop{\prod_{b=1}^R\prod_{j=1}^{M_b}}_{(b,j)\neq(a,k)}
S_{a,b}(u_{a,k},u_{b,j})
\quad\mbox{for } a=1,\ldots,R \mbox{ and }k=1,\ldots,M_a.
\label{eq:bethegen}
\end{equation}
The momentum function $p(t,u)$ and the scattering factors 
$S_{a,b}(u)$ have been given in the previous subsection.

Note that these Bethe equations are merely \emph{asymptotic}
\cite{Sutherland:1978aa,Serban:2004jf}, they are valid only up to a 
certain perturbative order in the moduli. 
This is because the scattering factor is an IR 
quantity and the distinction between IR and UV 
is given by the range of the Hamiltonian. 
As long as the range of one of the conserved charges
does not exceed the length of the chain $N$,
the above asymptotic Bethe equations give the correct spectrum
for this particular charge. Otherwise there can be
some UV contributions that we have not taken into account 
in the above derivation.
The range of $\charge_r$ is discussed in \Secref{sec:intrange}.
If one uses the one-parameter deformation 
$\charge_r(\lambda)=\sum_n \lambda^n \charge_r^{(n)}$
then the range of $\charge_r^{(n)}$ is $r+n$. 
Consequently, the Bethe ansatz gives the correct
eigenvalue at order $\lambda^{N-r}$
while the order $\lambda^{N-r+1}$ is not properly defined.
See also \Secref{sec:inhomeigen} for a different approach,
but with similar conclusions on the validity of the results.

Here we have taken the point of view that the 
dispersion relation and the scattering matrix have been deformed. 
There is another point of view that deserves being mentioned: 
The generating equation for our long-range integrable system
\eqref{eq:deformationX}
shows that the latter is obtained by a similarity transformation
on a nearest-neighbor model:
\[
\charge_r(\xi)=\transfer(\xi) \charge\supups{NN}_r \transfer(\xi)^{-1}.
\]
It implies that the spectra of the operators must be identical.
This is indeed true for infinite chains where the spectrum
is continuous. 
On a finite system, however, the proposed deformations 
need not be defined consistently. 
For example, the boost operator \eqref{eq:boostdef} requires the definition 
of a spin chain origin. 
However, the origin is not equivalent to the site shifted by $L$ steps. 
In other words, the definition of the boost deformation is not
compatible with closed boundary conditions. 
We list the various compatibilities between boundary conditions
and deformations in \Tabref{tab:boundcomp}.
Whenever a boundary condition is compatible with a deformation,
we can apply the above similarity transformation 
without deforming the spectrum. 
For compatible deformations the corresponding modulus 
will not appear in the Bethe equations. 
In particular, this was observed in \cite{Beisert:2008cf}
for the deformation $\biop_{2r,2s}$ on an open chain
and led to the discovery of the generating equation
by reversing the argument.

\begin{table}\centering
\begin{tabular}{|c|c|c|c|}\hline		
deformation&infinite&closed&open\\
generator&chain&chain&chain\\\hline
$\boostop_k$&compatible&incompatible&incompatible\\
$\biop_{2r+1,2s}$&compatible&incompatible&incompatible\\
$\biop_{2r+1,2s+1}$&compatible&incompatible&not integrable\\
$\biop_{2r,2s}$&compatible&incompatible&compatible\\
$\rotgen_{m,n}$&compatible&compatible&compatible\\
$\loc_l$&compatible&compatible&compatible\\\hline
\end{tabular}
\caption{Different types of deformations and their compatibility with
the infinite, closed and open boundary conditions. Incompatibility
corresponds to a twist of the boundary conditions. The Bethe equations
remain undeformed for compatible transformations. On an open chain,
deformations by bilocal operators $\biop_{2r+1,2s+1}$ composed of only
odd charges violate the boundary Yang-Baxter equation
\cite{Beisert:2008cf}.}
\label{tab:boundcomp}
\end{table}

\begin{figure}\centering
\includegraphics[scale=1.5]{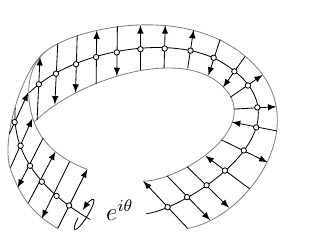}
\caption{The long-range deformations presented in this 
work can also be viewed as twists of the boundary conditions.}
\label{fig:twistlala}
\end{figure}

The alternative point of view is that an incompatible deformation 
twists the boundary conditions (\Figref{fig:twistlala}). 
Consequently, a closed long-range 
spin chain is nothing but a nearest-neighbor model with 
unusual boundary conditions.
For that purpose one would write the above asymptotic Bethe equations \eqref{eq:bethegen}
as the Bethe equations for a nearest-neighbor chain with a twist
\begin{equation}
\exp\bigbrk{ip\supups{NN}(t_a,u_{a,k})N}
=
\exp\bigbrk{i\phi_{a,b}(\set{u_{b,j}})}
 \mathop{\prod_{b=1}^R\prod_{j=1}^{M_b}}_{(b,j)\neq(a,k)}
S\supups{NN}_{a,b}(u_{a,k},u_{b,j}).
\end{equation}
The twist phase depends on all Bethe roots and 
for the long-range system it reads
\[
\phi_{a,k}(\set{u_{b,j}})=
\sum_{b=1}^R\sum_{j=1}^{M_b}2\theta_{a,b}(u_{a,j},u_{b,j})
-\bigbrk{p(t_a,u_{a,k})-p\supups{NN}(t_a,u_{a,k})}N.
\]
Written in this fashion,
the above Bethe equation is reminiscent of equations 
for finite-size spectra \cite{Teschner:2007ng}, 
see also \cite{Gromov:2009tv}.

\section{Alternating Spin Chains}
\label{sec:altchains}
The construction of integrable long-range deformations presented in this
paper is actually not limited to standard nearest-neighbor spin chains. 
A possible generalization of these models is given by \emph{alternating}
spin chain models (cf.\ \cite{Faddeev:1985qu,Destri:1989aa,deVega:1991rc,Abad:1997aa,Martins:1999aa}) which
recently attracted notice in the context of gauge/string dualities 
following the work \cite{Minahan:2008hf}. 

\paragraph{Long-Range Deformations.}

An alternating spin
chain is given by a tensor product of modules transforming in
alternating representations of the symmetry algebra:
\begin{equation}
\ket{\dots,v_k,\dots,v_{k+\ell},\dots}
\in\dots\,\otimes\spc{V}_\mathrm{e}\otimes\spc{V}_\mathrm{o}\otimes\spc{V}_\mathrm{e}\otimes\spc{V}_\mathrm{o}\otimes\,\dots.
\end{equation}
In fact, such a model can be brought to the form of
a standard homogeneous chain by combining
two adjacent modules into a larger one, 
$\spc{V}_\mathrm{e}\otimes\spc{V}_\mathrm{o}\to\spc{V}$.
Consequently the nearest-neighbor interactions 
in terms of $\spc{V}$ turn into next-to-nearest neighbor 
interactions in terms of the alternating modules $\spc{V}\indup{e/o}$.
With regard to our work, we would like to understand the range 
of the deformations in the alternating chain.

Particularly interesting is the case when the 
symmetry algebra splits into two components 
$\alg{g}=\alg{g}\indup{e}\oplus\alg{g}\indup{o}$.
Then the even spin sites can transform in a representation of 
$\alg{g}_\mathrm{e}$ while the odd sites transform in a
representation of $\alg{g}_\mathrm{o}$. 
Hence, we have two independent sets of commuting charges -- 
one for each of the two subchains:%
\footnote{In principle nothing prevents us from considering more
than two different alternating algebras or representations,
respectively. For illustration purposes we restrict to two different
symmetries here. In the general case one finds one set of conserved
charges for each connected component of the symmetry algebras' Dynkin
diagram.}
\begin{equation}
\charge_{n\mathrm{e}}^{(0)},\qquad
\charge_{n\mathrm{o}}^{(0)}.
\end{equation}
The action of $\charge_{n\mathrm{e}}^{(0)}$ and
$\charge_{n\mathrm{o}}^{(0)}$ is restricted to either of the two subchains
while the other subchain remains untouched. The same holds for the
corresponding boosts $\boost{\charge_{2\mathrm{e}}^{(0)}}$ and
$\boost{\charge_{2\mathrm{o}}^{(0)}}$. Hence, one obtains two sets
$\charge_{n\mathrm{e}}^{(0)}$, $\charge_{n\mathrm{o}}^{(0)}$ of
mutually commuting leading-order charges via \eqref{eq:nncharges}.

We can now use the two independent and mutually commuting sets of charges 
$\charge_{n\mathrm{e}}^{(0)}$, $\charge_{n\mathrm{o}}^{(0)}$
to construct deformations:
The boosts again exclusively act on the even/odd subchain, which in turn
implies that even/odd charges can only be deformed by even/odd boosts. 
Therefore there are two sets of boost deformation parameters
$\alpha_{k\mathrm{e}}$, $\alpha_{k\mathrm{o}}$ that correspond to
deformations by $\boost{\charge_{k,\mathrm{e}/\mathrm{o}}}$. 
Since even/odd boosts commute with odd/even charges, i.e.\
\begin{equation}
\bigcomm{\boost{\charge_{k\mathrm{x}}}}{\charge_{k\mathrm{y}}}=0,\qquad
\mathrm{x}\neq\mathrm{y}\in\{\mathrm{e},\mathrm{o}\},
\end{equation}
all charge terms that are multiplied by mixed powers of the $\alpha_k$ coefficients, 
such as $\alpha_{k\mathrm{e}}\alpha_{l\mathrm{o}}$ 
or $\alpha_{k\mathrm{e}}\alpha_{k\mathrm{o}}$, vanish. 
To summarize, all leading-order charges as well as all boost deformations 
on the two subchains completely decouple from each other. One obtains two
independent sets of boost-deformed models with symmetry
$\alg{g}_\mathrm{e}$ and $\alg{g}_\mathrm{o}$ and corresponding
parameters $\alpha_{k\mathrm{e}}$ and $\alpha_{k\mathrm{o}}$. The same
holds for bilocal deformations induced by operators composed of
\emph{only even} or \emph{only odd} charges
$\biloc{\charge_{r\mathrm{x}}}{\charge_{s\mathrm{x}}}$,
$\mathrm{x}\in\{\mathrm{e},\mathrm{o}\}$, with parameters
$\beta_{r\mathrm{x},s\mathrm{x}}$. So far the idea of an alternating
spin chain is only a notational issue: Two long-range chains of
symmetry $\alg{g}_\mathrm{e}$ and $\alg{g}_\mathrm{o}$ are written as
one alternating chain.

New structures appear when we introduce mixed bilocal deformations
$\biloc{\charge_{r\mathrm{x}}}{\charge_{s\mathrm{y}}}$,
$\mathrm{x}\neq\mathrm{y}\in\{\mathrm{e},\mathrm{o}\}$, corresponding
to new degrees of freedom $\beta_{r\mathrm{x},s\mathrm{y}}$. These
operators deform both, the $\mathrm{even}$ \emph{and} $\mathrm{odd}$
charge terms since bilocal operators that are composed of even and odd
charges generally neither commute with $\charge_{n\mathrm{e}}$ nor
with $\charge_{n\mathrm{o}}$. Therefore they result in
structures that couple the two subchains. Also these new structures
are fully described by the construction in the above sections.

\paragraph{Closed Chain Bethe Equations.}

The general Bethe equations \eqref{eq:bethegen} presented in
\Secref{sec:betheansatz} also apply to the alternating spin chain.
Considering two different alternating symmetries, the Bethe equations
have to be specified to the case of a product group, i.e.\ to the
direct sum $\alg{g}_\mathrm{e}\oplus\alg{g}_\mathrm{o}$. The Cartan
matrix $C_{a,b}$ in \eqref{eq:SNN} is then of block diagonal form
\begin{equation}
     C=
     \begin{pmatrix}
     C\indup{e}&0\\
     0&C\indup{o}
     \end{pmatrix} ,
\end{equation}
such that \eqref{eq:bethegen} splits into two sets of Bethe equations.
These two sets are only coupled by the dressing phases
$\theta_{a\mathrm{e},b\mathrm{o}}$ and
$\theta_{a\mathrm{o},b\mathrm{e}}$ \eqref{eq:dressinggen} with odd
\emph{and} even indices. Boost deformations do not induce a coupling
of the subchains. 

One gets two sets of magnons moving on either of the two chains
$\{u_{a\mathrm{e},k},u_{a\mathrm{o},k}\}$. Since the odd integrable
charges do not see magnons on the even spin chain and vice versa, 
\begin{equation}
Q_{n\mathrm{x}}=
\sum_{a=1}^{R\indup{x}}
\sum_{k=1}^{M_{a\mathrm{x}}}
q_{n\mathrm{x}}(t_{a\mathrm{x}},u_{a\mathrm{x},k})
,\qquad\mathrm{x}\in\{\mathrm{e,o}\},
\end{equation}
we have to distinguish two sets of one-magnon charge eigenvalues
\begin{equation}
q_{n\mathrm{x}}(t,u)
=\frac{i}{n-1}\left(
	 \frac{1}{x_\mathrm{x}(u+\ihalf t)^{n-1}}
	-\frac{1}{x_\mathrm{x}(u-\ihalf t)^{n-1}}
	\right),\qquad\mathrm{x}\in\{\mathrm{e,o}\}.
\end{equation}
Here, the rapidity map $x_{\mathrm{e/o}}(u)$ is parametrized by the
moduli $\alpha_{k\mathrm{e}}$ or $\alpha_{k\mathrm{o}}$, respectively. 
 
As an example consider the alternating spin chain with spins
transforming in the fundamental and anti-fundamental representation of
$\alg{su}(2)$, i.e.\
$\alg{g}_\mathrm{e}\oplus\alg{g}_\mathrm{o}=\alg{su}(2)\oplus\alg{su}(2)$
and
\begin{equation}
\begin{pmatrix}
	C_{\mathrm{e},\mathrm{e}}&C_{\mathrm{e},\mathrm{o}}\\
	C_{\mathrm{o},\mathrm{e}}&C_{\mathrm{o},\mathrm{o}}
\end{pmatrix}=
\begin{pmatrix}
	2&0\\
	0&2
\end{pmatrix}.
\end{equation}
The Bethe equations are given by
\begin{align}
\frac{x_\mathrm{e}(u_{\mathrm{e},k}+\ihalf)^N}{x_\mathrm{e}(u_{\mathrm{e},k}-\ihalf)^N}
&=\prod_{\textstyle\atopfrac{j=1}{j\neq k}}^{M_\mathrm{e}}
  \frac{u_{\mathrm{e},k}-u_{\mathrm{e},j}+i}{u_{\mathrm{e},k}-u_{\mathrm{e},j}-i}\,
  \exp\bigbrk{2i\theta_{\mathrm{e},\mathrm{e}}(u_{\mathrm{e},k},u_{\mathrm{e},j})}
  \prod_{\textstyle\atopfrac{\ell=1}{}}^{M_\mathrm{o}}
  \exp\bigbrk{2i\theta_{\mathrm{e},\mathrm{o}}(u_{\mathrm{e},k},u_{\mathrm{o},\ell})}
\,,\nln
\frac{x_\mathrm{o}(u_{\mathrm{o},k}+\ihalf)^N}{x_\mathrm{o}(u_{\mathrm{o},k}-\ihalf)^N}
&=\prod_{\textstyle\atopfrac{j=1}{j\neq k}}^{M_\mathrm{o}}
  \frac{u_{\mathrm{o},k}-u_{\mathrm{o},j}+i}{u_{\mathrm{o},k}-u_{\mathrm{o},j}-i}
  \exp\bigbrk{2i\theta_{\mathrm{o},\mathrm{o}}(u_{\mathrm{o},k},u_{\mathrm{o},j})}
  \prod_{\textstyle\atopfrac{\ell=1}{}}^{M_\mathrm{e}}
  \exp\bigbrk{2i\theta_{\mathrm{o},\mathrm{e}}(u_{\mathrm{o},k},u_{\mathrm{e},\ell})}\,.
\end{align}
The mixed dressing phase takes the
form
\[
\theta_{\mathrm{x,y}}(u,u')
=-\theta_{\mathrm{y,x}}(u',u)
=\sum_{s,r=2}^\infty\beta_{r\mathrm{x},s\mathrm{y}}\,
q_{r\mathrm{x}}(u)\,q_{s\mathrm{y}}(u'),
\qquad
\mathrm{x}\neq\mathrm{y}\in\{\mathrm{e,o}\}\,.
\]
The two sets of Bethe equations are only coupled by this phase. 

\paragraph{The \texorpdfstring{$\alg{gl}(K\indup{e})\oplus\alg{gl}(K\indup{o})$}{gl(Ke)+gl(Ko)} Alternating Spin Chain.}

In order to explicitly study the $\alg{gl}(K\indup{e})\oplus\alg{gl}(K\indup{o})$ chain,
we introduce two new types of permutation symbols whose first leg acts
on even/odd spin chain sites only
\begin{equation}
\mathcal{L}_\mathrm{e}=[\pi(1),\pi(2),\dots,\pi(n)]_\mathrm{e}\,,\quad
\mathcal{L}_\mathrm{o}=[\pi(1),\pi(2),\dots,\pi(m)]_\mathrm{o}.
\end{equation}
The action of $[3,\gray{2},1]_\mathrm{e}$ for instance is given by
\begin{equation}
[3,\gray{2},1]_\mathrm{e}\,\ket{\dots,\stackrel{e}{1},\gray{\stackrel{\mathrm{o}}{2}},\stackrel{\mathrm{e}}{3},\gray{\stackrel{\mathrm{o}}{4}},\stackrel{\mathrm{e}}{5},\dots}
=\,\,\dots
\,+\,\ket{\dots,\stackrel{\mathrm{e}}{3},\gray{\stackrel{\mathrm{o}}{2}},\stackrel{\mathrm{e}}{1},\gray{\stackrel{\mathrm{o}}{4}},\stackrel{\mathrm{e}}{5},\dots}
\,+\,\ket{\dots,\stackrel{\mathrm{e}}{1},\gray{\stackrel{\mathrm{o}}{2}},\stackrel{\mathrm{e}}{5},\gray{\stackrel{\mathrm{o}}{4}},\stackrel{\mathrm{e}}{3},\dots}\,+\,\dots.
\end{equation}

\begin{figure}\centering
\includegraphicsbox{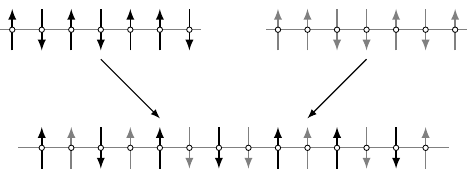}\qquad
\includegraphicsbox[scale=.7]{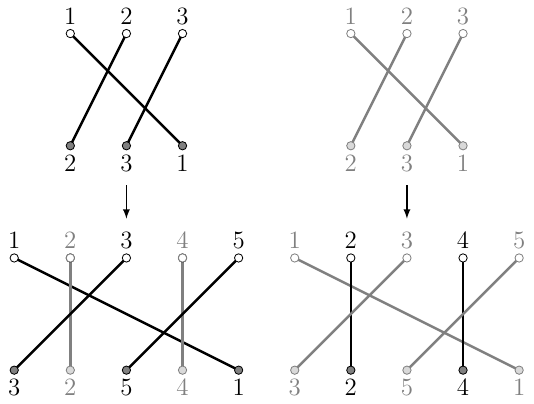}
\caption{The alternating 
$\alg{gl}(K\indup{e})\oplus\alg{gl}(K\indup{o})$ spin chain is given by two staggered versions of the standard
$\alg{gl}(K)$ spin chain. We can use the same
operator notation as before if we insert identity legs at every second
position of the permutation operators and let them act on even or odd positions of
the chain only.}
\label{fig:N6spinchain}
\end{figure}

The alternating $\alg{gl}(K\indup{e})\oplus\alg{gl}(K\indup{o})$ model
has two operators of minimum range 3 being 
the standard $\alg{gl}(K)$ Hamiltonian on the staggered subchains
\[
\charge_{2\mathrm{e}}^{(0)}=[1]_\mathrm{e}-[3,\gray{2},1]_\mathrm{e},\quad
\charge_{2\mathrm{o}}^{(0)}=[\gray{1}]_\mathrm{o}-[\gray{3},2,\gray{1}]_\mathrm{o}.
\label{eq:altham}
\]
Here we have simply taken the leading order $\alg{gl}(K)$ Hamiltonian
\eqref{eq:nncharges}, stretched it by a central identity leg and
restricted the action of the permutation symbol to even or odd sites,
respectively. For better readability we denote the part acting on
the odd subchain in gray. Both copies of the resulting Hamiltonian commute with
each other and thus describe a system of two staggered
nearest-neighbor spin chains. The form of the
Hamiltonian implies that also all leading-order charges
generated through \eqref{eq:nngen} have such an alternating structure
with identity legs on every other permutation site, e.g.\
\[
\charge_{3\mathrm{e}}^{(0)}=\ihalf\bigbrk{[5,\gray{2},1,\gray{4},3]_\mathrm{e}-[3,\gray{2},5,\gray{4},1]_\mathrm{e}},
\quad
\charge_{3\mathrm{o}}^{(0)}=\ihalf\bigbrk{[\gray{5},2,\gray{1},4,\gray{3}]_\mathrm{o}-[\gray{3},2,\gray{5},4,\gray{1}]_\mathrm{o}}.
\]
As discussed at the beginning of this section, this is due to the fact
that two local operators of this staggered form (including boosts)
acting on either even or odd sites only, commute with each other,
e.g.\
\begin{equation}
\bigcomm{[5,\gray{2},1,\gray{4},3]_\mathrm{e}}{[\gray{7},2,\gray{3},4,\gray{5},6,\gray{1}]_\mathrm{o}}=0 \,,
\quad\quad
\bigcomm{\boost{[\gray{3},2,\gray{1}]_\mathrm{o}}}{[3,\gray{2},5,\gray{4},1]_\mathrm{e}}=0 \,.
\end{equation}

Considering bilocal operators, we find the new
interaction terms described in the previous paragraphs. These
terms do not only act on one of the two staggered spin chains but on
both chains at the same time. Thus we obtain terms like
\begin{align}
\charge_{2\mathrm{e}}
&=\charge_{2\mathrm{e}}^{(0)}
 +\beta_{2\mathrm{e},3\mathrm{o}}\charge_{2\mathrm{e}}^{[2\mathrm{e}|3\mathrm{o}]}
 +\dots,
\quad\mbox{with}\\
\charge_{2\mathrm{e}}^{[2\mathrm{e}|3\mathrm{o}]}
&=2\,\bigbrk{-[3,\gray{4},5,\gray{6},1,\gray{2}]_\mathrm{e} +[3,\gray{6},5,\gray{2},1,\gray{4}]_\mathrm{e}
		+[5,\gray{4},1,\gray{6},3,\gray{2}]_\mathrm{e} -[5,\gray{6},1,\gray{2},3,\gray{4}]_\mathrm{e}\nln
	&\quad\quad -[\gray{3},4,\gray{5},6,\gray{1},2]_\mathrm{o} +[\gray{3},6,\gray{5},2,\gray{1},4]_\mathrm{o}
		+[\gray{5},4,\gray{1},6,\gray{3},2]_\mathrm{o} -[\gray{5},6,\gray{1},2,\gray{3},4]_\mathrm{o}} \,,\nn
\end{align}
which act nontrivially on both chains. Note, though, that even and
odd numbers are always on even and odd positions of the permutation
symbols, respectively. That is, also these interactions do not
interchange sites of the two staggered chains. Similarly, one finds
new structures proportional to mixed $\alpha\beta$ coefficients like
for instance $\alpha_{3\mathrm{e}}\beta_{2\mathrm{o},3\mathrm{e}}$. 

\paragraph{Interaction Range for \texorpdfstring{$\alg{gl}(K\indup{e})\oplus\alg{gl}(K\indup{o})$}{gl(Ke)+gl(Ko)}.}

All charge terms that emerge from deformations by boost or bilocal
operators acting on either the even or the odd chain exclusively are
given by stretched versions of the corresponding $\alg{gl}(K)$ terms.
Their interaction range is thus given by (cf.\
\eqref{eq:lengthofterm})
\begin{equation}
\bigrange{\charge_{n\mathrm{x}}^{[k_1\mathrm{x},\ldots,k_t\mathrm{x},r_1\mathrm{x}|s_1\mathrm{x},\ldots,r_u\mathrm{x}|s_u\mathrm{x}]}}
=2\biggbrk{n+\sum_{\ell=1}^t(k_\ell-2)+\sum_{\ell=1}^u(s_\ell-1)}-1\,,
\qquad\mathrm{x}\in\lrbrc{\mathrm{e},\mathrm{o}}.
\label{eq:rangealtchain}
\end{equation}
At leading order, the new interaction terms arising from deformations
by bilocal operators composed of one even and one odd charge
apparently have interaction range
\begin{equation}
\bigrange{\charge_{n\mathrm{x}}^{[r\mathrm{x}|s\mathrm{y}]}}
=
\begin{cases}
2n+2r-3\,, & r+1>s,\\
2n+2r-2\,, & r+1=s,\\
2n+2s-5\,, & r+1<s \,,
\end{cases}
\qquad\mathrm{x}\neq\mathrm{y}\in\lrbrc{\mathrm{e},\mathrm{o}}.
\label{eq:rangealtnewterms}
\end{equation}

\paragraph{Number of Crossings for \texorpdfstring{$\alg{gl}(2)\oplus\alg{gl}(2)$}{gl(2)+gl(2)}.}

Again, a particularly interesting case is the
$\alg{gl}(2)\oplus\alg{gl}(2)$ chain since it incorporates the
$\alg{su}(2)\oplus\alg{su}(2)$ sector spin chain of $\superN=6$
superconformal Chern-Simons theory. The natural building block (cf.\
the Hamiltonian \eqref{eq:altham}) corresponding to the term
``crossing'' for the alternating $\alg{gl}(2)$ spin chain is given by
the diagrams
\begin{equation}
\includegraphicsbox{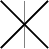}
\quad\text{and}\quad
\includegraphicsbox{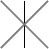}\,.
\label{eq:gl2cross}
\end{equation}
We take a look at the $\alg{gl}(K\indup{e})\oplus\alg{gl}(K\indup{o})$ Hamiltonian
printed in \Tabref{tab:N6Q2} at the end of this paper: The first
interesting operator is given by the permutations multiplied by
$\alpha_{3\mathrm{x}}$, $\mathrm{x}\in\{\mathrm{e,o}\}$. The term with
the highest number of elementary permutations is 
\begin{align}
 	&[5,\gray{2},3,\gray{4},1]_\mathrm{x}\,,\nln
	&\hspace{0.18cm}\includegraphics[scale=1]{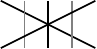}
\label{eq:mostelem}
\end{align}
which naively contains $7$ elementary permutations. This permutation
symbol, however, can be considered as the
combination of the two operators
\begin{align}
	&[3,2,1] \quad\text{and}\quad \gray{[1,2]}\,,\nln
	&\hspace{0.18cm}\includegraphics[scale=1]{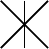}\hspace{2.1cm}\includegraphics[scale=1]{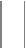}
\end{align}
acting on the two staggered spin chains simultaneously. As for the standard
$\alg{gl}(2)$ chain we can use the $\varepsilon_{ijk}$-identity
\eqref{eq:epsid3} for each of the alternating $\alg{gl}(2)$ symmetries
to reduce the number of crossings:
\begin{equation}
\begin{tabular}[b]{c@{\,$=$\,}c@{\,$+$\,}c@{\,$-2$\,}c@{\,$+$\,}c}
$[5,\gray{2},3,\gray{4},1]_\mathrm{x}$ &
$[3,\gray{2},5,\gray{4},1]_\mathrm{x}$ &
$[5,\gray{2},1,\gray{4},3]_\mathrm{x}$ &
$[3,\gray{2},1]_\mathrm{x}$ &
$[1]_\mathrm{x}$\,.\\[.15cm]
\includegraphicsbox{PicPermut52341stagg}\hspace{.15cm} &
\includegraphicsbox{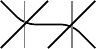}\hspace{.15cm} &
\includegraphicsbox{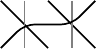}\hspace{.15cm} &
\includegraphicsbox{PicPermut321stagg}\hspace{.15cm} &
\includegraphicsbox{PicPermut1}\hspace{.35cm}
\end{tabular}
\end{equation}
In particular, we now have expressed the permutation symbol \eqref{eq:mostelem} 
in terms of the elementary crossings \eqref{eq:gl2cross}.  
Similarly, it appears to be possible to write the higher $\alpha_{k\mathrm{x}}$ 
as well as the  $\beta_{r\mathrm{x},s\mathrm{x}}$ structures 
in terms of elementary crossings without additional assumptions. 

The new terms corresponding to deformations with mixed bilocal operators 
$\biloc{\charge_{r\mathrm{x}}}{\charge_{s\mathrm{y}}}$ composed of charges
acting on the odd and even chain are given by the permutation operators 
multiplied by $\beta_{r\mathrm{x},s\mathrm{y}}$ (cf.\ \Tabref{tab:N6Q2}). 
An example is given by the following operator proportional to $\beta_{2x,3y}$:
\begin{align}
 	&[3,\gray{6},5,\gray{2},1,\gray{4}]_\mathrm{x}\,.\nln
	&\hspace{.2cm}\includegraphics[scale=1]{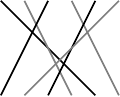}
\end{align}
This structure can be considered as the combination of
\begin{align}
 	&[2,3,1]\quad\text{and}\quad \gray{[3,1,2]}\,,\nln
	&\hspace{.2cm}\includegraphics[scale=1]{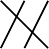}\hspace{1.9cm}\includegraphics[scale=1]{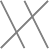}
\end{align}
such that using the identity
\begin{equation}
\includegraphicsbox{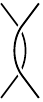}=\;\includegraphicsbox{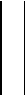}\;\,,
\end{equation}
we can again explicitly rewrite the diagram in terms of the
six-vertices \eqref{eq:gl2cross}:
\begin{equation}
\includegraphicsbox{PicPermut365214stagg}=\;\includegraphicsbox{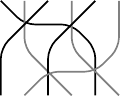}\;\;.
\end{equation}
Our experiments with the $\alg{gl}(K\indup{e})\oplus\alg{gl}(K\indup{o})$ Hamiltonian in
\Tabref{tab:N6Q2} therefore suggest a formula for the number of
crossings in analogy to \eqref{eq:crossingsallgl2}:
\begin{equation}
\crossings{\charge_{t\mathrm{x}}^{[k_1,\ldots,k_t,r_1|s_1,\ldots,r_u|s_u]}}_{\alg{gl}(2)\oplus\alg{gl}(2)}=t-1+\sum_{\ell=1}^u(r_\ell+s_\ell-2)+\sum_{\ell=1}^t(k_\ell-2).
\end{equation}

\paragraph{Parameter Restrictions from the
\texorpdfstring{$\alg{su}(2)\oplus\alg{su}(2)$}{su(2)+su(2)} Sector of
\texorpdfstring{$\superN=6$}{N=6} Superconformal Chern-Simons Theory.}

Considering restrictions from gauge theory, 
the main difference to the $\superN=4$ case 
is the scalar interaction
$\sim \lambda^2 Y^A Y_A^\dagger Y^B Y_B^\dagger Y^C Y_C^\dagger$ 
in the $\superN=6$ Lagrangian (cf.\ \cite{Minahan:2008hf}). 
Disregarding other interactions,
this translates to the vertex correspondence
\begin{equation}
\includegraphicsbox{PicPermut321stagg}
\quad\text{or}\quad
\includegraphicsbox{PicPermut321staggcol}\;\sim\lambda^2\,.
\label{eq:sixvertex}
\end{equation}
Accordingly, only even powers of the coupling $\lambda$ contribute
to the integrable charges and the spin chain and gauge theory Hamiltonian are related by
\begin{equation}
\ham(\lambda^2)=\ham^{(0)}
+\lambda^2\charge_{2\mathrm{e}}(\lambda^2) 
+\lambda^2\charge_{2\mathrm{e}}(\lambda^2) 
\end{equation}
We thus find the same gauge theory restrictions on the parameter functions 
as for the $\superN=4$ chain:
\begin{align}
\alpha_{k\mathrm{x}}(\lambda^2)&=\order{\lambda^{2(k-2)}},\nln
\beta_{r\mathrm{x},s\mathrm{y}}(\lambda^2)&=\order{\lambda^{2(r+s-2)}},\quad
\mathrm{x},\mathrm{y}\in\{\mathrm{e},\mathrm{o}\}.\label{eq:N6param1}
\end{align}
Similar to the $\superN=4$ case, we believe these restrictions to hold
also if one includes non-scalar interactions from the $\superN=6$
Lagrangian.

The staggered $\alg{gl}(K\indup{e})\oplus\alg{gl}(K\indup{o})$
Hamiltonian including all deformation degrees of freedom 
is printed at the end of this paper up to order $\order{\lambda^6}$.

\section{Inhomogeneous vs.\ Long-Range Spin Chains}
\label{sec:inhchain}

All of the above considerations were based on the 
assumption of infinite chains. 
Even the Bethe ansatz for the spectrum of closed finite chains 
in \Secref{sec:betheansatz} 
retains the notion of infinite extent:
It is asymptotic and valid only 
as long as the range of the interactions does not exceed the
length of the chain, that is only to finite order in the deformations
\cite{Sutherland:1978aa,Serban:2004jf} 
(see also \cite{Beisert:2004hm,Kotikov:2007cy} 
for explicit considerations of the failure beyond this order).

In the following we reconsider a relation 
of long-range to inhomogeneous integrable spin chains
without dressing phase, 
i.e.\ in the absence of bilocal deformations,
cf.\ \cite{Beisert:2004hm}
\begin{equation}
 \beta_{r,s}=0.
\end{equation}
It allows us to formulate consistently
a large class of integrable models on finite chains.
Unfortunately, the problem of undefined wrapping interactions
enters in this class of models as well.

For presentation purposes, we will only consider the simplest example 
of spins transforming in the fundamental representation of $\alg{su}(2)$,
a Heisenberg XXX$_{1/2}$ chain.
This is sufficient because only the momentum-carrying Bethe roots of the first level
will be important;
the auxiliary levels of the Bethe ansatz for a higher-rank symmetry algebra 
are manifestly identical on both sides of the relation.

\subsection{Bethe Equations}\label{sec:InhBethe}

The standard Bethe equations of a closed Heisenberg XXX$_{1/2}$ chain 
with inhomogeneities are given by
\begin{equation}\label{eq:inhBethe}
\frac{P_N(u_k+\ihalf)}{P_N(u_k-\ihalf)}=
\prod_{\textstyle\atopfrac{j=1}{j\neq k}}^M 
\frac{u_k-u_j+i}{u_k-u_j-i}\,,
\qquad\qquad 
P_N(u)=\prod_{k=1}^N (u-\mu_k).
\end{equation}
Here $P_N$ is a polynomial of degree $N$ incorporating the
inhomogeneities $\mu_k$. A homogeneous chain is obtained 
by setting all parameters $\mu_k$ to be the same, conventionally $\mu_k=0$.
The above equations are reminiscent of the BDS equations \cite{Beisert:2004hm}
for a closed long-range XXX$_{1/2}$ chain 
without bilocal deformations, $\beta_{r,s}=0$, 
\begin{equation}\label{eq:lrBethe}
\frac{x(u_k+\ihalf)^N}{x(u_k-\ihalf)^N}=
\prod_{\textstyle\atopfrac{j=1}{j\neq k}}^M 
\frac{u_k-u_j+i}{u_k-u_j-i}\,,
\qquad\qquad 
x(u)=u-\frac{\alpha_3}{u}+\dots.
\end{equation}
The boost parameters $\alpha_k$, however, are chosen generically
\begin{equation}
\label{eq:simpleboost}
u=x+\sum_{k=3}^\infty
\frac{\alpha_k}{u^{k-2}}\,,\qquad
 \alpha_k=\order{\lambda^{k-2}}.
\end{equation}

Now it has been observed that remarkably $x(u)^N$ 
is a polynomial of degree $N$ in $u$ 
up to terms of order $\order{\lambda^N}$ \cite{Beisert:2004hm}
\begin{equation}
\label{eq:Pisx}
 P_N(u)=x(u)^N+\order{\lambda^N}.
\end{equation}
There is a simple proof for this observation:
It is obvious from \eqref{eq:lrBethe}
that the expansion of $x(u)^N$ 
in powers of $\lambda$ contains 
only terms $u^k$ with $k\leq N$.
One then needs to show that 
all the terms with $k<0$ vanish
up to $\order{\lambda^N}$. This is equivalent to showing 
\[
\oint_{u=0} x(u)^N \,f(u)\,\dd u=\order{\lambda^N}
\]
for all functions $f(u)$ which are analytic at $u=0$. 
We perform a change of variable to $x$
\[
\oint_{x\approx 0} x^N \,f(u(x))\, u'(x)\,\dd x\stackrel{?}{=}\order{\lambda^N}.
\]
We note that the transformation \eqref{eq:simpleboost}
relates the point $u=0$ to two points $x=\order{\lambda^{1/2}}$
as well as several points $x=\order{\lambda}$ which all
lie near $x=0$.
We can thus expand the integrand around $x=0$ and 
pick out the residue at $x=0$.
The transformation \eqref{eq:simpleboost}
states that $u(x)$ consists of terms $\lambda^{k}/x^{k}$
and $u'(x)$ of terms $\lambda^{k}/x^{k+1}$.
Altogether the term $x^{N-N-1}$ in the integrand must be of order $\lambda^{N}$
which proves the above statement \eqref{eq:Pisx}.

The Bethe equations for the long-range chain are asymptotic,
they yield results which are valid up to terms of 
at most $\order{\lambda^N}$. Therefore it is safe
to replace $x(u)^N$ by a suitable $P_N(u)$ 
in \eqref{eq:lrBethe} and obtain 
exactly the inhomogeneous Bethe equations \eqref{eq:inhBethe}.
The latter are completely well-defined equations
even for finite chains.
For example it is safe to assume that
they are \emph{complete}, 
i.e.\ they have precisely $2^N$ physical solutions
reproducing the dimension of the Hilbert space
of a Heisenberg chain of length $N$.
Consequently the map 
appears to open a window to finite-size corrections
within long-range chains. 

\subsection{Charge Eigenvalues}\label{sec:inhomeigen}

For a solution of the above inhomogeneous Bethe equations, the eigenvalues
of the standard transfer matrix are given by the relation
\begin{equation}
\label{eq:baxinh}
T(u)=P_N(u+\ihalf)\prod_{k=1}^M\frac{u-u_k-i}{u-u_k}+P_N(u-\ihalf)\prod_{k=1}^M\frac{u-u_k+i}{u-u_k}\,.
\end{equation}
Alternatively this equation can be viewed as a Baxter equation: Absence of 
poles at $u=u_k$ is equivalent to the Bethe equations.
Consequently $T(u)$ is a polynomial of degree $N$
encoding the $N$ independent charge eigenvalues for this state.

As already seen, the eigenvalues of the long-range charges take the form
\begin{equation}\label{eq:qeigen}
 Q_r=\sum_{k=1}^M q_r(u_k)+\order{\lambda^{N+1-r}},\qquad 
q_r(u) = \frac{i}{r-1}\left( \frac{1}{x(u+\ihalf )^{r-1}} -\frac{1}{x(u-\ihalf )^{r-1}} \right). 
\end{equation}
Unfortunately, these charges make explicit reference to the magnons
whereas in standard integrable spin chains all
observable charges follow directly from 
transfer matrix eigenvalues $T(u)$.
This would also be preferable from the analytic Baxter-type equation 
point of view. 
Indeed we find a way to extract the above $Q_r$ directly 
from $T(u)$ by the following residue integral
\begin{equation}\label{eq:eigenintegral}
 Q_r=\mp\frac{1}{2\pi i}\oint\limits_{u=\pm i/2}\frac{i}{(r-1)x(u\mp \ihalf)^{r-1}}
\,\dd \log\frac{T(u)}{P_N(u\pm \ihalf)}+\order{\lambda^{N+1-r}}.
\end{equation}
Here the different signs $\pm$ indicate two possible choices yielding the same result. 
The contour of the integral is meant to encircle the small branch cut 
of $x(u\mp\ihalf)=u\mp\ihalf+\order{\lambda}$ near $u=\pm \ihalf$. 
Our curious observation is that \eqref{eq:eigenintegral} 
holds up to wrapping order in $\lambda$.

The equality can be proven as follows: 
There are two terms in $T(u+\ihalf)/P_N(u+i)$, cf.\ \eqref{eq:baxinh}.
If there was only the first term, 
\[
Q_r=-\frac{1}{2\pi i}\oint\limits_{u=0}\frac{i}{(r-1)x(u)^{r-1}}
\,\dd \log \prod_{k=1}^M\frac{u-u_k-\ihalf}{u-u_k+\ihalf}\,,
\]
the equality of \eqref{eq:qeigen}
and \eqref{eq:eigenintegral} follows simply by inverting the integration contour
and summing over the poles at $u=u_k\pm\ihalf$.
The remainder thus takes the form
\[
\mathrm{\Delta} Q_r
=-\frac{1}{2\pi}\oint\limits_{u=0}\frac{1}{(r-1)x(u)^{r-1}}
\,\dd \log\lrbrk{
1+\frac{P_N(u)}{P_N(u+i)}
\prod_{k=1}^M\frac{u-u_k+\sfrac{3i}{2}}{u-u_k-\ihalf}
}.
\]
To the order we are working at we can substitute $P_N(u)=x(u)^N$ which allows
us to write
\[
\mathrm{\Delta} Q_r
=
-\frac{1}{2\pi}\oint\limits_{u=0}\frac{\dd \log\lrbrk{1+x(u)^N F(u)}}{(r-1)x(u)^{r-1}}
=
-\frac{1}{2\pi}\oint\limits_{u=0}\frac{\dd \lrbrk{x(u)^N F(u)}}{(r-1)x(u)^{r-1}\bigbrk{1+x(u)^N F(u)}}
\,.
\]
Here $F(u)$ is some function analytic at $u=0$,
a fact we can use to simplify further
\[
\mathrm{\Delta} Q_r
=
-\frac{1}{2\pi}\oint\limits_{u=0}
\dd u\, F(u)\, x(u)^{N-r}x'(u)
=
\frac{1}{2\pi}\oint\limits_{u=0}
\frac{x(u)^{N+1-r}}{N+1-r}\,\dd F(u)
=\order{\lambda^{N+1-r}}.
\]
The point is that $x^{N+1-r}(u)$ is a polynomial 
\eqref{eq:Pisx} up to terms
of order $\order{\lambda^{N+1-r}}$ and hence the residue
integral must be trivial at this order.

\subsection{Charge Operators}

Above we have seen that the charge eigenvalues $Q_r$ for a 
long-range chain can be obtained from the transfer matrix
eigenvalues of an inhomogeneous chain. 
This statement on the spectra of commuting operators
can be lifted to a statement of the operators themselves:
Up to a similarity transformation, 
the long-range charges $\charge_r$ must equal 
\begin{equation}
\label{eq:chargeopint}
\charge_r\simeq -\frac{1}{2\pi i}\oint\limits_{u=0}\frac{i}{(r-1)x(u)^{r-1}}
\,\dd \log\frac{\transfer(u+\ihalf)}{P_N(u+i)}+\order{\lambda^{N+1-r}}.
\end{equation}
As an example we now want to see explicitly
how to obtain the first long-range deformation of the Hamiltonian $\charge_2$
from an inhomogeneous spin chain. In particular, we will
have to find a suitable similarity transformation.

We first expand the transfer matrix according to the
usual relation into a sequence of charges 
\begin{equation}
\label{eq:localexp}
\frac{\transfer(u+\ihalf)}{P_N(u+i)}=\shift \exp\lrbrk{ \sum_{r=2}^{\infty}i u^{r-1} \bar\charge_r}.
\end{equation}
Here $\shift$ denotes the operator that shifts the sites of the spin chain
by one unit. 
The above relation determines the long-range Hamiltonian $\charge_2$ 
in terms of the inhomogeneous charges $\bar\charge_r$
\[
\charge_2\simeq \bar\charge_2+3\lambda\bar\charge_4+\order{\lambda^2}.
\]
This is in fact precisely the same relation as derived in 
\Secref{sec:range.boost}.
The transfer matrix $\transfer$ is defined as 
the trace of a product of R-matrices 
(see \cite{Beisert:2004ry} for a review of the R-matrix formalism)
\begin{equation}
 \transfer(u+\ihalf)=\Tr\nolimits_0\rmat_{0,1}(u-\mu_1)\rmat_{0,2}(u-\mu_2)\dots\rmat_{0,N}(u-\mu_N).
\end{equation}
Up to an overall factor the $\alg{su}(2)$ R-matrix%
\footnote{In fact the considerations apply equally well to $\alg{su}(K)$
without modifications.}
in the fundamental 
representation takes the form
\begin{equation}
  \rmat_{k,l}(u)=u\,\idop_{k,l}+i\,\permop_{k,l}\,,
\end{equation}
where $\idop_{k,l}$ is the identity operator acting on the spin
sites $k$ and $l$, while $\permop_{k,l}$ exchanges the two spins.
The resulting inhomogeneous Hamiltonian $\bar\charge_2$ is given by
\<
\bar\charge_2
\eq
\sum_{k=1}^N \bigbrk{\PTerm{1}_k-\PTerm{2,1}_k}
+i\sum_{k=1}^N \mu_k \bigbrk{\PTerm{3,1,2}_{k-1}-\PTerm{2,3,1}_{k-1}}
\nl
+\sum_{k=1}^N\mu_k^2\bigbrk{-\PTerm{1}_k+\PTerm{2,1}_k+\PTerm{2,1}_{k-1}-\PTerm{3,2,1}_{k-1}}
\nl
+\sum_{k=1}^N \mu_{k-1}\mu_{k}
\bigbrk{\PTerm{2,3,4,1}_{k-2}-\PTerm{2,4,1,3}_{k-2}+\PTerm{4,1,2,3}_{k-2}-\PTerm{3,1,4,2}_{k-2}}
\nl
+\order{\lambda^{3/2}}.
\>
Note that the inhomogeneities $\mu_k$ are small and thus
to leading order the charges $\bar\charge_r$ coincide
with the charges $\charge\supup{NN}_r$ of a nearest neighbor chain;
only the higher orders in $\lambda$ contain inhomogeneous terms.

We now have to find a similarity transformation that
turns the inhomogeneous terms into homogeneous long-range interactions
\[
\label{eq:q2barsim}
\charge_2=\op{S}\bigbrk{\bar\charge_2+3\lambda\bar\charge_4+\ldots} \op{S}^{-1}.
\]
An ansatz for the similarity transformation $\op{S}$ relating the two
spin chain Hamiltonians is
\[
\label{eq:q2barsimop}
\op{S}=
\exp\lrbrk{
i\sum_{k=1}^N \nu_k \bigbrk{\PTerm{1}_k-\PTerm{2,1}_k}
-\sum_{k=1}^N\rho_k \bigbrk{\PTerm{3,1,2}_{k-1}-\PTerm{2,3,1}_{k-1}}+\ldots
}.
\]
In order to find suitable coefficients $\nu_k$ and $\rho_k$ 
we have to be more specific about the inhomogeneities $\mu_k$.
The factorization of the long-range polynomial $P_N(u)$ in \eqref{eq:inhBethe}
yields the following set of inhomogeneities \cite{Beisert:2004hm}
\begin{equation}\label{eq:inhomogen}
 \mu_k=2 \sqrt{\alpha_3} \cos{\frac{\pi (2k-1)}{2N}}+\frac{\alpha_4}{\alpha_3}\,
\cos{\frac{\pi (2k-1)}{N}}+\ldots\,.
\end{equation}
Note that the order of the inhomogeneities matters only to a certain extent;
different orderings are related by similarity transformations.
We choose the natural ordering for which the 
similarity transformation to the long-range model is the simplest.
The parameters $\nu_k$ and $\rho_k$ then have to be chosen as
\begin{equation}
 \nu_k=
\sqrt{\alpha_3}\,\frac{\sin{\frac{\pi k}{N}}}{\sin{\frac{\pi}{2N}}}
+
\frac{\alpha_4}{\alpha_3}\,\frac{\sin{\frac{2\pi k}{N}}}{2\sin{\frac{\pi}{N}}}
+\ldots
\,,\qquad 
\rho_k=\alpha_3\,\frac{\sin{\frac{\pi (2k-1)}{N}}}{2\sin{\frac{\pi}{N}}}+\ldots\,.
\end{equation}
This transformation \eqref{eq:q2barsim,eq:q2barsimop}
cancels the inhomogeneous $\sqrt{\alpha_3}$ term 
as well as the inhomogeneous terms at $\order{\alpha_3}$.
We then find the well known form of the
first-order long-range Hamiltonian \eqref{eq:leadboostdef}
\<
\charge_2\eq
\sum_{k=1}^N\bigbrk{\PTerm{1}_k-\PTerm{2,1}_k}
+\alpha_3\sum_{k=1}^N\bigbrk{-3\PTerm{1}_{k}+4\PTerm{2,1}_{k}-\PTerm{3,2,1}_{k-1}}
+\order{\lambda^{3/2}}
\nln
\eq\PTerm{1}-\PTerm{2,1}+\alpha_3\bigbrk{-3\PTerm{1}+4\PTerm{2,1}-\PTerm{3,2,1}}+\order{\lambda^{3/2}}.
\>

\subsection{Beyond Wrapping}

The asymptotic Bethe ansatz for long-range chains holds
only as long as the range of the interactions 
does not exceed the length of the chain. 
Beyond this so-called wrapping order, not only the Bethe equations fail
to provide consistent results \cite{Beisert:2004hm}, 
but also the construction allows for arbitrary corrections
to the spectrum \cite{Beisert:2004ry,Beisert:2008cf}.
With the above considerations 
one might contemplate to resolve these problems:
We could simply \emph{define} the undetermined wrapping interactions 
such that the mapping between inhomogeneous and long-range chains becomes exact.
While this solves the problem of consistency, it unfortunately does
not cure the arbitrariness as we shall see below.

First we replace the asymptotic Bethe equations \eqref{eq:lrBethe} 
by the inhomogeneous ones \eqref{eq:inhBethe}.
This is beneficial because the standard Bethe ansatz
is believed to be complete:
To every multiplet of eigenstates there corresponds precisely one solution 
of the Bethe equations. 
There is no doubt that this also applies for arbitrary inhomogeneities.
Conversely, the asymptotic Bethe equations \eqref{eq:lrBethe}
are apparently not complete because the number of 
acceptable solutions is different for small and for large $\lambda$.%
\footnote{We thank Didina Serban for pointing out this property
and for discussions.}
Thus we have gained an exact definition 
of the spectrum of $\op{T}(u)$ even for finite $\lambda$.

Secondly we choose the inhomogeneities $\mu_k$ 
in \eqref{eq:inhomogen} such that \eqref{eq:Pisx} holds.
However, this relation defines the inhomogeneities only up to a certain order,
it is not clear how to continue \eqref{eq:inhomogen} to all orders.
Nevertheless, this arbitrariness is only mild, because 
it introduces $N$ new degrees of freedom parametrizing 
a spectrum of $2^N$ states.
Furthermore, for the special choice of $\alpha_{k>3}=0$ \cite{Beisert:2004hm,Beisert:2005tm},
which is relevant to the gauge theory models discussed above,
there is a natural exact choice for $\mu_k$ 
consisting only in the first term in \eqref{eq:inhomogen}.

Finally we have to extract the charge eigenvalues $Q_r$ from
the transfer matrix eigenvalue $T(u)$. 
This is done via the integral \eqref{eq:eigenintegral}
which we can also define to be exact in $\lambda$.
There is nevertheless a residual ambiguity in this definition:
The function $x(u)$ contains a branch cut and the
contour of integration encircles it tightly. 
As long as the branch cut is infinitesimally small at $\lambda\approx 0$, 
there is a canonical way to define the contour. If at finite $\lambda$ 
the branch cut has finite extent, however, it can be deformed
and the result of the contour integral will depend on the shape.

Furthermore there is the option to add corrections at $\order{\lambda^{N-r+1}}$
to the expression for $Q_r$. 
These corrections introduce a huge arbitrariness
in that each eigenvalue can be deformed independently. This can be seen as follows:
The definition \eqref{eq:eigenintegral} is a map from the
transfer matrix eigenvalues $T(u)$ to the charges $Q_r$. 
The $T(u)$ are normalized polynomials of degree $N$ 
and can thus be viewed as elements of $\Complex^N$, i.e.\ 
we are interested in analytic functions $\Complex^N\to\Complex$.
It is straight-forward to construct an analytic function which 
maps a given set of $2^N$ vectors (the solutions to the Bethe equations) 
to a given set of $N$ numbers (the spectrum). 
Thus with a suitable definition for the map $T(u)\mapsto Q_r$ we 
can construct any desired spectrum for $Q_r$. 

If any desired spectrum can be obtained from an integrable spin chain model,
we may wonder what is special about integrability. 
The point is that the map $T(u)\mapsto Q_r$ must enjoy additional properties:
The expression \eqref{eq:eigenintegral} ensures that
only one of the two terms in the transfer matrix eigenvalue \eqref{eq:baxinh}
contributes to $Q_r$ up to terms of order $\order{\lambda^{N-r+1}}$.
This is the equivalent of the \emph{locality} property 
for the corresponding spin chain charges.
The logarithm of the transfer matrix $\log\transfer(u)$ 
has an expansion in terms of \emph{local} operators $\bar\charge_r$
according to \eqref{eq:localexp}.
Only linear combinations of the $\bar\charge_r$
lead to local operators while 
non-linear maps generate non-local terms.
And indeed the map $\log T(u)\mapsto Q_r$ in \eqref{eq:eigenintegral} is linear.
Unfortunately at wrapping order the notion of locality,
or analogously the irrelevance of one of the two terms in \eqref{eq:baxinh},
becomes meaningless. One would have to replace it
by a suitable property for the map $T(u)\mapsto Q_r$.

In conclusion,
the spectrum of long-range spin chains can be adjusted
arbitrarily beyond wrapping order,
at least in the absence of further insights or constraints
for the map $T(u)\mapsto Q_r$.
Let us merely present a possible scenario how the
arbitrariness could be cured: 
Ideally there would be a unique way to consistently 
complete the map $T(u)\mapsto Q_r$ beyond wrapping order 
which also removes the ambiguity of the integration contour. 
Whether or not this works remains to be seen.
It would also be interesting to find out whether
the bilocal deformation alias the dressing phase can somehow
be incorporated into the above framework. 

Let us comment on two interesting models 
which could be of help in resolving the above issue.
The first is the Inozemtsev spin chain
which can be viewed as a long-range model 
in a certain limit for the coupling constant \cite{Serban:2004jf}. 
What is nice about it is that the Hamiltonian
is well-defined for all couplings and thus
the spectrum beyond wrapping can be computed unambiguously. 
This model has an asymptotic Bethe ansatz
which fails already at half wrapping.
If one succeeds in completing the above transformation 
to an inhomogeneous Heisenberg chain, one would perhaps gain some
inspiration for different models. 
A similar model is the Hubbard chain 
whose spectrum is known for all values of its coupling constant.
A twisted sector was shown in \cite{Rej:2005qt} to be equivalent to 
the BDS chain \eqref{eq:lrBethe,eq:qeigen} \cite{Beisert:2004hm} up to wrapping terms. 
Therefore we can again reproduce the spectrum
by an inhomogeneous Heisenberg chain. 
In this case it is clear, however, that the two models cannot be 
equivalent at finite coupling: 
The separation of the above mentioned sector in the Hubbard model
cannot be formulated at finite coupling. 
In other words the Hilbert spaces are different at finite coupling;
the one of the Hubbard model is considerably larger.
Similarly, the Lieb--Wu equations for the Hubbard model 
are manifestly different from the Bethe equations
for an inhomogeneous Heisenberg chain.
Consequently, studying the Inozemtsev model will be 
more helpful for our purposes than the Hubbard model.

\section{Conclusions and Outlook}
In this paper we have presented a general framework for the construction of long-range integrable 
spin chain models. The most important results can be summarized as follows:
\begin{itemize}
\item On infinite chains an integrable long-range spin chain model
can be constructed to all orders in the deformation parameters.
\item In the long-range Bethe ansatz 
the boost and bilocal deformations twist the boundary conditions 
which results in the rapidity map \eqref{eq:rapmap,eq:rapdefgen} 
\begin{equation}
\lrbrk{\frac{u_{k}+\ihalf t}{u_{k}-\ihalf t}}^L
\to\lrbrk{\frac{x(u_{k}+\ihalf t)}{x(u_{k}-\ihalf t)}}^L\,,\qquad
u(x)=x+\sum_{k=3}^\infty\frac{\alpha_k}{x^{k-2}}\,,
\end{equation}
and the phase \eqref{eq:sdeformgen,eq:dressinggen}
dressing the otherwise undeformed nearest-neighbor scattering factors 
$S\supups{NN}(u_{k}-u_{j})\to S\supups{NN}(u_{k}-u_{j})\exp(2i\theta(u_{k},u_{j}))$:
\begin{equation}
\theta (u_{k},u_{j})
=\sum_{s>r=2}^\infty
\beta_{r,s}\,\bigbrk{q_r(u_{k})q_s(u_{j})
-q_s(u_{k})q_r(u_{j})}.
\end{equation}
\item The presented recursion relation provides a recipe to explicitly construct the 
long-range interactions of the integrable charges.
\end{itemize}

We shortly recapitulate the paper: Starting with an arbitrary set of 
integrable short-range charges on infinite spin chains we have defined long-range charges 
by parallel transport in moduli space with respect to a connection
$\defform$. The resulting operators are manifestly integrable.
Four different types of deformation operators were specified 
and the moduli space determined in \cite{Beisert:2005wv} was recovered. 
In particular, the rapidity map parameters $\alpha_k$ 
and the dressing phase parameters $\beta_{r,s}$ were related 
to the boost and bilocal deformations. 
They change the one-magnon momenta and the two-magnon scattering
phase, respectively, which is reflected in the Bethe equations. 
These appear to form a complete set of admissible long-range deformations of the
closed Bethe equations compatible with the gauge/string structure and
integrability.

A canonical basis of the charge algebra was chosen at each point of
the moduli space by associating a certain change of basis to the boost
deformations. This basis provides normalized charges with a minimal
interaction range that grows additively in the order of each
deformation parameter $\alpha_k$, $\beta_{r,s}$. In addition, the
chosen change of basis renders the connection $\defform$ on moduli space flat.

Applying our method to alternating spin chains, we have explicitly
constructed a generalized Hamiltonian for the fundamental 
$\alg{gl}(K\indup{e})\oplus\alg{gl}(K\indup{o})$ spin chain. The Bethe equations for the
alternating $\alg{su}(2)\oplus\alg{su}(2)$ chain were specified and
gauge theory restrictions on the moduli were discussed.

Finally we have presented a correspondence between long-range and 
inhomogeneous spin chains. 
Here, it was shown how the boost deformed long-range spin chain can be written 
in terms of an inhomogeneous chain. The inhomogeneity of the boost operator 
was thus reflected by a manifestly inhomogeneous chain. 

Several questions arise in this context:

The minimization of the interaction range of the charge operators in \Secref{sec:intrange} 
was based on extensive experiments with the $\alg{gl}(K)$ spin chain. 
It would be desirable to obtain a rigorous derivation 
of the minimizing conditions \eqref{eq:fixedforms}. 
This requires a better understanding of the range of 
the generated operators and its dependence on the moduli.
Presumably, the conditions can be derived by methods similar to those
used in \Appref{app:cancellation}.

For open long-range integrable spin chains the moduli space looks
different \cite{Beisert:2008cf}. Most notably, one finds an additional
parameter $\delta_k$ corresponding to a phase for reflections at the boundaries of
the chain.
Can the recursion presented in this paper be extended to the 
case of open spin chains?
For instance, it might seem natural to associate this phase
to bilocal deformations with the boundary-counting operator 
($\PTerm{1}-\PTerm{1,2}$ in our $\alg{gl}(K)$ notation)
on one leg of the bilocal operator.
Can we reproduce the moduli space for open chains in this fashion?

Other interesting integrable models include quantum-deformed 
spin chains. Again it is obvious that our construction
works for these models as well. However, there certainly are  
further permissible deformations once only the
Cartan algebra of $\alg{g}$ needs to be respected. 
These include the Reshetikhin twist deformations 
\cite{Reshetikhin:1990ep}
considered in 
\cite{Roiban:2003dw,Berenstein:2004ys,Frolov:2005ty,Beisert:2005if}.

The map between long-range and inhomogeneous spin chains was 
restricted to boost deformations; bilocal deformations were not included. 
This might have a natural origin in the inhomogeneity of the boost operator. 
Is it possible to map the bilocal long-range deformations to 
some other well-understood model of finite spin chains?

Trying to extend the presented method to sectors of $\superN=4$ or
$\superN=6$ gauge theory beyond the $\alg{su}(2)$ sectors, two main
obstacles appear: Firstly, the underlying algebra $\alg{psu}(2,2|4)$
or $\alg{osp}(6|4)$ is no longer manifest but must receive
long-range deformations in the same way as the charge operators. Even
more, the Hamiltonian becomes a part of the symmetry
algebra. To address these problems (cf.\ \Secref{sec:symgen}),
one might think about joining our
method with the ideas to deform the algebra of the $\alg{su}(1,1|2)$
sector of $\superN=4$ SYM theory presented in \cite{Zwiebel:2008gr}.
Secondly, the length of the spin chain in other gauge theory sectors
is no longer conserved. The quantum numbers of different numbers of
fields coincide such that length fluctuations are admissible; the spin
chain becomes dynamic \cite{Beisert:2003ys}. A starting point to
tackle this issue might be a transformation to an undynamic spin
chain as presented in \cite{Beisert:2008qy} for the
$\alg{su}(2|3)$ sector of $\superN=4$ SYM. Is it possible to extend the 
method presented in this paper to the whole gauge/string theory algebra?

\appendix

\section{Cancellation of Longest Terms}
\label{app:cancellation}

In the construction of deformations that minimized
the range of the charges, we made use of the relation
\eqref{eq:longestmatch}, which states that, up to an overall factor,
the longest-range terms of $\comm{\boost{\charge_s}}{\charge_r}$
and $\charge_{s+r-1}$ are the same:
\begin{equation}
 i(s-1)M\bigbrk{\comm{\boost{\charge_s}}{\charge_r}}
=-(s+r-2)M\brk{\charge_{s+r-1}}\,.
\label{eq:longestmatch2}
\end{equation}
Here, we denote by $M(\loc_k)$ the part of the local operator $\loc_k$
that has the longest range. In the following, we will prove the
statement \eqref{eq:longestmatch2} for the undeformed charges
$\charge_r\equiv\charge_r\supups{NN}$ that are generated by
\eqref{eq:nngen}. In this case, \eqref{eq:longestmatch2} reduces to
\begin{equation}
(s-1)M\bigbrk{\comm{\boost{\charge_s}}{\charge_r}}=M\bigbrk{\comm{\boost{\charge_2}}{\charge_{s+r-2}}}
\label{eq:longestmatch3}
\end{equation}
In order to understand this equality, we first note that the
longest-range terms of $(s-1)\comm{\boost{\charge_s}}{\charge_r}$ equal
the ones of $(r-1)\comm{\boost{\charge_r}}{\charge_s}$. Namely, the
longest-range terms of $\comm{\boost{\charge_s}}{\charge_r}$ are
given by
\begin{equation}
 M\bigbrk{\comm{\boost{\charge_s}}{\charge_r}}
=(r-1)\sum_k\bigcomm{\charge_s(k+r-1)}{\charge_r(k)}\,,
\label{eq:longterms}
\end{equation}
which follows directly from the definition \eqref{eq:boostdef} and the
fact that $\comm{\charge_s}{\charge_r}=0$. Using \eqref{eq:longterms},
one finds that indeed
\begin{equation}
 (s-1)M\bigbrk{\comm{\boost{\charge_s}}{\charge_r}}
-(r-1)M\bigbrk{\comm{\boost{\charge_r}}{\charge_s}}
=(s-1)(r-1)M\bigbrk{\comm{\charge_s}{\charge_r}}
=0\,.
\label{eq:longtermsequal}
\end{equation}
With the help of \eqref{eq:longtermsequal}, we can now prove
\eqref{eq:longestmatch3} inductively. For $r=2$, the equality follows directly
from \eqref{eq:longtermsequal}. Assuming that it holds for any given
$r=r_0$, we can show that it also holds for $r=r_0+1$. Namely, by
\eqref{eq:longtermsequal}, one finds
\begin{equation}
 (s-1)M\bigbrk{\comm{\boost{\charge_s}}{\charge_{r_0+1}}}
=r_0M\bigbrk{\comm{\boost{\charge_{r_0+1}}}{\charge_s}}\,.
\label{eq:matchleft}
\end{equation}
With the help of \eqref{eq:nngen} and \eqref{eq:longterms}, this can
be written as
\begin{align}
\label{eq:matchleft2}
 &\frac{i}{r_0-1}M\bigbrk{\comm{\boost{\charge_s}}{\charge_{r_0+1}}}\\
&=\frac{1}{(s-1)(r_0-1)}M\bigbrk{\comm{\boost{\comm{\boost{\charge_2}}{\charge_{r_0}}}}{\charge_s}}\nln
&=\sum_k\bigcomm{\comm{\charge_2(k+s+r_0-2)}{\charge_{r_0}(k+s-1)}}{\charge_s(k)}\nln[-0.2cm]
&=\sum_k\includegraphicsbox{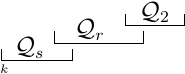}
  \,-\,\includegraphicsbox{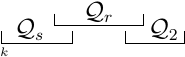}
  \,-\,\includegraphicsbox{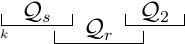}
  \,+\,\includegraphicsbox{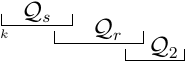}\,.\nn
\end{align}
The operators in the double commutator overlap only
by one site.
On the other hand, by the assumption that \eqref{eq:longestmatch2}
holds for $r=r_0$ and again using \eqref{eq:longterms}, one finds that
\begin{align}
\label{eq:matchright}
 &\frac{i}{(s-1)(r_0-1)}M\bigbrk{\comm{\boost{\charge_2}}{\charge_{s+r_0-1}}}\\
&=\frac{i}{(s-1)(r_0-1)}M\bigbrk{\comm{\boost{\charge_2}}{M\brk{\charge_{s+r_0-1}}}}\nln
&=\frac{1}{(r_0-1)(s+r_0-2)}M\bigbrk{\comm{\boost{\charge_2}}{M\bigbrk{\comm{\boost{\charge_s}}{\charge_{r_0}}}}}\nln
&=\sum_k\bigcomm{\charge_2(k+s+r_0-2)}{\comm{\charge_s(k+r_0-1)}{\charge_{r_0}(k)}}\nln[-0.2cm]
&=\sum_k\includegraphicsbox{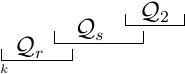}
  \,-\,\includegraphicsbox{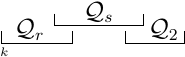}
  \,-\,\includegraphicsbox{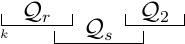}
  \,+\,\includegraphicsbox{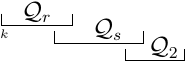}\,.\nn
\end{align}
Since in the commutator under the sum in \eqref{eq:matchleft2} all
operators overlap only by one site, the terms can be reshuffled to
yield
\begin{align}
  (s-1)&M\bigbrk{\comm{\boost{\charge_s}}{\charge_{r_0+1}}}
 -M\bigbrk{\comm{\boost{\charge_2}}{\charge_{s+r_0-1}}}\nln
&=-i(s-1)(r_0-1)\sum_k\big[\charge_2(k+s+r_0-2),\nln[-0.4cm]
&\mspace{210mu}\comm{\charge_{r_0}(k+s-1)}{\charge_s(k)}+\comm{\charge_{r_0}(k)}{\charge_s(k+r-1)}\big]\nln
&=-i(s-1)(r_0-1)\sum_k\bigcomm{\charge_2(k+s+r_0-2)}{M\brk{\comm{\charge_r}{\charge_s}}(k)}\,.
\label{eq:matchboth}
\end{align}
This expression vanishes, since
$M\brk{\comm{\charge_r}{\charge_s}}=0$. Hence,
\eqref{eq:longestmatch3} also holds for $r=r_0+1$, wich proves
\eqref{eq:longestmatch2} for arbitrary $r$.

\section{Reading Off the Charges}
\label{app:example}

As discussed in \Secref{sec:Range.Flatness},
the parametrization \eqref{eq:fixedforms} of the moduli space yields a
flat connection $\defform$ \eqref{eq:conn}. Hence the charges $\charge_r$
can be directly expanded in the deformation moduli $\alpha_k$,
$\beta_{r,s}$ \eqref{eq:expansionalphabeta} (up to local similarity
transformations, i.e.\ deformations \eqref{eq:deformationepsilon}
by local operators). Taking the generating equation
\eqref{eq:deformationall} as well as the redefinitions
\eqref{eq:fixedforms} into account, one can iteratively determine all
higher order terms $\charge_r^{[\ldots]}$ of the deformed charges in
terms of the undeformed charges $\charge_r^{(0)}$. By virtue of
equation \eqref{eq:nngen},
\begin{equation}
\charge_{r+1}^{(0)}=-\frac{i}{r}\bigcomm{\boost{\charge_2^{(0)}}}{\charge_r^{(0)}}\,,
\end{equation}
these can in turn be generated from the charge $\charge_2^{(0)}$
alone.

In the following we will explicitly carry out the expansion and
express the first few terms of
\eqref{eq:expansionalphabeta} in terms of the undeformed
charges $\charge_r^{(0)}$.
For the $\alg{gl}(K)$ chain, the resulting Hamiltonian $\charge_2$ is
given in \Tabref{tab:N4Q2}.
Let us start with the term proportional to
$\alpha_3$: Equation \eqref{eq:deformallgammafix} together with the
parametrization \eqref{eq:sf} yields
\begin{equation}
\frac{\partial\charge}{\partial\alpha_3}
=i\comm{\tilde\boostop_3}{\charge}+\ldots
=2i\comm{\boostop_3}{\charge}+i\comm{\rotgen_3}{\charge}+\ldots\,,
\end{equation}
which implies
\begin{align}
\charge_r&=\charge_r^{(0)}+\alpha_3 \charge_r^{[3]}+\dots\,,\nln
\charge_r^{[3]}&=2i\comm{\boost{\charge_3^{(0)}}}{\charge_r^{(0)}}+(r+1)\charge_{r+2}^{(0)}\,.
\end{align}
Similarly one finds all other terms proportional to $\alpha_k$:
\begin{align}
\charge_r&=\charge_r^{(0)}+\sum_{k=3}^\infty\alpha_k\charge_r^{[k]}+\dots\,,\nln
\charge_r^{[k]}&=(k-1)i\comm{\boost{\charge_k^{(0)}}}{\charge_r^{(0)}}+(r+k-2)\charge_{r+k-1}^{(0)}\,.
\label{eq:firstchargeorder}
\end{align}
The next thing to do is to consider terms proportional to higher powers in
$\alpha_k$. So
let us start with the term multiplied by $\alpha_3\alpha_4$:
Equation \eqref{eq:sf} shows that
\begin{align}
\frac{\partial\charge}{\partial\alpha_3}
&=i\comm{\tilde\boostop_3}{\charge}
+(4-2)\alpha_4i\comm{\tilde\boostop_{3+4-1}}{\charge}
+\dots\,,\nln
\frac{\partial\charge}{\partial\alpha_4}
&=i\comm{\tilde\boostop_4}{\charge}
+(3-2)\alpha_3 i\comm{\tilde\boostop_{4+3-1}}{\charge}
+\dots\,.
\end{align}
Due to the flatness of the connection, the coefficient
$\charge_r^{[4,3]}$ of $\alpha_4$ in
$\partial\charge/\partial\alpha_3$ equals the coefficient
$\charge_r^{[3,4]}$ of $\alpha_3$ in
$\partial\charge/\partial\alpha_4$ (up to local similarity transformations),
i.e.\
\begin{align}
\charge_r
&=\charge_r^{(0)}+\alpha_k\charge_r^{[k]}+\alpha_3\alpha_4\charge_r^{[4,3]}+\dots\nln
&=\charge_r^{(0)}+\alpha_k\charge_r^{[k]}+\alpha_4\alpha_3\charge_r^{[3,4]}+\dots\,,
\end{align}
where $\charge_r^{[34]}\simeq\charge_r^{[43]}$ and
\begin{align}
\charge_r^{[4,3]}
&=i\comm{\tilde\boostop_3^{[4]}}{\charge_r^{(0)}}
 +i\comm{\tilde\boostop_3^{(0)}}{\charge_r^{[4]}}
 +(4-2)i\comm{\tilde\boostop_{3+4-1}^{(0)}}{\charge_r^{(0)}}\nln
&=i\comm{2\boostop_3^{[4]}}{\charge_r^{(0)}}
 +i\comm{2\boostop_3^{(0)}}{\charge_r^{[4]}}
 +i\comm{\rotgen_3}{\charge^{[4]}}_r
 +2i\comm{5\boostop_6^{(0)}}{\charge_r^{(0)}}
 +2i\comm{\rotgen_6}{\charge^{(0)}}_r\nln
&=2i\comm{\boostop_3^{[4]}}{\charge_r^{(0)}}
 +2i\comm{\boostop_3^{(0)}}{\charge_r^{[4]}}
 +(r+1)\charge_{r+2}^{[4]}
 +10i\comm{\boostop_6^{(0)}}{\charge_r^{(0)}}
 +2(r+4)\charge_{r+5}^{(0)}\,,\nln
\charge_r^{[3,4]}
&=i\comm{\tilde\boostop_4^{[3]}}{\charge_r^{(0)}}
 +i\comm{\tilde\boostop_4^{(0)}}{\charge_r^{[3]}}
 +(3-2)i\comm{\tilde\boostop_{4+3-1}^{(0)}}{\charge_r^{(0)}}\nln
&=i\comm{3\boostop_4^{[3]}}{\charge_r^{(0)}}
 +i\comm{3\boostop_4^{(0)}}{\charge_r^{[3]}}
 +i\comm{\rotgen_4}{\charge^{[3]}}_r
 +i\comm{5\boostop_6^{(0)}}{\charge_r^{(0)}}
 +i\comm{\rotgen_6}{\charge^{(0)}}_r\nln
&=3i\comm{\boostop_4^{[3]}}{\charge_r^{(0)}}
 +3i\comm{\boostop_4^{(0)}}{\charge_r^{[3]}}
 +(r+2)\charge_{r+3}^{[3]}
 +5i\comm{\boostop_6^{(0)}}{\charge_r^{(0)}}
 +(r+4)\charge_{r+5}^{(0)}\,.
\end{align}
In other words, the terms $\charge_r^{[4,3]}$ and $\charge_r^{[3,4]}$
determine the coefficient of $\alpha_3\alpha_4$ equally well and
differ only by a local deformation \eqref{eq:deformationepsilon}.
Since we have
already determined $\charge_r^{[k]}$, we can plug in the result
\eqref{eq:firstchargeorder} to find
\begin{align}
\charge_r^{[4,3]}
&=2i\bigcomm{\boost{3i\comm{\boostop_4^{(0)}}{\charge_3^{(0)}}+5\charge_6^{(0)}}}{\charge_r^{(0)}}
 +2i\bigcomm{\boostop_3^{(0)}}{3i\comm{\boostop_4^{(0)}}{\charge_r^{(0)}}+(r+2)\charge_{r+3}^{(0)}}
\nln&\quad
 +(r+1)\bigbrk{3i\comm{\boostop_4^{(0)}}{\charge_{r+2}^{(0)}}+(r+4)\charge_{r+5}^{(0)}}
 +10i\bigcomm{\boostop_6^{(0)}}{\charge_r^{(0)}}
 +2(r+4)\charge_{r+5}^{(0)}\,,\nln
\charge_r^{[3,4]}
&=3i\bigcomm{\boost{2i\comm{\boostop_3^{(0)}}{\charge_4^{(0)}}+5\charge_6^{(0)}}}{\charge_r^{(0)}}
 +3i\bigcomm{\boostop_4^{(0)}}{2i\comm{\boostop_3^{(0)}}{\charge_r^{(0)}}+(r+1)\charge_{r+2}^{(0)}}
\nln&\quad
 +(r+2)\bigbrk{2i\comm{\boostop_3^{(0)}}{\charge_{r+3}^{(0)}}+(r+4)\charge_{r+5}^{(0)}}
 +5i\bigcomm{\boostop_6^{(0)}}{\charge_r^{(0)}}
 +(r+4)\charge_{r+5}^{(0)}\,.
\label{eq:terma34}
\end{align}
Now the whole structure is
expressed in terms of zeroth order charges $\charge_r^{(0)}$ which can
in turn be expressed in terms of $\charge_2^{(0)}$ by means of
\eqref{eq:nngen}. Similarly we can continue to higher orders in the
parameters $\alpha_k$ and will find operators that can all
be expressed in terms of the simple short-range Hamiltonian.

The next step is to consider operators proportional to the free
parameters $\beta_{r,s}$: Equation \eqref{eq:betamap} yields
\begin{equation}
\frac{\partial\charge}{\partial\beta_{r,s}}
=2i\bigcomm{\biop_{r,s}}{\charge}+\ldots
=2i\bigcomm{\biloc{\charge_r}{\charge_s}}{\charge}+\ldots\,,
\end{equation}
such that we find
\begin{equation}
\charge_t=\charge_t^{(0)}+\beta_{r,s}\charge_t^{[r|s]}+\dots,	
\end{equation}
with
\begin{equation}
\charge_t^{[r|s]}=2i\bigcomm{\biloc{\charge_r^{(0)}}{\charge_s^{(0)}}}{\charge_t^{(0)}}.
\label{eq:firstorderbi}
\end{equation}
Higher orders in the parameters $\beta_{r,s}$ as in 
\begin{equation}
\charge_t
=\charge_t^{(0)}
+\beta_{r,s}\charge_t^{[r|s]}
+\beta_{r,s}\beta_{p,q}\charge_t^{[r|s,p|q]}
+\dots,	
\end{equation}
are then given by 
\begin{equation}
\charge_t^{[r|s,p|q]}
=2i\bigcomm{\biloc{\charge_r^{[p|q]}}{\charge_s^{(0)}}}{\charge_t^{(0)}}
+2i\bigcomm{\biloc{\charge_r^{(0)}}{\charge_s^{[p|q]}}}{\charge_t^{(0)}}
+2i\bigcomm{\biloc{\charge_r^{(0)}}{\charge_s^{(0)}}}{\charge_t^{[p|q]}}\,,
\end{equation}
where again the flatness of the connection
implies that up to local similarity transformations
this term is equal to
\begin{equation}
\charge_t^{[p|q,r|s]}
=2i\bigcomm{\biloc{\charge_p^{[r|s]}}{\charge_q^{(0)}}}{\charge_t^{(0)}}
+2i\bigcomm{\biloc{\charge_p^{(0)}}{\charge_q^{[r|s]}}}{\charge_t^{(0)}}
+2i\bigcomm{\biloc{\charge_p^{(0)}}{\charge_q^{(0)}}}{\charge_t^{[r|s]}}\,.
\end{equation}
Again, we could plug in the solution for the first orders
\eqref{eq:firstorderbi} and express the whole operator in terms of
the undeformed charges, and then in terms of $\charge_2^{(0)}$ by
\eqref{eq:nngen}.

We are now only missing the terms $\charge_t^{[k,r|s]}$
proportional to mixed powers
$\alpha_k\beta_{r,s}$ of the free parameters in 
\begin{equation}
\charge_t
=\charge_t^{(0)}+\alpha_k\charge_t^{[k]}
+\alpha_k\alpha_l\charge_t^{[k,l]}
+\beta_{r,s}\charge_t^{[r|s]}
+\beta_{r,s}\beta_{p,q}\charge_t^{[r|s,p|q]}
+\alpha_k\beta_{r,s}\charge_t^{[k,r|s]}
+\dots\,.
\end{equation}
Again, we can determine these either by plugging the
$\beta_{r,s}$-deformation into the $\alpha_k$-deformation equation or
vice versa. Due to flatness, both results are equal up to local
deformations \eqref{eq:deformationepsilon}.
The $\alpha_k$-equation \eqref{eq:sf}
together with \eqref{eq:boostopfix} yields 
\begin{align}
\frac{\partial\charge}{\partial\alpha_k}
&=i\comm{\tilde\boostop_k}{\charge}+\ldots
\nln
&=(k-1)i\comm{\boostop_k}{\charge}
 +i\comm{\rotgen_k}{\charge}
 -2\beta_{r,s}\bigcomm{\biloc{\comm{\rotgen_k}{\charge}_r}{\charge_s}+\biloc{\charge_r}{\comm{\rotgen_k}{\charge}_s}}{\charge}
 +\ldots,
\end{align}
which results in 
\begin{align}
\charge_t^{[r|s,k]}
&=i(k-1)\bigcomm{\boostop_k^{[r|s]}}{\charge_t^{(0)}}
 +i(k-1)\bigcomm{\boostop_k^{(0)}}{\charge_t^{[r|s]}}
 +(t+k-2)\charge_{t+k-1}^{[r|s]}
\nln&\quad
 +2i\bigcomm{\biloc{(r+k-2)\charge_{r+k-1}^{(0)}}{\charge_s^{(0)}}
            +\biloc{\charge_r^{(0)}}{(s+k-2)\charge_{s+k-1}^{(0)}}}{\charge_t^{(0)}}\,.
\label{eq:betaalpha1}
\end{align}
The $\beta_{r,s}$-equation \eqref{eq:bilocsf} in contrast yields
\begin{equation}
\charge_t^{[k,r|s]}
=2i\bigcomm{\biloc{\charge_r^{[k]}}{\charge_s^{(0)}}}{\charge_t^{(0)}}
+2i\bigcomm{\biloc{\charge_r^{(0)}}{\charge_s^{[k]}}}{\charge_t^{(0)}}
+2i\bigcomm{\biloc{\charge_r^{(0)}}{\charge_s^{(0)}}}{\charge_t^{[k]}}\,.
\label{eq:betaalpha2}
\end{equation}
Again, due to flatness of the connection,
the expressions \eqref{eq:betaalpha1} and
\eqref{eq:betaalpha2} are equal up to local similarity transformations.

Note that the explicit expansions given in this appendix not always
yield terms with the desired interaction range \eqref{eq:lengthofterm}
straightforwardly. In order to obtain terms of this minimal range, one
generically has to add deformations \eqref{eq:deformationepsilon} by
some local operators. For example, when one explicitly calculates the
terms \eqref{eq:terma34} for the $\alg{gl}(K)$ chain, one finds that
$\charge_r^{[3,4]}$ has the minimal interaction range $(r+3)$ as given
in \eqref{eq:lengthofterm}, while the term $\charge_r^{[4,3]}$ has
range $(r+4)$. Adding a deformation
$\locform=\loc_{4,3}\alpha_4\dd\alpha_3$ by some local operator
$\loc_{4,3}$ however results in $\charge_r^{[4,3]}=\charge_r^{[3,4]}$.
Similarly, for all terms we calculated for the $\alg{gl}(K)$ chain,
\eqref{eq:betaalpha2} yields the correct range \eqref{eq:lengthofterm},
while \eqref{eq:betaalpha1} needs to be corrected by a local
deformation $\locform=\loc_{r|s,k}\beta_{r,s}\dd\alpha_k$.

For explicit results for the $\alg{gl}(K)$ and
$\alg{gl}(K_\mathrm{e})\otimes\alg{gl}(K_\mathrm{o})$ chains, see
\Tabref{tab:N4Q2} and \Tabref{tab:N6Q2}.


\section*{List of Symbols}
\addcontentsline{toc}{section}{List of Symbols}

\newcommand{\losline}[2]{#1\,&#2\\}

{\noindent\tabcolsep0pt 
\begin{tabular}{ll}
\losline{$\loc$}
	{Local (homogeneous) operator, \eqref{eq:homogeneousaction}}
\losline{$\loc_{\mathrm{e/o}}$}
	{Local operator $\loc$ acting on even/odd sites of the spin chain only}
\losline{$\range{\loc}$}
	{Interaction range of the local operator $\loc$}
\losline{$\vev{\loc}$}
	{Ferromagnetic vacuum expectation value of $\loc$, \eqref{eq:ibtb,eq:ibtbi}}
\losline{$\crossings{\loc}$}
	{Number of elementary permutations contained in $\loc$, \eqref{eq:crossings}}
\losline{$\perm{a_1,\ldots,a_k}$}
	{Permutation that acts homogeneously on the spin chain,
	\eqref{eq:local}}
\losline{$\liegen_a$}
	{Lie-symmetry generator, \Secref{sec:symgen}}
\losline{$\yang_a$}
	{Yangian generator, \Secref{sec:symgen}}
\losline{$\charge_r$}
	{Conserved charge/observable, \eqref{eq:intcond}}
\losline{$\charge_r\supups{NN}=\charge_r^{(0)}=\charge_r(0)$}
	{Undeformed nearest-neighbor charges, \eqref{eq:nngen}}
\losline{$\charge_r^{(k)}$}
	{Coefficient of $\lambda^k$ in the expansion \eqref{eq:expansionlambda}}
\losline{$\ham$}
	{Hamiltonian, $\ham=\charge_2$, \Secref{sec:nnmodels}}
\losline{$\specleg$}
	{Spectator leg, \eqref{eq:spectator}}
\losline{$\idop$}
	{Identity operator, \eqref{eq:ibtb}}
\losline{$\lengthop$}
	{Length operator, \eqref{eq:ibtb}}
\losline{$\defop$}
	{General deformation operator, \eqref{eq:deformationl}}
\losline{$\boost{\loc}$}
	{Boost of the local operator $\loc$, \eqref{eq:boostdef}}
\losline{$\boostop_k$}
	{Boost of the conserved charge $\charge_k$,
	\eqref{eq:deformationboost}}
\losline{$\tilde\boostop_k$}
	{Modified boost operator,
	\eqref{eq:boostconnwithgamma,eq:boostopfix}}
\losline{$\biloc{\loc_k}
	{\loc_\ell}$}{Bilocal operator composed of the two local operators
	$\loc_k$, $\loc_\ell$, \eqref{eq:defbiloc}}
\losline{$\biop_{r,s}$}
	{Bilocal operator composed of the two charges $\charge_r$,
	$\charge_s$, \eqref{eq:deformationbiloc}}
\losline{$\charge$}
	{Vector of conserved charges $\charge_r$, \Secref{sec:rotations}}
\losline{$\basgen_k$}
	{Basis vector in the space of conserved charges $\charge_r$,
	\Secref{sec:rotations}}
\losline{$\rotgen_{r,s}$}
	{Generator of changes of the basis of conserved charges,
	\eqref{eq:rotgen,eq:rotgenfix}}
\losline{$\defform=\defform_k\xi_k$}
	{General connection that generates deformations,
	\eqref{eq:connectall}}
\losline{$\boostform=\boostop_k\boostform_k$}
	{Connection that generates boost deformations,
	\eqref{eq:deformationall,eq:connectall}}
\losline{$\tilde\boostform=\tilde\boostop_k\tilde\boostform_k$}
	{Modified boost connection, \eqref{eq:boostconnwithgamma,eq:sf}}
\losline{$\biform=\biop_{r,s}\biform_{r,s}$}
	{Connection that generates bilocal deformations,
	\eqref{eq:deformationall,eq:connectall}}
\losline{$\rotform=\rotgen_{r,s}\rotform_{r,s}$}
	{Connection that generates changes of basis,
	\eqref{eq:deformationall,eq:connectall,eq:gammafix}}
\losline{$\locform=\loc_k\dd\varepsilon_k$}
	{Connection that generates local similarity transformations,
	\eqref{eq:deformationall,eq:connectall}}
\losline{$\cder$}
	{Covariant derivative in moduli space, \eqref{eq:covder}}
\losline{$x(u)$}
	{Rapidity map, inverse of $u(x)$, \eqref{eq:rapmap}}
\losline{$\alpha_k$}
	{Moduli parametrizing boost deformations,
	\eqref{eq:sf}}
\losline{$\beta_{r,s}$}
	{Moduli parametrizing bilocal deformations,
	\eqref{eq:betamap}}
\losline{$\gamma_{m,n}$}
	{Moduli parametrizing a basis change of the charges,
	\eqref{eq:postrot}}
\losline{$\varepsilon_{l}$}
	{Moduli parametrizing local similarity transformations,
	\eqref{eq:betamap}}
\losline{$\charge_r^{[k_1,\ldots,k_m,r_1|s_1,\ldots,r_n|s_n]}$}
	{Coefficient of $\alpha_{k_1}\ldots\alpha_{k_m}\beta_{r_1,s_1}\ldots\beta_{r_n,s_n}$ in \eqref{eq:expansionalphabeta}}
\end{tabular}}


\bibliography{boostlong}
\bibliographystyle{nb}

\begin{table}\footnotesize
\begin{align}
\charge_2(\lambda)
&=[1] -[2,1]\nn\\[7pt]
&+\alpha_3\,(-3[1] +4[2,1] -[3,2,1])\nn\\[7pt]
&+\alpha_3^2\,(20[1] -29[2,1] +10[3,2,1] -[2,3,4,1] +[2,4,1,3] +[3,1,4,2] -[4,1,2,3] -[4,2,3,1])\nln
&+\ihalf\alpha_4\,(6[2,3,1] -6[3,1,2] -[2,4,3,1] -[3,2,4,1] +[4,1,3,2] +[4,2,1,3])\nln
&+\half\beta_{23}\,(-4[1] +8[2,1] -2[2,3,1] -2[3,1,2] -2[2,1,4,3] -2[2,3,4,1] +2[2,4,1,3] +[2,4,3,1]\nln
	&\quad\quad +2[3,1,4,2] +[3,2,4,1] -2[3,4,1,2] -2[4,1,2,3] +[4,1,3,2] +[4,2,1,3])\nn\\[7pt]
&+\sfrac{1}{3}\alpha_3^3\,(-525[1] +792[2,1] -308[3,2,1] +60[2,3,4,1] -62[2,4,1,3] -62[3,1,4,2] +2[3,4,2,1]\nln
	&\quad\quad +60[4,1,2,3] +44[4,2,3,1] +2[4,3,1,2] -5[2,3,5,4,1] -2[2,4,3,5,1] +5[2,5,1,4,3]\nln
	&\quad\quad +2[2,5,3,1,4] +5[3,1,5,4,2] -5[3,2,4,5,1] +5[3,2,5,1,4] +2[4,1,3,5,2] +5[4,2,1,5,3]\nln
	&\quad\quad -5[5,1,2,4,3] -2[5,1,3,2,4] -5[5,2,1,3,4] -3[5,2,3,4,1])\nln
&+\sfrac{i}{3}\alpha_3\alpha_4\,(-138[2,3,1] +138[3,1,2] +36[2,4,3,1] +36[3,2,4,1] -36[4,1,3,2] -36[4,2,1,3]\nln
	&\quad\quad -6[2,3,4,5,1] +6[2,3,5,1,4] +6[2,4,1,5,3] -6[2,5,1,3,4] -3[2,5,3,4,1] +6[3,1,4,5,2]\nln
	&\quad\quad -6[3,1,5,2,4] -3[3,2,5,4,1] -6[4,1,2,5,3] -3[4,2,3,5,1] +6[5,1,2,3,4] +3[5,1,3,4,2]\nln
	&\quad\quad +3[5,2,1,4,3] +3[5,2,3,1,4])\nln
&+\half\alpha_5\,(10[1] -18[2,1] +10[3,2,1] -6[2,3,4,1] +6[2,4,1,3] +6[3,1,4,2] -6[4,1,2,3]\nln
	&\quad\quad -2[4,2,3,1] +[2,3,5,4,1] -[2,5,1,4,3] -[3,1,5,4,2] +[3,2,4,5,1]\nln
	&\quad\quad -[3,2,5,1,4] -[4,2,1,5,3] +[5,1,2,4,3] +[5,2,1,3,4])\nln
&+\sfrac{i}{6}\beta_{2,4}\,(4[2,3,1] -4[3,1,2] -4[2,3,4,1] +2[2,4,3,1] +2[3,2,4,1] -2[3,4,2,1] +4[4,1,2,3]\nln
	&\quad\quad -2[4,1,3,2] -2[4,2,1,3] +2[4,3,1,2] -2[2,1,4,5,3] +2[2,1,5,3,4] -2[2,3,1,5,4]\nln
	&\quad\quad -4[2,3,4,5,1] +4[2,3,5,1,4] +[2,3,5,4,1] +4[2,4,1,5,3] +2[2,4,3,5,1] -2[2,4,5,1,3]\nln
	&\quad\quad -4[2,5,1,3,4] +[2,5,1,4,3] +2[3,1,2,5,4] +4[3,1,4,5,2] -4[3,1,5,2,4] -[3,1,5,4,2]\nln
	&\quad\quad +[3,2,4,5,1] -[3,2,5,1,4] -2[3,4,1,5,2] +2[3,5,1,2,4] -4[4,1,2,5,3] +2[4,1,5,2,3]\nln
	&\quad\quad +[4,2,1,5,3] +4[5,1,2,3,4] -[5,1,2,4,3] -2[5,1,3,2,4] -[5,2,1,3,4])\nln\
&+\sfrac{1}{12}\beta_{3,4}\,(-8[1] +24[2,1] -8[3,2,1] -8[2,1,4,3] +4[2,3,4,1] -4[2,4,1,3] -2[2,4,3,1]\nln
	&\quad\quad -4[3,1,4,2] -2[3,2,4,1] +8[3,4,1,2] +4[4,1,2,3] -2[4,1,3,2] -2[4,2,1,3] -8[2,1,3,5,4]\nln
	&\quad\quad +4[2,1,5,4,3] +2[2,3,5,1,4] -2[2,4,5,3,1] +2[2,5,1,3,4] +2[2,5,3,4,1] -2[2,5,4,1,3]\nln
	&\quad\quad +2[3,1,4,5,2] +4[3,2,1,5,4] -2[3,4,2,5,1] +4[3,4,5,1,2] -4[3,5,1,4,2] -2[3,5,2,1,4]\nln
	&\quad\quad +2[4,1,2,5,3] -2[4,1,5,3,2] +2[4,2,3,5,1] -4[4,2,5,1,3] -2[4,3,1,5,2] +4[4,5,1,2,3]\nln
	&\quad\quad +2[5,1,3,4,2] -2[5,1,4,2,3] +2[5,2,3,1,4] -2[5,3,1,2,4])\nn\\[7pt]
&+\sfrac{1}{2}\alpha_3\beta_{2,3}\,(52[1] -112[2,1] +26[2,3,1] +26[3,1,2] +8[3,2,1] +36[2,1,4,3] +26[2,3,4,1]\nln
	&\quad\quad -28[2,4,1,3] -15[2,4,3,1] -28[3,1,4,2] -15[3,2,4,1] +24[3,4,1,2] +2[3,4,2,1]\nln
	&\quad\quad +26[4,1,2,3] -15[4,1,3,2] -15[4,2,1,3] +2[4,3,1,2] -2[2,1,5,4,3] +2[2,3,4,5,1]\nln
	&\quad\quad -[2,3,5,1,4] -3[2,3,5,4,1] -2[2,4,3,5,1] +[2,4,5,3,1] -[2,5,1,3,4] +3[2,5,1,4,3]\nln
	&\quad\quad +2[2,5,3,1,4] +[2,5,3,4,1] -[2,5,4,1,3] -[3,1,4,5,2] +3[3,1,5,4,2] -2[3,2,1,5,4]\nln
	&\quad\quad -3[3,2,4,5,1] +3[3,2,5,1,4] +2[3,2,5,4,1] +[3,4,2,5,1] -2[3,5,1,4,2] -[3,5,2,1,4]\nln
	&\quad\quad -[4,1,2,5,3] +2[4,1,3,5,2] -[4,1,5,3,2] +3[4,2,1,5,3] +[4,2,3,5,1] -2[4,2,5,1,3]\nln
	&\quad\quad -[4,3,1,5,2] +2[5,1,2,3,4] -3[5,1,2,4,3] -2[5,1,3,2,4] +[5,1,3,4,2] +[5,1,4,2,3]\nln
	&\quad\quad -3[5,2,1,3,4] +2[5,2,1,4,3] +[5,2,3,1,4] +[5,3,1,2,4])\nn\\
&+\order{\lambda^4}\nn
\end{align}
\caption{Long-range Hamiltonian $\charge_2$ for the $\alg{gl}(K)$ spin
chain at order $\lambda^3$ printed up to basis changes and local
deformations.}
\label{tab:N4Q2}
\end{table}

\begin{table}
\begin{align}
\charge_{2\mathrm{x}}(\lambda)
&=[1]_\mathrm{x}-[3,\gray{2},1]_\mathrm{x}\nn\\[7pt]
&+\alpha_{3\mathrm{x}}\,(-12\,[1]_\mathrm{x} +16\,[3,\gray{2},1]_\mathrm{x} +4\,[5,\gray{2},3,\gray{4},1]_\mathrm{x})\nn\\[7pt]
&+16\alpha_{3\mathrm{x}}^2\,(20\,[1]_\mathrm{x} -29\,[3,\gray{2},1]_\mathrm{x}
		+10\,[5,\gray{2},3,\gray{4},1]_\mathrm{x} -[3,\gray{2},5,\gray{4},7,\gray{6},1]_\mathrm{x}\nln
	&\quad\quad +[3,\gray{2},7,\gray{4},1,\gray{6},5]_\mathrm{x} +[5,\gray{2},1,\gray{4},7,\gray{6},3]_\mathrm{x}\nln
	&\quad\quad -[7,\gray{2},1,\gray{4},3,\gray{6},5]_\mathrm{x} -[7,\gray{2},3,\gray{4},5,\gray{6},1]_\mathrm{x})\nln
&+4i\alpha_{4\mathrm{x}}\,(6\,[3,\gray{2},5,\gray{4},1]_\mathrm{x}-6\,[5,\gray{2},1,\gray{4},3]_\mathrm{x}
		-[3,\gray{2},7,\gray{4},5,\gray{6},1]_\mathrm{x}\nln
	&\quad\quad -[5,\gray{2},3,\gray{4},7,\gray{6},1]_\mathrm{x} +[7,\gray{2},1,\gray{4},5,\gray{6},3]_\mathrm{x}
		+[7,\gray{2},3,\gray{4},1,\gray{6},5]_\mathrm{x})\nn\\[7pt]
&+\beta_{2\mathrm{x},3\mathrm{x}}(-4\,[1]_\mathrm{x} +8\,[3,\gray{2},1]_\mathrm{x} -2[3,\gray{2},5,\gray{4},1]_\mathrm{x}
		-2\,[5,\gray{2},1,\gray{4},3]_\mathrm{x} -2\,[3,\gray{2},1,\gray{4},7,\gray{6},5]_\mathrm{x}\nln
	&\quad\quad -2\,[3,\gray{2},5,\gray{4},7,\gray{6},1]_\mathrm{x} +2\,[3,\gray{2},7,\gray{4},1,\gray{6},5]_\mathrm{x}
		+[3,\gray{2},7,\gray{4},5,\gray{6},1]_\mathrm{x}\nln
	&\quad\quad +2\,[5,\gray{2},1,\gray{4},7,\gray{6},3]_\mathrm{x} +[5,\gray{2},3,\gray{4},7,\gray{6},1]_\mathrm{x}
		-2\,[5,\gray{2},7,\gray{4},1,\gray{6},3]_\mathrm{x}\nln
	&\quad\quad -2\,[7,\gray{2},1,\gray{4},3,\gray{6},5]_\mathrm{x} +[7,\gray{2},1,\gray{4},5,\gray{6},3]_\mathrm{x}
		+[7,\gray{2},3,\gray{4},1,\gray{6},5]_\mathrm{x})\nn\\[10pt]
&+2i\beta_{2\mathrm{x},2\mathrm{y}}([3,\gray{4},5,\gray{2},1]_\mathrm{x} -[3,\gray{2},5,\gray{4},1]_\mathrm{x}
		+[5,\gray{2},1,\gray{4},3]_\mathrm{x} -[5,\gray{4},1,\gray{2},3]_\mathrm{x})\nn\\[7pt]
&+\beta_{2\mathrm{x},3\mathrm{y}}\,(-[3,\gray{4},5,\gray{6},1,\gray{2}]_\mathrm{x} +[3,\gray{6},5,\gray{2},1,\gray{4}]_\mathrm{x}
		+[5,\gray{4},1,\gray{6},3,\gray{2}]_\mathrm{x} -[5,\gray{6},1,\gray{2},3,\gray{4}]_\mathrm{x}\nln
	&\quad\quad -[\gray{3},4,\gray{5},6,\gray{1},2]_\mathrm{y} +[\gray{3},6,\gray{5},2,\gray{1},4]_\mathrm{y}
		+[\gray{5},4,\gray{1},6,\gray{3},2]_\mathrm{y} -[\gray{5},6,\gray{1},2,\gray{3},4]_\mathrm{y})\nln
&+\beta_{2\mathrm{y},3\mathrm{x}}\,(-2\,[1]_\mathrm{x} +4\,[3,\gray{2},1]_\mathrm{x} -2\,[3,\gray{4},1,\gray{2}]_\mathrm{x}
		-2\,[3,\gray{2},5,\gray{4},7,\gray{6},1]_\mathrm{x}\nln
	&\quad\quad +[3,\gray{2},5,\gray{6},7,\gray{4},1]_\mathrm{x} +2\,[3,\gray{2},7,\gray{4},1,\gray{6},5]_\mathrm{x}
		-[3,\gray{2},7,\gray{6},1,\gray{4},5]_\mathrm{x}\nln
	&\quad\quad +[3,\gray{4},5,\gray{2},7,\gray{6},1]_\mathrm{x} -[3,\gray{4},7,\gray{2},1,\gray{6},5]_\mathrm{x}
		+2\,[5,\gray{2},1,\gray{4},7,\gray{6},3]_\mathrm{x}\nln
	&\quad\quad -[5,\gray{2},1,\gray{6},7,\gray{4},3]_\mathrm{x} -[5,\gray{4},1,\gray{2},7,\gray{6},3]_\mathrm{x}
		-2\,[7,\gray{2},1,\gray{4},3,\gray{6},5]_\mathrm{x}\nln
	&\quad\quad +[7,\gray{2},1,\gray{6},3,\gray{4},5]_\mathrm{x} +[7,\gray{4},1,\gray{2},3,\gray{6},5]_\mathrm{x}
		-2\,[\gray{1}]_\mathrm{y} +4\,[\gray{3},2,\gray{1}]_\mathrm{y} -2\,[\gray{3},4,\gray{1},2]_\mathrm{y})\nln
&+2i\beta_{3\mathrm{x},3\mathrm{y}}\,(-[3,\gray{4},5,\gray{6},7,\gray{2},1]_\mathrm{x}
		+[3,\gray{4},7,\gray{6},1,\gray{2},5]_\mathrm{x} +[3,\gray{6},5,\gray{2},7,\gray{4},1]_\mathrm{x}\nln
	&\quad\quad -[3,\gray{6},7,\gray{2},1,\gray{4},5]_\mathrm{x} +[5,\gray{4},1,\gray{6},7,\gray{2},3]_\mathrm{x}
		-[5,\gray{6},1,\gray{2},7,\gray{4},3]_\mathrm{x}\nln
	&\quad\quad -[7,\gray{4},1,\gray{6},3,\gray{2},5]_\mathrm{x} +[7,\gray{6},1,\gray{2},3,\gray{4},5]_\mathrm{x}
		-2\,[\gray{3},2,\gray{5},4,\gray{1}]_\mathrm{y} \nln
	&\quad\quad +2\,[\gray{3},4,\gray{5},2,\gray{1}]_\mathrm{y} +2\,[\gray{5},2,\gray{1},4,\gray{3}]_\mathrm{y}
		-2\,[\gray{5},4,\gray{1},2,\gray{3}]_\mathrm{y})\nn\\[7pt]
&+\order{\lambda^6}\nn
\mspace{430mu}\mathrm{x}\neq\mathrm{y}\in\{\mathrm{e},\mathrm{o}\}
\end{align} 
\caption{Long-range Hamiltonian for the alternating spin chain printed
up to basis changes and local deformations. The first part gives two
stretched copies of the $\alg{gl}(K)$ Hamiltonian
(\protect\Tabref{tab:N4Q2}) acting on odd and even spin sites
respectively. The second part contains novel interaction symbols which
act nontrivially on the odd and even copy of the chain at the same
time. The Hamiltonian of the integrable spin chain supposedly describing the
$\alg{su}(2)\times\alg{su}(2)$ sector of $\superN =6$ superconformal
Chern--Simons theory is given by a specific choice of the free parameters.}
\label{tab:N6Q2}
\end{table}

\end{document}